\documentclass[fleqn,usenatbib,useAMS]{mnras}
\usepackage{graphicx}
\usepackage{amsmath}
\usepackage{amssymb}
\usepackage{multicol} 
\usepackage{bm}	
\usepackage{pdflscape}
\usepackage{pdfpages}
\usepackage{caption}
\usepackage{floatrow}
\usepackage{paralist}
\usepackage{verbatim}
\usepackage{caption}
\usepackage{subcaption}
\usepackage{float}
\usepackage{rotating}
\usepackage{times}
\usepackage[T1]{fontenc}
\usepackage{ae,aecompl}
\usepackage{newtxtext,newtxmath}

\newcommand{\vect}[1]{\vec{#1}}

\newcommand{\lorb}{\cal{L}_\mathrm{orb}}
\newcommand{\lmod}{\cal{L}_\mathrm{mod}}
\newcommand{\ldat}{\cal{L}_\mathrm{data}}

\newcommand{\norb}{N_\mathrm{orbit}}
\newcommand{\nkin}{N_\mathrm{data}}
\newcommand{\nbin}{N_\mathrm{losvd}}
\newcommand{\nphot}{N_\mathrm{phot}}
\newcommand{\nvel}{N_\mathrm{vlos}}

\newcommand{\kdat}{\vect{\cal{L}}_\mathrm{data}}
\newcommand{\kmod}{\vect{\cal{L}}_\mathrm{mod}}
\newcommand{\korb}{\cal{L}_\mathrm{orb}}

\newcommand{\phdat}{\vect{p}_\mathrm{data}}
\newcommand{\phmod}{\vect{p}_\mathrm{mod}}
\newcommand{\phorb}{P_\mathrm{orb}}

\title[Triaxial Orbit Models]{\textit{SMART}: A new implementation of Schwarzschild's Orbit Superposition technique for triaxial galaxies and its application to an N-body merger simulation}

\author[B. Neureiter]{B. Neureiter$^{1}$\thanks{E-mail: \href{mailto:mn@ras.org.uk}{bneu@mpe.mpg.de}}
\\
$^{1}$Royal Astronomical Society, Burlington House, Piccadilly, London W1J 0BQ, UK}

\author[B. Neureiter et al.]{
B. Neureiter$^{1,2}$\thanks{E-mail:\href{mailto:bneu@mpe.mpg.de}{bneu@mpe.mpg.de}},
J. Thomas$^{1,2}$,
R. Saglia$^{1,2}$,
R. Bender$^{1,2}$,
F. Finozzi,
A. Krukau$^{3,4}$,
\newauthor
T. Naab$^{5}$,
A. Rantala$^{5}$
and M. Frigo$^{5,6}$
\\
$^{1}$Max-Planck-Institut f\"ur Extraterrestriche Physik, Giessenbach-Str. 1, D-85748, Garching, Germany\\
$^{2}$Universit\"ats-Sternwarte M\"unchen, Scheinerstrasse 1, D-81679 M\"unchen, Germany \\
$^{3}$Leibniz-Rechenzentrum (LRZ), Boltzmannstrasse 1, D-85748 Garching, Germany \\
$^{4}$Excellence Cluster Universe, Boltzmannstrasse 2r, D-85748 Garching, Germany \\
$^{5}$Max-Planck-Institut f\"ur Astrophysik, Karl-Schwarzschild-Str. 1, D-85748, Garching, Germany \\
$^{6}$Excellence Cluster ORIGINS, Boltzmannstrasse 2, D-85748 Garching, Germany}
\date{Last updated XXX; in original form YYY}
\pubyear{2020}

\begin{document}
\label{firstpage}
\pagerange{\pageref{firstpage}--\pageref{lastpage}}
\maketitle

\begin{abstract}
We present \texttt{SMART}, a new 3D implementation of the Schwarzschild Method and its application to a triaxial $N$-body merger simulation. \texttt{SMART} fits full line-of-sight velocity distributions (LOSVDs) to determine the viewing angles, black hole, stellar and dark matter (DM) masses and the stellar orbit distribution of galaxies. Our model uses a 5D orbital starting space to ensure a representative set of stellar trajectories adaptable to the integrals-of-motion space and it is designed to deal with non-parametric stellar and DM densities. \texttt{SMART}'s efficiency is demonstrated by application to a realistic $N$-body merger simulation including supermassive black holes which we model from five different projections. When providing the true viewing angles, 3D stellar luminosity profile and normalized DM halo, we can (i) reproduce the intrinsic velocity moments and anisotropy profile with a precision of $\sim1\%$ and (ii) recover the black hole mass, stellar mass-to-light ratio and DM normalization to better than a few percent accuracy. 
This precision is smaller than the currently discussed differences between initial-stellar-mass functions and scatter in black hole scaling relations. 
Further tests with toy models suggest that the recovery of the anisotropy in triaxial galaxies is almost unique when the potential is known and full LOSVDs are fitted.
We show that orbit models even allow the reconstruction of full intrinsic velocity distributions, which contain more information than the classical anisotropy parameter. Surprisingly, the orbit library for the analysed $N$-body simulation's gravitational potential contains orbits with net rotation around the intermediate axis that is stable over some Gyrs.
\end{abstract}

\begin{keywords}
galaxies: elliptical and lenticular, cD -- stars: kinematics and dynamics -- galaxies: evolution -- galaxies: structure -- galaxies: supermassive black holes -- methods: numerical 
\end{keywords}

\begingroup
\let\clearpage\relax
\endgroup
\newpage

\section{Introduction}
Early-type galaxies (ETGs) have long been believed to emerge from collisions between other smaller progenitor galaxies (first proposed by \citealt{Toomre72}) but nowadays it is clear that their formation history is more complex \citep[e.g.][]{Oser10}. 
Their structural and kinematic properties divide them into (1) fainter (absolute magnitude $M_B > -20.5$) and coreless fast rotators which are nearly axisymmetric and have discy-distorted isophotes and (2) brighter and more massive slow rotators with flat cores, which are moderately triaxial and have boxy-distorted isophotes 
(\citealt{Faber87,Bender88_a,Bender89,Kormendy96,Cappellari07,Emsellem07}).
For the formation of fainter elliptical galaxies dissipational processes are believed to be important (e.g. \citealt{Bender92,Barnes96,Genzel01,Tacconi05,Cappellari07,Hopkins08, Johansson09}), whereas the latest evolutionary phases in the formation of massive ellipticals are dominated by collisionless processes \citep[e.g.][]{Naab06,Cappellari16,Naab17,Moster19}. \\
In general, these merging events modify the potential structure and populate a rich diversity of stellar orbits \citep{Roettgers14}. The intrinsic shape and orbital structure of such galaxies are not directly observable. Instead, sophisticated dynamical models are needed to process kinematic and photometric observational data to extract all the information about the orbital structure and internal composition of the galaxy. \\
Dynamical models are based on the collisionless Boltzmann equation which governs the motion of stars in elliptical galaxies. Dynamical models that go beyond the recovery of velocity moments and aim at reconstructing the entire galaxy structure, additionally take advantage of the Jeans theorem \citep[e.g.][]{Binney08}. This implies that the distribution function, which is the most general description of a system of stars, is constant along individual trajectories in phase-space.
In this regard, \citet{Schwarzschild79} pioneered an orbit superposition technique, where the equations of motion are numerically integrated for a finite number of stellar trajectories embedded in an assumed gravitational potential with contributions from the stars and possibly dark components. The weighted superposition of the orbits is determined for which the surface brightness and projected velocity distributions of the model match the observed ones in a least squares sense \citep[e.g.][]{Richstone84}. Besides the orbital weights, all unknown quantities like the central black hole mass and dark matter distribution are varied between different models. The model producing the best fit to the projected velocity distributions is then associated with the correct model parameters.
Any galaxy in a steady state can be modeled by Schwarzschild’s orbit superposition technique.
In order to determine both, the mass and internal motions of the stars, and solve an underlying mass-anisotropy-entanglement one needs to describe the deviation of the observed absorption lines from a Gaussian profile by additional Gauss-Hermite functions of at least third and fourth order (\citealt{Gerhard93,vanderMarel93,Bender94}), or, preferably if the signal-to-noise ration permits, measure the line shape non-parametrically (e.g. \citealt{Mehrgan19}). \\
Early applications of Schwarzschild's orbit superposition technique concentrated on spherical models (e.g. \citealt{Richstone85,Rix97}). Since the simplified assumption of spherical symmetry is not true for most galaxies, later applications of Schwarzschild's orbit superposition technique assumed axisymmetry (e.g. \citealt{vanderMarel98,Cretton99,Gebhardt00,Thomas04,Valluri04}). \\
However, it is nowadays known that the most massive galaxies are neither spherical nor axisymmetric but triaxial objects.
Observational indications are provided by isophotal twists in the surface brightness distribution, velocity anisotropy, minor axis rotation, kinematically decoupled cores and the statistical distribution of the ellipticity of the isophotes 
(\citealt{Illingworth77,Bertola78,Bertola79,Williams79,Schechter78,Bender88_b,Franx88,Vincent05}). 
 \citet{Schwarzschild79} proved the existence of self-consistent triaxial stellar systems in dynamical equilibrium with numerical orbit superposition models. 
 Also, $N$-body simulations supported the idea of triaxial ellipsoidal stellar bulges and dark matter halos (e.g. \citealt{Aarseth78,Hohl79,Miller79,Barnes92,Jing02,Naab03,Bailin05}). \\
High mass galaxies are of particular interest in several astrophysical aspects, e.g., the stellar initial mass function (IMF) is discussed to vary among galaxies, with the largest excess stellar mass compared to the locally measured Kroupa \citep{Kroupa01} or Chabrier \citep{Chabrier03} IMF occuring in the most massive galaxies (e.g. \citealt{vanDokkum10,Cappellari12,Vazdekis15,Parikh18,Thomas11,Posacki15,Treu10,Smith15}). Moreover, different growth models for supermassive black holes (SMBH) predict different amounts of intrinsic scatter at the high-mass end of SMBH scaling relations (e.g. \citealt{Peng07,Jahnke11,Somerville15,Naab17}). 
In order to address these questions precision dynamical mass measurements of the stars and SMBHs are required and these are directly linked to triaxial modeling to avoid artificial scatter introduced by wrong symmetry assumptions. 
\citet{Thomas07} showed that the stellar mass-to-light-ratio (and, thus, indirect inferences about the IMF) can be biased by up to 50\% in extreme cases when using axisymmetric models for a maximally triaxial galaxy. Moreover, a wrongly assumed mass-to-light ratio influences the determination of the mass of the central black hole in the model. The work by \citet{vandenBosch10} suggests that the assumption of axisymmetry may bias black-hole measurements in massive ellipticals. 
They find that the best-fitting black hole mass estimate doubles when modeling NGC 3379 with their triaxial code \citep{vandenBosch08} in comparison to axisymmetric models. Triaxial dynamical modeling routines are therefore required to recover unbiased stellar mass-to-light ratios and black hole masses with the best possible accuracy.\\
To understand the uncertainties and ambiguities of triaxial modeling one has to understand the following three essentially different effects:
\begin{itemize}
	\item[ 1)] the intrinsic uncertainty of the applied dynamical modeling algorithm, which can only be tested under circumstances where the solution is designed to be unique. This is one of the aspects covered in this paper.
	\item[ 2)] the uncertainty in the reconstruction of orbital and mass parameters from typical observational data given the right deprojection, which is also addressed in this paper,
	\item[ 3)] the uncertainty of the deprojection routine. This topic is covered in \citealt{deNicola20}.
\end{itemize}	
All previously described effects need to be combined to evaluate the uncertainties in the whole modeling process. This will be investigated in a future paper. \\

So far, there exist two dynamical modeling codes using Schwarzschild's orbit superposition technique dealing with triaxiality by \citet{vandenBosch08} and \citet{Vasiliev19}. Their estimated precision and efficiency will be later mentioned in the discussion. 
In this paper, we present our newly developed triaxial Schwarzschild code called \verb'SMART' ("Structure and MAss Recovery of Triaxial galaxies") and test the code on a realistic high-resolution numerical merger simulation including supermassive black holes by \citet{Rantala18}. \\
This paper is structured as follows: We introduce \texttt{SMART} and describe its specific benefits in Section~\ref{sec:Triaxial Schwarzschild code SMART}. We then discuss the most important aspects for choosing this particular simulation in Section~\ref{sec:The N-body simulation}. In this section we furthermore explain all relevant steps to extract the data needed for modeling the simulation. In Section~\ref{sec:Results} we will show the results of these models. Section~\ref{sec:The quasi-uniqueness of the anisotropy reconstruction when fitting full LOSVDs} deals with the quasi-uniqueness of the anisotropy recovery when fitting full line-of-sight velocity distributions. This is followed by a short discussion about remaining sources of systematics and comparison to other triaxial modeling codes in Section~\ref{sec:Discussion}. We summarize our results and conclusions in Section~\ref{sec:Summary and Conclusion}.

\section{Triaxial Schwarzschild code \texttt{SMART}}
\label{sec:Triaxial Schwarzschild code SMART}
\begin{figure*}
    \centering
    \includegraphics[width=0.8\textwidth]{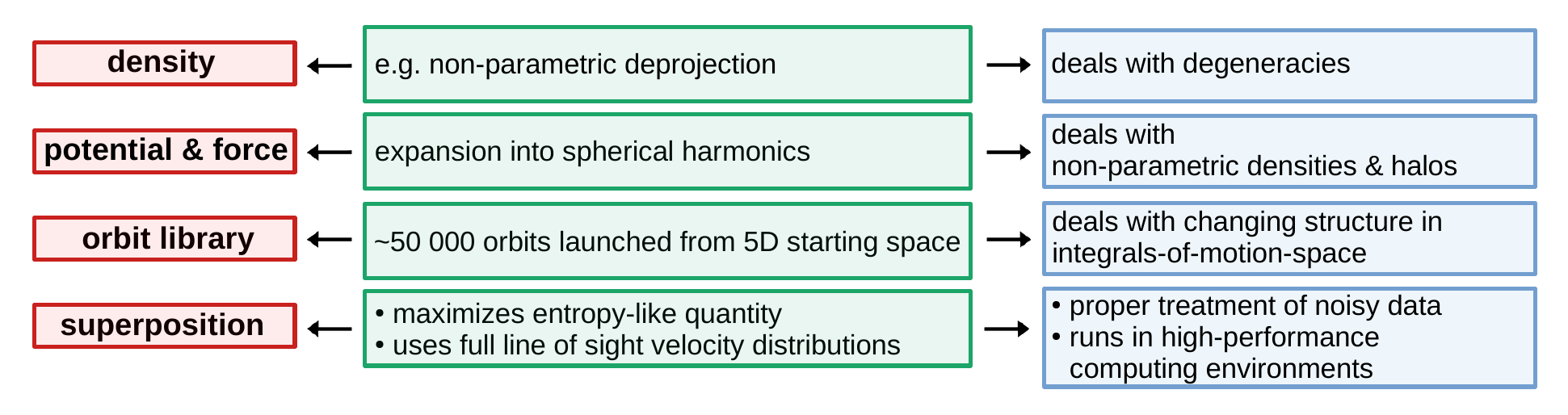}
    \caption{Our implemented triaxial Schwarzschild Modeling code \texttt{SMART} follows the classical Schwarzschild Modeling routine (red panels) but it is unique in calculating the potential by expansion into spherical harmonics, setting up an adaptive orbit library and computing the orbit superposition by maximizing an entropy-like quantity (green panels). This results in specific advantages (blue panels), e.g. that our code is able to deal with realistic changes in the gravitational potential, e.g. when the SMBH causes a more spherical potential in the center. Our orbit library is adaptive and responds to changes in the integrals of motion. \texttt{SMART} can process density output from any deprojection routine, e.g. a non-parametric deprojection dealing with degeneracies.}
    \label{fig:Figure1}
\end{figure*}
\verb'SMART' ("Structure and MAss Recovery of Triaxial galaxies") is a fully 3D orbit superposition code based on the axisymmetric code of \citet{Thomas04} and its original extension to three dimensions and non-axisymmetric densities by \citet{Finozzi18}. It is written in FORTRAN 90/95 \citep{Brainerd96}. 
\texttt{SMART} follows the classical Schwarzschild method consisting of the computation of the potential and forces for a given density, the setting up of an orbit library and the subsequent superposition of the orbits. \\
In the following sections we will explain in more detail how \texttt{SMART} creates self-consistent density-, potential- and orbit-configurations and how we weight the orbits in order to fit the input density and velocity structure. One main feature of our code is that it uses a five dimensional starting space for the orbit library to adapt to potentials with a radially varying structure of the integrals-of-motion space. 
Fig.~\ref{fig:Figure1} gives a schematic overview of the code's main modules and the benefits of their specific implementation.

\subsection{Coordinate systems and binning}
\label{sec:Coordinate systems and binning}
We use two different coordinate systems to describe the intrinsic and projected properties of a galaxy. To transform between the intrinsic coordinates ($x$, $y$, $z$), adapted to the symmetry of the object, and the coordinates ($x^{\prime}$, $y^{\prime}$, $z^{\prime}$) adapted to the sky projection, two matrices $P$ and $R$ are used: 
\begin{equation}
\left(\begin{array}{l}{x^{\prime}} \\ {y^{\prime}} \\ {z^{\prime}}\end{array}\right)=R \cdot P \cdot\left(\begin{array}{l}{x} \\ {y} \\ {z}\end{array}\right),
\end{equation}
with
\begin{equation}
\label{eq:R matrix}
R =\left(\begin{array}{ccc}{\sin \psi} & {-\cos \psi} & {0} \\ {\cos \psi} & {\sin \psi} & {0} \\ {0} & {0} & {1}\end{array}\right)
\end{equation}
and
\begin{equation}
P =\left(\begin{array}{ccc}{-\sin \varphi} & {\cos \varphi} & {0} \\ {-\cos \vartheta \cos \varphi} & {-\cos \vartheta \sin \varphi} & {\sin \vartheta} \\ {\sin \vartheta \cos \varphi} & {\sin \vartheta \sin \varphi} & {\cos \vartheta}\end{array}\right).
\end{equation}
$P$ and its corresponding viewing angles $\vartheta$ and $\varphi$ project to the plane of the sky with $z^{\prime}$ being the line of sight. $R$ and its corresponding rotation angle $\psi$ rotate the coordinates $x^{\prime}$ and $y^{\prime}$ in the plane of the sky along $z^{\prime}$. If not stated otherwise, the intrinsic long axis is hereafter assumed to coincide with $x$, the intrinsic intermediate axis with $y$ and the intrinsic short axis with $z$.

\texttt{SMART} works with a cell structure based on spherical coordinates. Intrinsic properties like the stellar or dark matter distribution or individual orbital properties are integrated over small cells in configuration and/or velocity space. We use a linear sampling for the longitude $\theta \in [-90^{\circ},90^{\circ}]$ and the azimuth $\phi \in [0^{\circ},360^{\circ}]$. Radial bins are spaced in even intervals of the radial binning index 
\begin{equation}
    \label{eq:radbin_cval_celz}
    i_r=\frac{1}{a} \log \left(c+\frac{a}{b} r\right).
\end{equation} 
The constant $c$ allows to adapt the central binning scheme from logarithmic ($c=0$) to linear ($c=1$). The constants $a$ and $b$ are determined once the radial extent of the library $r_\mathrm{min} \le r \le r_\mathrm{max}$ and the number of radial bins $N_r$ are set and they are chosen so that the minimum radius $r_\mathrm{min}$ lies within the first radial bin and the maximum radius $r_\mathrm{max}$ in the last one \citep[see also ][]{Siopis09}. \\
Similar to the spatial properties, the line-of-sight velocity distributions (LOSVDs) are integrated over small cells in phase space given by the spatial pattern of the observations (e.g. Voronoi bins; \citealt{Cappellari03}) and the velocity resolution of the LOSVD data. Like its axisymmetric predecessor \citep{Thomas04}, the code uses the entire information contained in the full LOSVDs. See Sections~\ref{sec:Spatial binning} and~\ref{sec:Kinematic Data and velocity binning} for more details.

\subsection{Density and potential}
The total gravitational potential
\begin{equation}
\Phi=\Phi_{*}+\Phi_{\mathrm{DM}}+\Phi_{\mathrm{SMBH}}
\end{equation}
is composed as the sum of a Keplerian contribution from a super-massive black hole ($\Phi_{\mathrm{SMBH}}$) and the contributions from the stars ($\Phi_{\mathrm{*}}$) and dark matter ($\Phi_{\mathrm{DM}}$). \texttt{SMART} allows to use non-parametric densities for both the stars and the dark matter. The stellar density is generally assumed to be provided in 3D tabulated form (as for example returned from a non-parametric deprojection, e.g. \citealt{deNicola20}). The same holds for the dark matter halo. However, the code can also run with quasi-parametric deprojections (e.g. Multi Gaussian Expansion or MGE models, \citealt{Monnet92,Emsellem94,Cappellari02}). It can also run with parametric dark matter halos (e.g. NFW profiles; \citealt{Navarro96}). \\
The solution to the Poisson equation is obtained with the help of an expansion in spherical harmonics \citep{Binney08}.
For this, the stellar and dark matter density are individually interpolated by first performing a bi-linear interpolation among the elevation and azimuthal angle bins and afterwards a linear interpolation among the logarithm of the radial bins. For integrating these interpolated densities we use a 10-point Gaussian quadrature algorithm from \citet{Press07}. 
The advantage of calculating the potential by expansion into spherical harmonics in comparison to other techniques, e.g. by using the MGE method, is its ability to deal with non-parametric densities and halos.

\subsection{Orbit library}
Every orbit in a gravitational potential is uniquely defined by its integrals of motion (e.g.~\citet{Binney08}). The number of (isolating) integrals of motion depends on the given potential. Furthermore, every integral of motion reduces the dimensionality of the trajectories of the stars in the galaxy. Regular potentials admit in general three integrals of motion, one of which being the energy $E$. In the axisymmetric case another classical integral is known explicitly: the z-component of the angular momentum, $L_z$. The third integral, $I_3$, is usually not known explicitly. In the general triaxial case, only the energy is given explicitly. Near the central SMBH, however, the potential becomes more and more Keplerian and the number of isolating integrals of motion in a Keplerian potential is five. An example for a system in an almost Keplerian potential that is described by a 5-integral distribution function is the asymmetric disc in the center of M31 (\citealt{Bender05,Brown13}). Since our code is not restricted to axisymmetric or triaxial symmetries, we aim for a 5D starting space for the stellar orbits, which we gain by systematically sampling $E$, $L_z$, $v_r$, $r$ and $\phi$. The details of the initial conditions sampling technique are described in Section~\ref{sec:Orbital Representation of the Phase Space}. \\
In total, \texttt{SMART} sets up and integrates $\sim50000$ orbits for 100 surfaces of section (SOS) crossings, i.e. for 100 crossings of the equatorial plane in upwards direction (for a more detailed description of SOS see App.~\ref{sec:Orbital Representation of the Phase Space}). 
This number of orbits was intentionally chosen to be higher than in the axisymmetric predecessor code of \citet{Thomas04} to address all necessary complexities given by, e.g., a radially changing structure in the integrals-of-motion space. Moreover, this amount of orbits proves to be sufficient to directly recover the (phase-space) density without any dithering of orbits in the starting space. \\
If the potential at hand is, e.g., axisymmetric, then all orbits conserve $L_z$ and precess around the rotation axis. In this case, the orbits will fill the $\phi$ dimension automatically. Likewise, orbits will be represented by invariant curves in the ($r$,$v_\theta$)-plane, due to the conservation of $I_3$. Hence, when sequentially sampling the orbital launch conditions, the dimension of the submanifold containing all the orbital initial conditions that are not yet represented shrinks automatically, according to the number of integrals of motion provided by the gravitational potential under study. \\
Since, in general, triaxial potentials have three integrals of motion, a 2D starting space at a given energy (\citealt{deZeeuw85,Schwarzschild93}) would provide a sufficient orbit sampling: 
One could sample initial conditions from the (x,z)-plane producing mainly tube orbits and compensate this with launching additional box orbits from the equipotential surfaces \citep{vandenBosch08}. However, it is not clear whether the distribution function of realistic triaxial galaxies requires a 5D starting space near the SMBH in the center. With our choice of a 5D starting space we guarantee that our set of orbits adapts to the actual complexity of the integrals-of-motion space. In a realistic triaxial galaxy, like in the studied simulation, it changes from a more spherical center (requiring at least four integrals of motion) into nearly prolate outskirts. Furthermore, it allows us to model systems like eccentric discs with distribution functions that obviously depend on more than three integrals of motion. In Fig.~\ref{fig:Figure23} we show that our implemented orbit sampling and integration routine (cf. Section~\ref{sec:Orbit integration and classification}) yields a homogeneous and dense coverage of phase space.

\subsubsection{Orbit integration and classification}
\label{sec:Orbit integration and classification}
\texttt{SMART} integrates the orbital equations of motions $ \frac{d\vec{v_{i}}}{dt} = - \vec{\nabla} \phi(\vec{x_i})$, where $i$ denotes the orbit index, in cartesian coordinates by means of the Cash-Karp algorithm \citep{Cash90}. The 5th order Runge-Kutta method is implemented by using an adaptive integration step-size (see \citealt{Press07}). The default integration time of the individual orbits corresponds to 100 SOS-crossings. \\ 
At each integrated time-step the contribution of orbit i to the luminosity, internal velocity moments and projected LOSVDs is calculated as the fraction of time the orbit spends in the corresponding bins. Projected quantities are  convolved with the relevant PSF (point spread function) in every time step and before binning. The PSF can either be provided as a parametrised two dimensional Gaussian or in terms of a PSF image. The convolution is performed via Monte-Carlo method by randomly perturbing the coordinates $x^{\prime}(t),y^{\prime}(t)$ (cf. section~\ref{sec:Coordinate systems and binning}) in dependence of the respective PSF. \\
Modeling a galaxy with \texttt{SMART} does not require an exact orbit classification analysis. However, we built in an approximate classification method. For this, \texttt{SMART} checks the sign conservation of the angular momentum in x-, y- and z-direction for every SOS-crossing event. If the sign of $L_x$ is conserved for the whole integration time and if this is not true for $L_y$ and $L_z$, the orbit gets classified as x-tube. The same applies to the other directions (cf.~\citealt{Barnes92}). If the 100 SOS crossings do not hold an angular momentum sign conservation along any direction, the orbit gets classified as box/chaotic orbit. If the sign conservation is true for every direction or if the orbit shows no radial and azimuthal change during the integration time, the orbit is classified as spherical/Kepler orbit. \\

\subsection{Orbit superposition}
\label{sec:Orbit superposition}
The orbital weights $w_i$, which are decisive for the consistency between the observed
and modeled luminosity as well as for the projected velocity profiles, are iteratively
changed until the difference $\chi^2$ between the observed LOSVDs $\ldat$ and modeled LOSVDs $\lmod$ is minimal:
\begin{equation}
    \begin{aligned} \chi^{2}=& \sum_{j^{\prime}}^{N_{\text {losvd }}} \sum_{k}^{N_{\text {vlos }}}\left(\frac{{\ldat}^{j^{\prime} k}-{\lmod}^{j^{\prime} k}}{\Delta {\ldat}^{j^{\prime} k}}\right)^{2}\end{aligned}.
    \label{chi squared}
\end{equation}
Here, $j^{\prime}$ describes the spatial bin index of the $N_{\text {losvd }}$ data cells and $k$ describes the velocity bin index of the $N_{\text {vlos }}$ velocity bins. $\Delta {\ldat}^{j^{\prime} k}$ is the error of the data in the specific bin. An advantage of \texttt{SMART} is that it uses the full information contained in the LOSVDs and not only the Gauss-Hermite parameters alone (cf., e.g., \citealt{Mehrgan19} for a discussion of the benefits of using non-parametric LOSVDs in measuring galaxy masses).
The luminosity density serves as a boundary condition for the choice of the orbital weights. \\
The problem of solving for the weights $w_i$ is usually underdetermined because the number of orbits is much larger than the number of data points. We therefore regularize our models by maximizing an entropy-like quantity
\begin{equation}
\label{eq:costfunc}
\hat{S} \equiv S - \alpha \, \chi^2,
\end{equation}
where
\begin{equation}
\label{eq:smoothing}
S = - \sum_i w_i 
\ln \left( \frac{w_i}{\omega_i} \right).
\end{equation}

In the absence of any other constraints, the entropy maximisation yields $w_i \propto \omega_i$ (cf. Section~\ref{sec:The quasi-uniqueness of the anisotropy reconstruction when fitting full LOSVDs}).
Thus, the $\omega_i$ are bias factors for the orbital weights $w_i$ and can be used to smooth the orbit model. Moreover, they can be used to construct orbit models with specific properties, e.g. orbit models dominated by certain families of orbits (cf. Section~\ref{sec:The quasi-uniqueness of the anisotropy reconstruction when fitting full LOSVDs} for examples).

The particular form of the entropy in Equation~\ref{eq:smoothing}
guarantees the positivity of the orbital weights. The maximum-entropy
technique is flexible, however.  Other choices for the entropy allow
for negative weights as well (\citet{Richstone88}).

Technically, for each regularization value $\alpha$, \texttt{SMART} maximises $\hat{S}$ by computing the relevant Lagrange multipliers. The iterative adjustment of the $w_i$'s is performed by using Newton's method. The implemented method is based on \cite{Richstone88}. A detailed description of the algorithm as well as tests demonstrating the high accuracy performance of \texttt{SMART} can be found in Appendix~\ref{sec:Optimization Algorithm and Testing}.

As we will describe in full detail in Section~\ref{sec:The quasi-uniqueness of the anisotropy reconstruction when fitting full LOSVDs}, the entropy term in equation \ref{eq:costfunc} makes the solution of the orbital weights unique. While this might be advantageous from an algorithmic point of view it comes in principle with the danger of a potential bias. As we will show, by varying the orbital bias factors $\omega_i$ the maximum-entropy technique allows in principle to reconstruct any of the potentially degenerate solutions for the orbital weights (cf. Section~\ref{sec:The quasi-uniqueness of the anisotropy reconstruction when fitting full LOSVDs}). However, the results of the following Sections imply that when fitting all the information contained in the full LOSVDs, the remaining degeneracies in the weight reconstruction have only little impact on the 'macroscopic' galaxy parameters of interest, like the anisotropy for example.

One natural choice for the orbital bias factors is $\omega_i = V_i$, where $V_i$ is the phase-space volume represented by orbit $i$  (cf. e.g. \citealt{Richstone88,Thomas04}). With this choice
\begin{equation}
    S= - \sum_{i} w_{i} \ln \left(\frac{w_{i}}{V_{i}}\right) = - \int f \ln (f) d^{3} r d^{3} v
\end{equation}
equals the Boltzmann entropy.
Since the Boltzmann entropy increases during dissipationless evolutionary processes due to phase mixing and violent relaxation, galaxy models with large Boltzmann entropy are more likely than those with a small one (\citealt{Richstone88,Thomas07}). However, in collisionless self-gravitating systems, every entropy-like functional is assumed to increase in phase-space, such as the generalized H-function \citep{Tremaine86}, the entropy of the ideal gas \citep{White87} or the Tsallis entropy \citep{Tsallis88}. The choice of $\omega_i$ is thus arbitrary to some degree. This and the fact that in case of a triaxial potential it is computationally expensive to calculate the correct phase-space volume $V_i$ for every orbit motivated us to set
\begin{equation}
    \omega_{i}= \mathrm{const}. = 1.
\end{equation}
This functional form was also tested by \citet{deLorenzi07} in a slightly different context of a made-to-measure (M2M; \citealt{Syer96,Bissantz04}) algorithm for 
$N$-body particle models.
Compared to the Boltzmann entropy, $\omega_i = \mathrm{const.}$ leads to relative preference of orbits with actually small $V_i$ while orbits with large $V_i$ are relatively suppressed.
With this choice of constant orbital bias factors $\omega_i$, the entropy equation (cf. eq. ~\ref{eq:smoothing}) resembles the Shannon entropy and yields the least "informed" set of orbital weights.

\subsection{Mass Optimisation with \texttt{SMART}}
\verb'SMART' is conceived to determine the viewing angles and the mass components, like the dark matter halo, the stellar mass-to-light ratio and black hole mass by looking for the model with the smallest $\chi^2$. 
To deal with this multi-dimensional parameter space, \verb'SMART' uses NOMAD (Nonlinear Optimisation by Mesh Adaptive Direct search), a software optimised for time-consuming constrained black-box optimisations (\citealt{Audet06,LeDigabel11}). NOMAD is able to optimise a noisy function with unknown derivatives to converge to the best fitting model by using a direct-search scheme.
\texttt{SMART} runs on multiple computer cores. The orbit processing (including the setup of the initial conditions, orbit integration, orbit classification and computation of the internal and projected velocity distributions) as well as the relevant linear algebra operations applied for their superposition are parallelised.

\section{\texorpdfstring{The $N$-body simulation}{The N-body simulation}}
\label{sec:The N-body simulation}
In order to test \texttt{SMART} on a realistic mock galaxy, we use a high-resolution collisionless numerical merger simulation by \citet{Rantala18}. The simulation is a single generation binary galaxy merger of two equal-mass elliptical galaxies with an effective radius of $7\mathrm{\,kpc}$ hosting a supermassive black hole of $8.5 \times 10^{9} M_{\odot}$ each and corresponds to the so-called $\gamma$-1.5-BH-6 simulation in \citet{Rantala18}. The two initial
galaxies are set up by using a spherically symmetric Dehnen density-potential \citep{Dehnen93} with an initial inner stellar density slope of $\rho \propto r^{-3 / 2}$ for both progenitor galaxies. The merger results in a remnant triaxial galaxy with a supermassive black hole of $1.7 \times 10^{10} M_{\odot}$ with a sphere of influence\footnote{We here use the definition of the sphere of influence as the radius within which the total stellar mass equals the black hole mass, i.e., $M_{*}(r_{\mathrm{SOI}})=M_{BH}.$} of $r_{\mathrm{SOI}} \sim1 \mathrm{\,kpc}$ and an effective radius of $r_e\sim14\mathrm{\,kpc}$. 
With that, the remnant resembles NGC 1600, a galaxy showing a very large core with a tangentially biased central stellar orbit distribution \citep{Thomas16}.
The simulation is based on the hybrid tree–N-body code KETJU (\citealt{Rantala17, Karl15}) which is able to accurately compute the dynamics close to the black hole due to the algorithmic chain regularization method AR-CHAIN (\citealt{Mikkola06,Mikkola08}). The computation of the global galactic dynamics is based on the tree code GADGET-3 \citep{Springel05}. 

We analyse a snapshot of the simulation $\sim$1.4 Gyr after the galaxy centers have merged, such that the remnant can be assumed to be in a steady state. 
At this stage, the actual distance of the two merging black holes in the simulation is
$5\mathrm{\,pc}$. The merger remnant shows a radially varying triaxiality parameter, being more oblate at smaller radii and increasingly more round in the centre with a maximum of triaxiality, i.e., $T = 0.5$, at about $3\mathrm{\,kpc}$ and more prolate outskirts.  
The simulation contains $8.3 \times 10^6$ stellar particles with masses of $10^{5} M_{\odot}$ each, leading to a total stellar mass of $8.3 \times 10^{11}M_{\odot}$. The mass ratio of one supermassive black hole and one stellar particle is $M_{BH}/M_*=8.5 \times 10^4$ and therefore sufficiently large to investigate a realistic interaction of the SMBH binary with the stars~\citep{Mikkola92}. 
The number of dark matter particles is $2 \times 10^7$ with masses of $7.5 \times 10^{6} M_{\odot}$ each, leading to a total dark matter mass of $1.5 \times 10^{14} M_{\odot}$.\\
The simulation is particularly suitable to test \texttt{SMART} because of its (1) very high resolution (including properly resolved black-hole dynamics); (2) realistic orbital structure and shape; (3) realistic mass composition (black hole, stars, and dark matter) with a realistic stellar density core (e.g. \citealt{Thomas09,Rantala19}). Finally, the number of stellar particles is large enough to measure fully resolved LOSVDs from the central sphere of influence of the black holes out to dark matter dominated regions. 

\subsection{Processing the simulation data}
\subsubsection{Orientation of the Simulation}
\label{sec:Orientation of the Simulation}
We aim at orienting our intrinsic coordinate system as closely as possible to the intrinsic symmetry axes of the merger remnant. However, the stellar and dark matter principal axes of the simulated remnant are not aligned. Hence, the orientation of the main axes depends on the radius and on the mass component for which the reduced inertia tensor (see e.g. \citealt{Bailin05}) is calculated. Such a shift between the stellar and dark matter halo axes is not unexpected for collisionless merger simulations (see e.g. \citealt{Novak06}). We decided to center the remnant on the stars and black holes and afterwards orientate it by using the reduced inertia tensor for stars and dark matter within $30\mathrm{\,kpc}$. With this, the stellar elliptical isophotes for the three different projections are well aligned with the projected principal axes within the field of view of $15\mathrm{\,kpc}$ x $15\mathrm{\,kpc}$. There is a negligible residual misalignment which is strongest for the major axis projection but nowhere larger than $\sim5^{\circ}$ (see Fig.~\ref{fig:Figure3} and~\ref{fig:Figure16}).

\subsubsection{Density}
Due to the good alignment and taking advantage of the nearly triaxial intrinsic symmetry of the merger remnant, we increase the resolution of the simulation for computing the density by a factor of 8 by folding all stellar and dark matter particles into one octant. 
The stars are then binned into concentric radial shells with 1000 stars in each
shell and the dark matter particles are binned into radial shells with 5000 dark
matter particles each. The single shells are subdivided into angular bins, such that the elevation angle $\theta \in [0^\circ, 90^\circ]$ increases in constant $\sin(\theta)$-steps of 0.1
and that the azimuthal angle $\phi \in [0^\circ, 90^\circ]$ increases in constant $\phi$-steps of $10^\circ$. \\
Within $r<0.28\mathrm{\,kpc}$ for the stellar and $r<10\mathrm{\,kpc}$ for the dark matter particles, the resolution is too low to extract a smooth density. Here, we extrapolate the logarithmic densities from the outer parts by a first order polynomial fit in the logarithm of the radius. 
For the stellar densities we use the slope and vertical intercept averaged over all angular bins. 
The dark matter density is extrapolated by using the slopes and vertical intercept values of the individual angular bins. 
We ensure that the total enclosed mass is well covered by the extrapolated density. 
To smooth the radial density profiles, SciPy's \texttt{gaussian\_filter()}-function \citep{Virtanen19} is used.

\subsubsection{Spatial binning}
\label{sec:Spatial binning}
For spatially binning the kinematic input data we use the Voronoi tessellation method of \citet{Cappellari03}. For each tested projection (see Section~\ref{sec:Applying SMART to the simulation}) we construct a separate set of Voronoi bins. To end up with a roughly constant number of stellar simulation particles $N_*$ in each bin, we define the signal-to-noise ratio as 
\begin{align}
    \frac{\mathrm{signal}}{\mathrm{noise}}=\sqrt{\mathrm{signal}}=\sqrt{N_{*}}= \begin{cases} 70 \textrm{ for } r<r_{\mathrm{SOI}} ,\\ 150 \textrm{ for } r_{\mathrm{SOI}}<r<15\mathrm{\,kpc} .  \end{cases}
\end{align}
When calculating the average over the five different projections, this results in a total number of $N_{\text {losvd }}=N_{\text {voronoi }}=227$ Voronoi bins within the whole field of view and 54 Voronoi bins within $r_\mathrm{SOI}$. \\
This resolution is chosen to conform with realistic observational data and does not exceed high-resolution wide-field spectral observations by, e.g., MUSE (cf. \citealt{Mehrgan19}) but still proves to be sufficiently high for this analysis. 

\subsubsection{Kinematic data and velocity binning}
\label{sec:Kinematic Data and velocity binning}
For each spatial bin we calculate the line-of-sight velocity distributions for $N_{\text {vlos }}=45$ equally-sized velocity bins with $v^{\mathrm{max}}_{\mathrm{min}} = \pm 1600 \mathrm{\,km\,s^{-1}}$, which is chosen so that it covers 
about 10 times the velocity dispersion. This results in a velocity resolution of $\Delta v_{\mathrm{vlos}} = 71.11 \mathrm{\,km\,s^{-1}}$. \\
It is not trivial which "error" for the kinematic input data of the simulation should be used since we do not have an error in the simulation in the sense that repeated measurements give the same results. 
However, the $\chi^2$-minimisation formally requires information about an "error". We tested several assumed "error-bars", such as
\begin{compactenum} 
    \item[-] the difference between two kinematic datasets, each determined by using half of the simulation particles,
    \item[-] the Poisson noise and
    \item[-] a constant absolute error for each LOSVD as 10 percent of the maximum per LOSVD.
\end{compactenum}
In order to prevent an underestimation of the relative error for the major axis projection holding more particles along the line of sight than the other projections, the constant absolute error proves to be the most suitable method and is used in the present analysis.
The choice of setting the error value to 10\% of the maximum value of each LOSVD corresponds to a
velocity uncertainty of $\Delta v = 13\mathrm{\,km\,s^{-1}}$, dispersion error of $\Delta \sigma = 13 \mathrm{\,km\,s^{-1}}$ (or 5\% relative error) and 
$\Delta h_n = 0.03$ for the higher-order Gauss-Hermite moments. This is a reasonable choice both in terms of real observational errors and of the scatter in the kinematic maps of the merger simulation (see Fig.~\ref{fig:Figure3} and \ref{fig:Figure19}).

\subsection{Applying \texttt{SMART} to the simulation}
\label{sec:Applying SMART to the simulation}
For the purpose of testing our code,
\verb'SMART' is provided with the correct viewing angles and with the 3D normalized stellar (i.e. luminosity) density $\rho_*$ from the simulation as well as with the normalized 3D dark matter density $\rho_{\mathrm{DM}}$. The DM scaling parameter $s_{\mathrm{DM}}$ is to be determined by \texttt{SMART}. We  skipped any surface brightness deprojection, since degeneracies in the deprojection (see e.g. \citealt{deNicola20}) would only hamper a correct evaluation of our code. We parameterise the density as 
\begin{equation}
\label{eq:densmodel}
\rho = M_{BH} \times \delta(r) + \Upsilon \cdot \rho_{*} + s_{\mathrm{DM}} \cdot \rho_{\mathrm{DM}},
\end{equation}
where our fit parameters are the 
black hole mass $M_{BH}$, stellar mass-to-light ratio $\Upsilon$ and the multiplication factor $s_\mathrm{DM}$ defining the magnitude of the dark matter density profile favored by \texttt{SMART}. We determine these parameters by finding the minimum in $\chi^2$. \\
We model and analyse five different projections with (1) $\vartheta=90^{\circ}$, $\varphi=0^{\circ}$, i.e., the major axis projection, (2) $\vartheta=90^{\circ}$, $\varphi=90^{\circ}$, i.e., the intermediate axis projection, (3) $\vartheta=0^{\circ}$, $\varphi=90^{\circ}$, i.e. the minor axis projection, (4) $\vartheta=90^{\circ}$, $\varphi=10^{\circ}$, i.e. a projection 10$^{\circ}$ off the major axis in azimuthal direction and (5) $\vartheta=90^{\circ}$, $\varphi=45^{\circ}$, i.e. a projection in between the major and intermediate axis projection. Without loss of generality, $\psi$ was set to $90^{\circ}$ for all three viewing directions, i.e., $R$ (see eq.~\ref{eq:R matrix}) equals the unit matrix.\\ 
The field of view is chosen to be $15 \mathrm{\,kpc}$ x $15 \mathrm{\,kpc}$. The minimum sampled starting radius is set to  $r_{\mathrm{min}}=0.05\mathrm{\,kpc}$ and the maximum sampled starting radius is set to $r_{\mathrm{max}}=80\mathrm{\,kpc}$. We find optimal results for a central binning with $c=0.5$ in Equation~\ref{eq:radbin_cval_celz} because this guarantees that in case of the simulated merger remnant the difference of the circular velocity within one radial bin equals the model's velocity resolution $\Delta v_{\mathrm{vlos}} = 71.11 \mathrm{\,km\,s^{-1}}$ at a radius of $r=0.16\mathrm{\,kpc}=0.16\cdot r_\mathrm{SOI}$. For $c=1$ this would be only reached at $r=0.38\mathrm{\,kpc}=0.38 \cdot r_\mathrm{SOI}$ resulting in a deteriorated black hole mass recovery by $\sim10\%$.\\
For each tested projection we model two halfs of the LOS kinematic data: After correct projection onto the plane of the sky (see Section~\ref{sec:Coordinate systems and binning}), we separately model the half of the kinematic data with positive $x'$-coordinates (hereafter called 'right half' of the galaxy) and the half of the kinematic data with negative $x'$-coordinates (hereafter called 'left half' of the galaxy). 

\section{Results}
\label{sec:Results}
\subsection{Choice of Regularization}
\label{sec:Choice of Regularization}
\begin{figure*}
    \centering
    \includegraphics[width=1.0\textwidth]{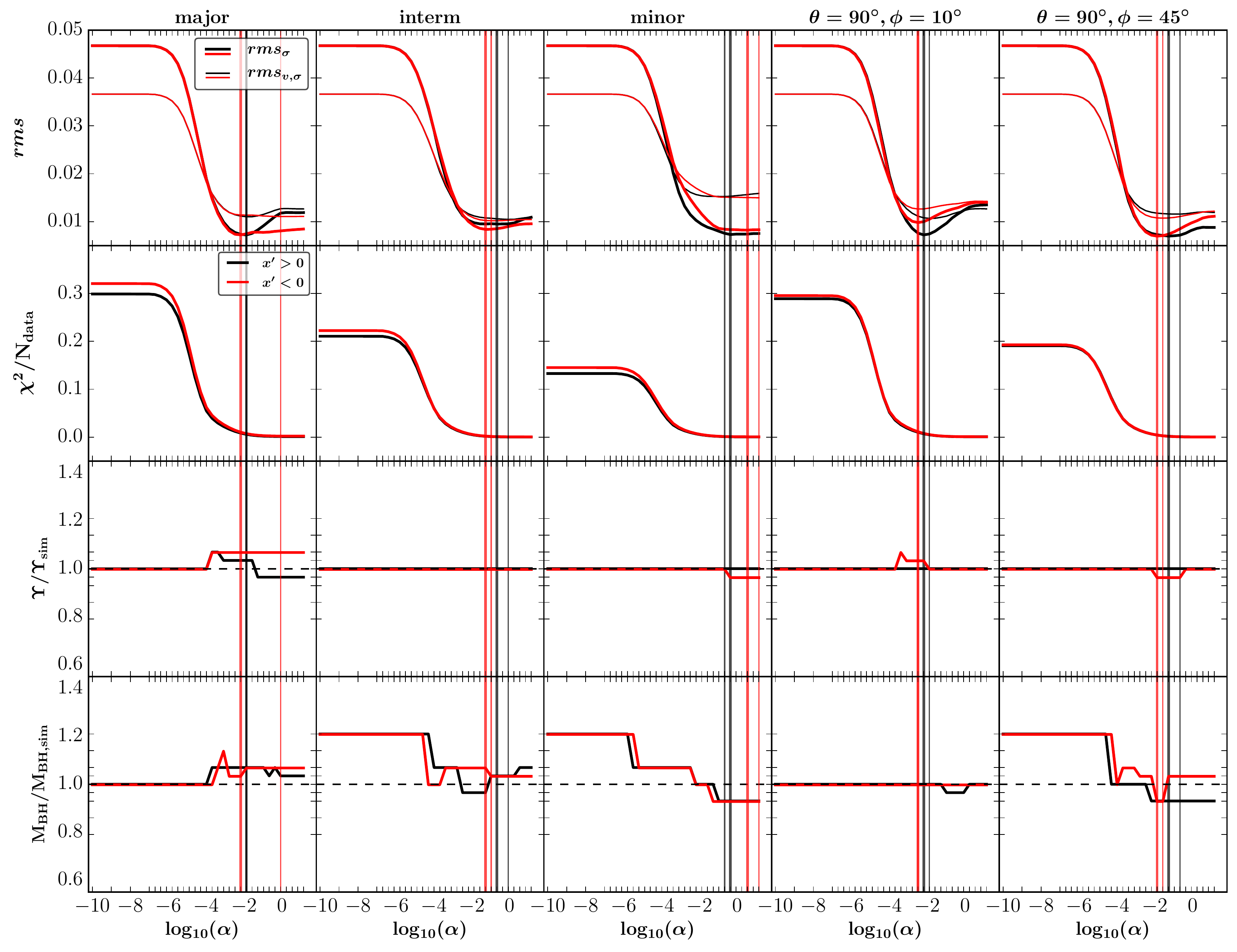}
    \caption{Choice of regularization and 1-dimensional mass recovery results for the five different projections (different columns). The first row shows the $rms_{\sigma}$- (thick line) and $rms_{v,\sigma}$-profile (thin line) for the modeled right half of the galaxy (black) and left half of the galaxy (red) with the correct black hole mass, stellar mass-to-light ratio and dark matter scale factor as input. All values are plotted against the increasing regularization value $\alpha$ in logarithmic units. 
    The x-axis-ticks thereby symbolise the tested $\alpha$-values. 
    The minima $\mathrm{min}(rms_{\sigma})$ (thick line) and $\mathrm{min}(rms_{v,\sigma})$ (thin line) are marked as vertical lines and suggest suitable regularization values. The thin black vertical line in the major axis panel thereby overlaps with the thick black vertical line and the thin red vertical line in the $\vartheta=90^{\circ}$, $\varphi=10^{\circ}$ panel overlaps with the thick red vertical line. The second row shows the corresponding $\chi^2/N_{\mathrm{data}}$ values. The third and fourth row show the 1-dimensional mass recovery results for the stellar mass-to-light ratio $\Upsilon/\Upsilon_{\mathrm{sim}}$ and black hole mass $M_{BH}/M_{BH,\mathrm{sim}}$ normalized over the correct values of the simulation.
    The y-axis-ticks for the third and fourth row symbolise the concrete masses which were tested and used as input values for the models. The black dotted line marks unity which is achieved when the model correctly recovers the mass. The stellar mass-to-light ratio and black hole mass were recovered with an accuracy better than 5\%. Such an intrinsic precision under similar conditions has not yet been demonstrated with other Schwarzschild modeling codes.}
    \label{fig:Figure2}
\end{figure*}

\begin{figure*}
    \centering
    \includegraphics[width=1.0\textwidth]{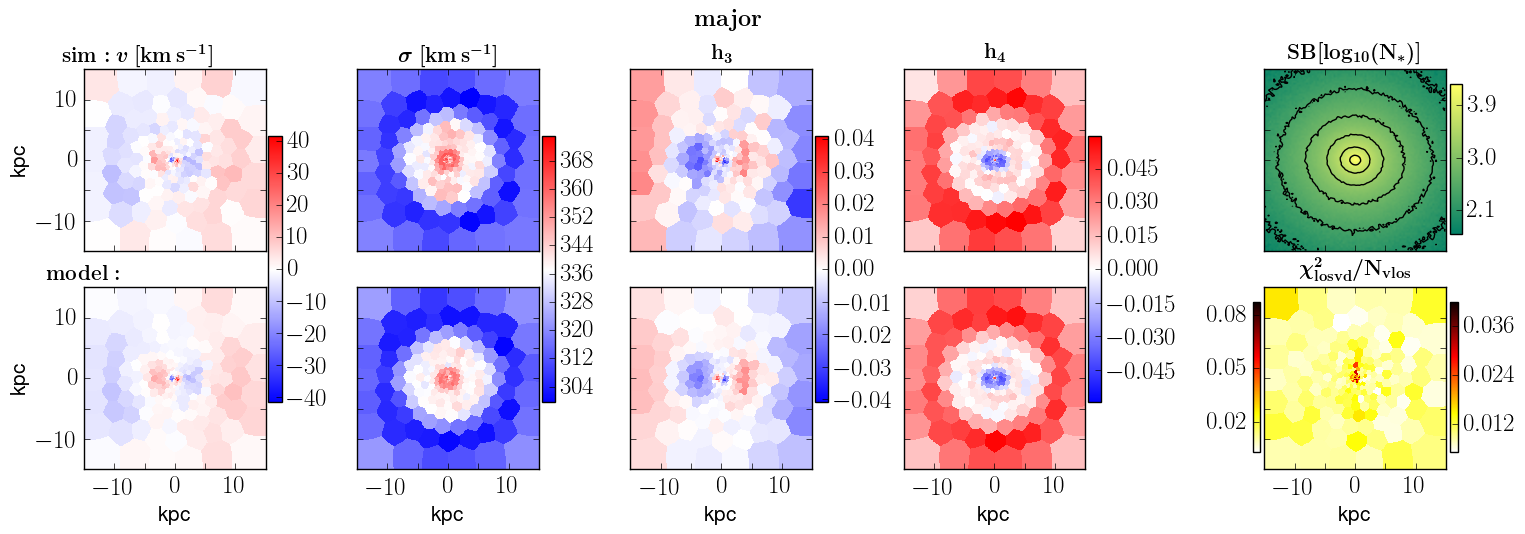} 
    \caption{Velocity maps of the simulation (top row) and of the model (bottom row) for the major axis projection, i.e. the projection with the line of sight being parallel to the major axis of the simulation based on the simulation's orientation described in Section~\ref{sec:Orientation of the Simulation} (for the other projections see Fig.~\ref{fig:Figure19}). The different panels show the velocity in $\mathrm{km\,s^{-1}}$ (first column), velocity dispersion in $\mathrm{km\,s^{-1}}$ (second column), the $h_3$-parameter (third column) and $h_4$-parameter (fourth column) plotted over the whole field of view. The fifth panel in the top row shows the corresponding surface brightness map from the simulation in units of logarithmic numbers of stellar particles $N_*$. The contour lines correspond to isodensity surfaces. The fifth panel in the bottom row shows $\chi_{\mathrm{losvd}}^2/N_{\mathrm{vlos}}$ as deviation from the kinematic input data with the modeled fit. We show the result for the model with the correct stellar mass-to-light ratio, black hole mass and dark matter scale factor as input parameters evaluated at the most suitable regularization parameter of $\alpha(min(rms_\sigma))$. The maps in the second row consist of the results of the modeled left and right half of the galaxy. Therefore, the $\chi^2$-maps show two colorbars for the two different halfs. }
    \label{fig:Figure3}
\end{figure*}

When applying the code to realistic noisy measurements, regularization becomes important to prevent the orbital weights to fit the noise in the data. The optimal regularization parameter $\alpha$ for a specific observational data set can be determined by running Monte-Carlo simulations on kinematic mock data \citep{Thomas05} and is given as the one providing the minimum deviation of the intrinsic properties (like the distribution function, or velocity moments, or mass parameters) in comparison to the default model.
When fitting noiseless ideal data, one would expect best results for $\chi^2 \to 0$, or $\alpha \to \infty$ (neglecting recovery degeneracies and assuming an "error" can be defined). Even in that case, however, due to residual systematics (like finite resolution of the orbit library etc.) and due to the intrinsic noise in the $N$-body simulation, we still expect that the best result may not necessarily be achieved asymptotically for very large $\alpha$, but already for some finite value of the regularization parameter. To take this into account, we split the code test into two phases: (1) We fit the orbit model with the correct mass parameters and determine that value of $\alpha$ for which the internal structure of the simulation is best recovered. Specifically, we use the 2nd order velocity moments for this comparison. (2) We then also vary the mass parameters and test how well they can be recovered. Our benchmark is the optimised $\alpha$ from the comparison of the moments, but we will discuss the results for all $\alpha$ to demonstrate their robustness. \\
For determining the deviation between the model's velocity dispersions $\sigma_r,\sigma_\theta$ and $\sigma_\phi$ and the real ones from the simulation we define
\begin{equation}
\label{eq: rms_sigma}
\begin{aligned}
\begin{split}
    rms_\sigma=\frac{1}{3} \sum_{i} rms_{\sigma_{\mathrm{i}}}= \frac{1}{3} \sum_{i} \sqrt{\frac{1}{N_{\mathrm{data}}} \sum_{j=1}^{N_{\mathrm{data}}}\left(\frac{\sigma_{i, \mathrm{data}}-\sigma_{i, \mathrm{mod}}}{\sigma_{i,\mathrm{data}}}\right)^{2}}, 
\end{split}    
\end{aligned}
\end{equation}
where the index $i$ denotes the three coordinates $r,\theta, \phi$. \\
These $rms_{\sigma}$-profiles of the five tested projections and their respective modeled halfs in dependence of the regularization parameter $\alpha$ are plotted as thick lines in the top row of Fig.~\ref{fig:Figure2}. The x-axis ticks thereby mark all $\alpha$-values which were tested in this analysis. \\
The second row in Fig.~\ref{fig:Figure2} shows the quality of the fit as $\chi^2/N_{\text{data}}$-profile (for definition of $\chi^2$ see formula~\ref{chi squared}) again plotted against the regularization parameter. $\chi^2$ as deviation from the kinematic input data with the modeled fit is here normalized over the number of input data $N_{\text{data}}$, which is composed of the number of Voronoi bins $N_{\text {voronoi }}$ times the number of kinematic bins $N_{\text{vlos}}$. As expected, the fit to the data is poor when $\alpha$ is low (high $\chi^2/N_{\mathrm{data}}$). In this regime, it is the entropy term which
is essentially maximised and the data (via $\alpha \cdot \chi^2$, cf. eq.~\ref{eq:costfunc}) have little influence on the fit. With this, the anisotropy strongly depends on the $\omega_i$ and, in our case, happens to be a poor representation of the internal moments of the merger (high $rms$ values, see Fig.~\ref{fig:Figure2}, first row).  With increasing $\alpha$, both the fit quality and the agreement with the merger structure improve. However, at very high $\alpha$-values further improvements of the fit do not make the internal moments better since we are dominated by the noise of the $N$-body simulation. 
The most suitable choices of regularization can be read off from the minima $\mathrm{min}(rms_{\sigma})$ of the $rms_{\sigma}$-profiles and are marked as thick vertical lines in Fig.~\ref{fig:Figure2}.
Their average value is $\alpha(\mathrm{min}(rms_{\sigma}))$=0.41. This value, however, depends on the specific implemented setup of \texttt{SMART} as well as of the input data. 
Evaluated at the individual most suitable regularization values and afterwards averaged over all ten models (5 projections and two halfs each) we get a minimum value of only $\mathrm{min}(rms_{\sigma})= 0.008$. This extremely good agreement demonstrates that our orbit sampling represents the phase space very well. 

We also test the influence of rotation on the comparison. Due to the overall small angular momentum and thus small absolute velocity amplitude, however, the relative errors in $v$ are sometimes large. The absolute error of the first order velocity moments, averaged over all angular and radial bins, is only $\Delta v=3.9\mathrm{\,km\,s^{-1}}$, but the maximum velocity over these bins is likewise only $23.4\mathrm{\,km\,s^{-1}}$.
We therefore here define $rms_{v,\sigma}$ as normalized deviation between the first internal moments and the velocity dispersions together as

\begin{equation}
\begin{aligned}
   rms_{v,\sigma}=\frac{1}{3} \sum_{i} rms_{v_i,\sigma_{\mathrm{i}}} =\frac{1}{3} \sum_{i} \sqrt{\frac{1}{2N_{\mathrm{data}}}\sum_{j=1}^{N_{\mathrm{data}}} 
    \left( \Delta^2_{v_i}+ \Delta^2_{\sigma_i} \right)}, \\
    \mathrm{with} \: \Delta_{v_i}=\frac{v_{i,\mathrm{data}}-v_{i,\mathrm{mod}}}{\sqrt{v^2_{i,\mathrm{data}} + \sigma^2_{i,\mathrm{data}}}} \:
    \mathrm{and} \: \Delta_{\sigma_i}=\frac{\sigma_{i,\mathrm{data}}-\sigma_{i,\mathrm{mod}}}{\sqrt{v^2_{i,\mathrm{data}} + \sigma^2_{i,\mathrm{data}}}},
\end{aligned}    
\end{equation}

where the index $i$ again denotes the three coordinates $r,\theta, \phi$. \\
The $rms_{v,\sigma}$-profiles are plotted against the regularization parameter as thin lines in the top row of Fig.~\ref{fig:Figure2}. As exptected, their minima appear at regularization values similar to the ones of the $rms_{\sigma}$-profiles. They are marked as thin vertical lines and their average value is $\alpha(\mathrm{min}(rms_{v,\sigma}))$=0.012. Again, this value depends on the specific setup. 
Evaluated at the individual $\alpha(\mathrm{min}(rms_{v,\sigma}))$-values and afterwards averaged over the five different projections and their respective modeled halfs we gain a value of only $\mathrm{min}(rms_{v,\sigma})=0.012$. \\
All values in the vicinity of $\alpha(min(rms_{\sigma}))$ and $\alpha(min(rms_{v,\sigma}))$ are good regularization choices. 
Within this regularization region \verb'SMART' is able to well fit the kinematic input data for each tested projection (see Fig.~\ref{fig:Figure3} and \ref{fig:Figure19}). Averaged over the five different projections and their respective modeled halfs we receive mean values and deviations of $\bar{v}=(0.11\pm2.09)\mathrm{\,km\,s^{-1}, } \text{ }\bar{\sigma}=(309.75\pm2.54)\mathrm{\,km\,s^{-1}, } \text{ }\bar{h_3}=0.00\pm0.01\text{ and }\bar{h_4}=0.01\pm0.01$, again evaluated at $\alpha(min(rms_\sigma))$. The $\chi^2$-maps in  Fig.~\ref{fig:Figure3} and \ref{fig:Figure19} show that the models for each projection are able to fit the kinematic input data homogeneously well over the field of view with slightly larger deviations in the center.
If not specifically annotated, the results shown in the further analysis are evaluated at $\alpha(min(rms_{\sigma}))$. However, the quality of the models does not strongly depend on the exact regularization value because a broader range of regularization values around these determined $\alpha$-values is sufficiently appropriate and results in equally good mass parameter reproductions within the overall scatter (see also third and fourth line in Fig.~\ref{fig:Figure2} which will be explained in Section~\ref{sec:Mass recovery}). \\

\subsection{Reproduction of internal moments and orbit structure}
\begin{figure}
    \centering
    \includegraphics[width=1.0\textwidth]{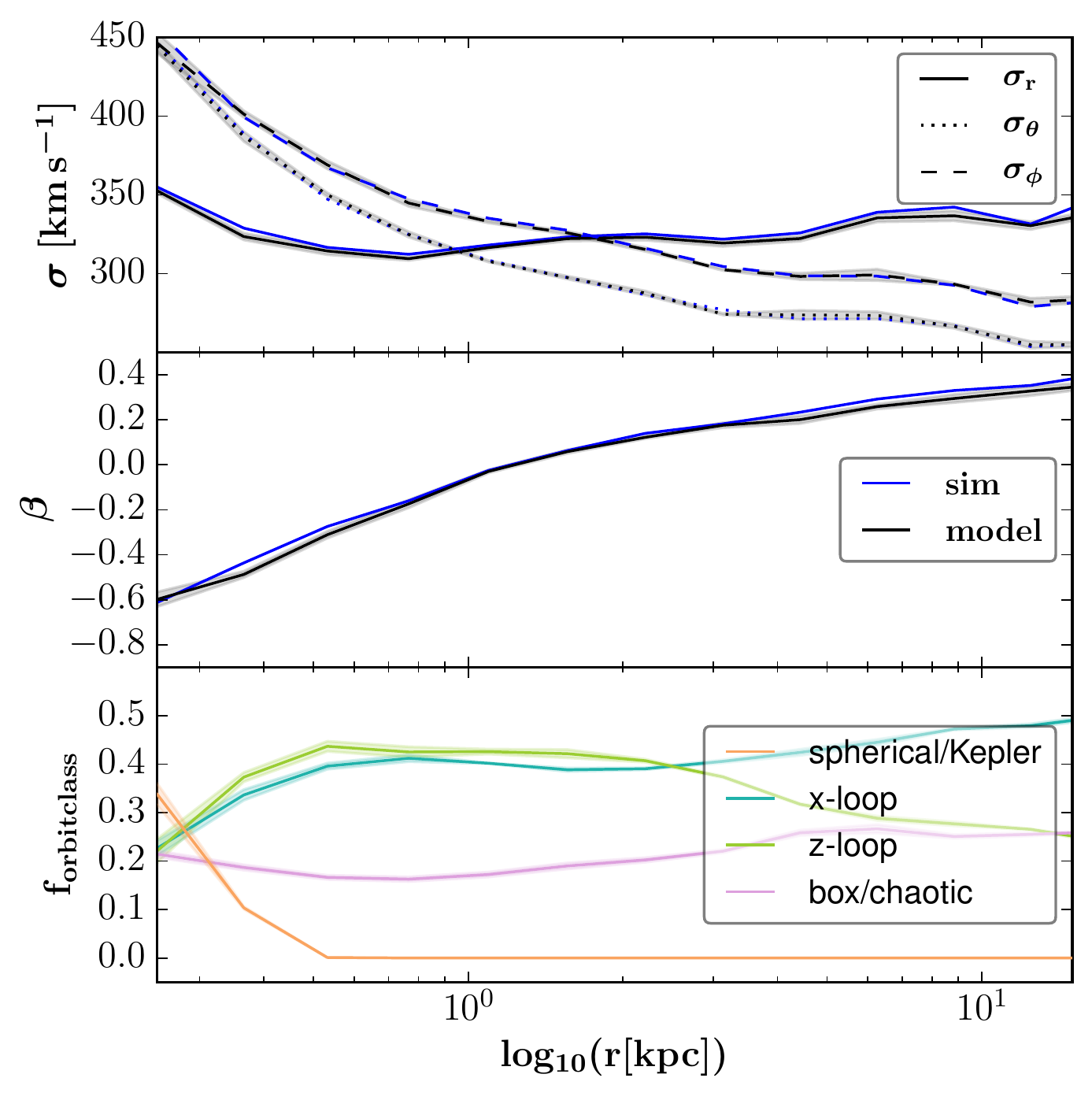}
    \caption{Reproduction of the internal properties within $r\in[0.25\mathrm{\,kpc}, 15\mathrm{\,kpc}]$. Top panel: The internal velocity dispersions in radial direction $\sigma_r$ (solid line), elevation direction $\sigma_{\theta}$ (dotted line) and azimuthal direction $\sigma_{\phi}$ (dashed line) from the model (black lines) averaged over the five different projections accurately follow the real ones from the simulation (blue lines) out to $15\mathrm{\,kpc}$ as field of view. The grey shaded lines mark the deviations between the different projections. Middle panel: Also the anisotropy parameter $\beta$ is well reproduced and represents the tangentially anisotropic orbit distribution ($\beta<0$) within the core radius $r_b=r_{\mathrm{SOI}}\sim1\mathrm{\,kpc}$ as well as the radially anisotropic orbit distribution ($\beta>0$) outside $r_b$. Bottom panel: Radial distribution of the orbit fractions $f_{\mathrm{orbitclass}}$ classified by \texttt{SMART}. }
    \label{fig:Figure4}
\end{figure}
The previously shown small $rms_\sigma$- and $rms_{v,\sigma}$-values already demonstrate the very good recovery of the internal moments by the model when providing the correct mass parameters and viewing angles. 
The high level of agreement between model and simulation is further illustrated in Fig.~\ref{fig:Figure4}.
The anisotropy parameter $\beta=1-\frac{\sigma_{\theta}^{2}+\sigma_{\phi}^{2}}{2 \sigma_{r}^{2}}$ of the model matches the profile of the simulated one (see Fig.~\ref{fig:Figure4}, middle panel). The model is also able to reproduce the negative $\beta$ within the core radius $r_b$, which equals the black hole sphere of influence \citep{Thomas16}, reflecting the tangential orbit distribution due to black hole 'core scouring': Within the sphere of influence ETGs at the high-mass end exhibit central regions which are fainter than an extrapolation of a Sérsic function \citep{Sersic63} as fit to the outer surface brightness profile would suggest. The commonly accepted theory for the formation of these 'cores' is a gravitational slingshot process of stars on radial orbits caused by SMBH binaries which were arised by galaxy mergers (e.g. \citealt{Begelman80, Hills80, Ebisuzaki91, Milosavljevic01,Merritt06, Rantala18}). \\ 

The classification of the integrated orbits is also plotted in Fig.~\ref{fig:Figure4} (bottom panel). Orbits in the immediate vicinity of the black hole are spherical and Keplerian orbits as expected due to the SMBH. Z-tubes are the predominant orbits within $0.3\mathrm{\,kpc}<r<3\mathrm{\,kpc}$ causing the more oblate shape of the galaxy in this range. For $r>3\mathrm{\,kpc}$ the majority of orbits are classified as x-tubes corresponding to the prolate shape in the outskirts of the simulation as described in Section~\ref{sec:The N-body simulation}. Our orbit classification analysis well matches the one by Frigo et al. in prep., which is done via an orbit frequency analysis of the simulation.

\subsection{Mass recovery}
\label{sec:Mass recovery}
So far, we have provided \texttt{SMART} with the correct mass parameters of the stellar mass-to-light ratio $\Upsilon_{\mathrm{sim}}$, black hole mass $M_{BH,\mathrm{sim}}$ and dark matter multiplication factor $s_{\mathrm{DM},\mathrm{sim}}$ of the simulation. Even though \texttt{SMART} can be provided with any type of dark matter profile, we here transfer the dark matter density profile shape of the simulation and concentrate on finding the correct mass multiplication scale factor $s_{\mathrm{DM}}$ of this predetermined halo shape (cf. eq.~\ref{eq:densmodel}). 

The following sections will show the one-dimensional mass reproduction results of individually determining the favored stellar mass-to-light ratio $\Upsilon$ or black hole mass $M_{BH}$ (Section~\ref{sec:1-dimensional mass recovery of upsilon and mbh}) as well as the two-dimensional mass recovery results of simultaneously determining the favored $\Upsilon$ and $M_{BH}$ (Section~\ref{sec:2-dimensional mass recovery of upsilon and mbh}) or $\Upsilon$ and $s_{\mathrm{DM}}$ (Section~\ref{sec:2-dimensional mass recovery of upsilon and mdm}). We skip any 3-dimensional mass parameter recovery since this would not provide more information than the combined 2-dimensional recoveries in the context of testing the orbit library.

\subsubsection{1-dimensional mass recovery of $\Upsilon$ and $M_{BH}$}
\label{sec:1-dimensional mass recovery of upsilon and mbh}
Fig.~\ref{fig:Figure2} shows the 1-dimensional mass recovery results of  $\Upsilon$ (third row) and $M_{BH}$ (fourth row) for the five different projections and their respective modeled halfs. For testing the recovery of the black hole mass, we provide the model with the correct $\Upsilon_{\mathrm{sim}}$- and $s_{\mathrm{DM},\mathrm{sim}}$-values and run nine models with different black hole masses within $M_{BH} \in [0.79 M_{BH,\mathrm{sim}}, 1.21 M_{BH,\mathrm{sim}}]$ including the correct one. The tested mass grid has a smaller grid size close to $M_{BH}/M_{BH,\mathrm{}sim}=1$ and the exact tested values can be read off from the ordinate ticks in Fig.~\ref{fig:Figure2}.  \\
For testing the 1-dimensional mass recovery of the stellar mass-to-light ratio we provide $M_{BH,\mathrm{sim}}$ and $s_{\mathrm{DM},\mathrm{sim}}$ and test the same $\Upsilon/\Upsilon_{\mathrm{sim}}$-values as for the black hole mass analysis. \\
Fig.~\ref{fig:Figure2} shows the favored mass parameters, i.e., the mass parameters where $\chi^2/N_{\text{data}}$ is smallest, as a function of $\alpha$. As one can see, for our fiducial choice of $\alpha$ (i.e. $\alpha=\alpha(\mathrm{min}(rms_{\sigma}))$ as the best recovery of the velocity dispersions), the \textit{average} mass recovery performs excellently, with $\Delta M_{BH}=5\%$ and $\Delta \Upsilon=2\%$. In fact, above $\log \alpha \gtrsim -3$, the results are very robust with little dependency on $\alpha$. Within the overall minor scatter, all models, independent from the chosen half of the galaxy or projection, show equally good fits and reproductions of the internal moments and mass parameters.

\subsubsection{2-dimensional mass recovery of $\Upsilon$ and $M_{BH}$}
\label{sec:2-dimensional mass recovery of upsilon and mbh}
\begin{figure*}
    \centering
    \includegraphics[width=1.0\textwidth]{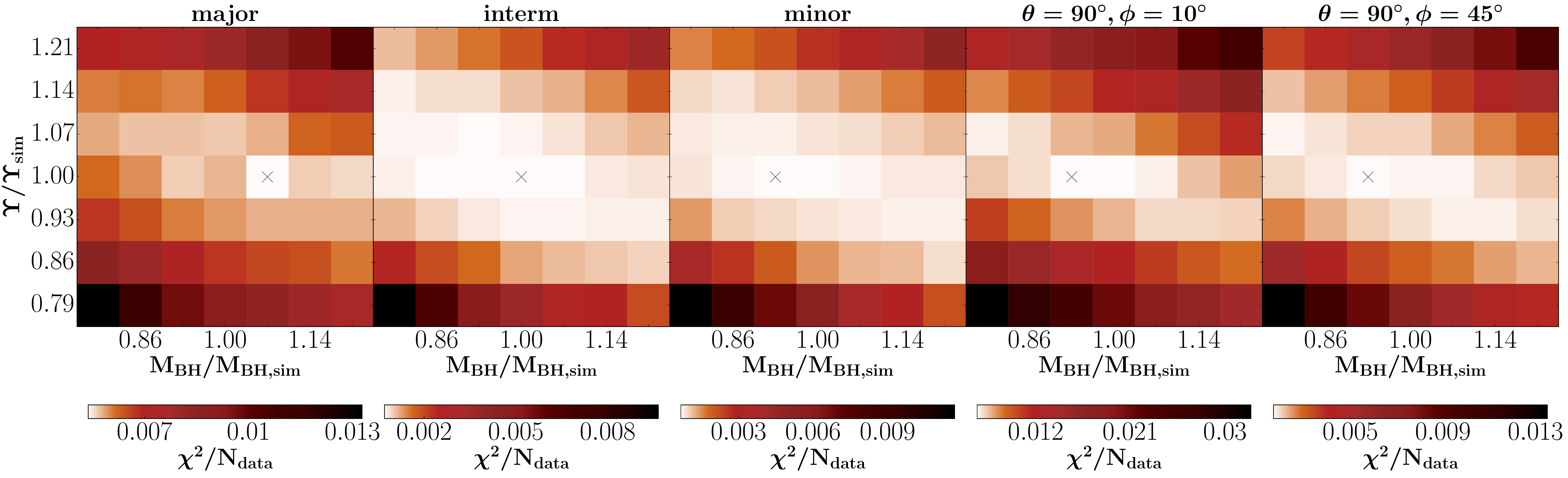}
    \caption{2-dimensional mass recovery results of $\Upsilon$ and $M_{BH}$ for the positive halfs of the five different projections. For each projection we evaluate 49 models with different $\Upsilon$- and $M_{BH}$-input-masses covering a 2-dimensional grid with a step size of 7\% around the correct mass parameter. Each plot contains the $\chi^2/N_{\mathrm{data}}$-colorbar for the individual projection. The favored models are marked with a grey cross. The model always finds the correct stellar mass-to-light ratio and the black hole mass with a minor averaged deviation of 6\%.}
    \label{fig:Figure5}
\end{figure*}

For simultaneously recovering $\Upsilon$ and $M_{BH}$ by \texttt{SMART} we sample a two-dimensional grid of input masses for 49 models per projection with $\Upsilon \in [0.79 \Upsilon_{\mathrm{sim}},1.21 \Upsilon_{\mathrm{sim}}]$ and $M_{BH}\in [0.79 M_{BH,\mathrm{sim}},1.21 M_{BH,\mathrm{sim}}]$. 
We again model all five different projections but fit only the right half of the galaxy since the 1-dimensional mass recovery showed no significant difference between the respective halfs of each projection. The results are plotted in Fig.~\ref{fig:Figure5}.
Evaluated at the regularization value $\alpha(min(rms_{\sigma}))$ the stellar mass-to-light ratio is in every case correctly recovered and the black hole mass is reproduced with an accuracy of $6\%$ averaged over the different projections.

\subsubsection{2-dimensional mass recovery of $\Upsilon$ and $s_{\mathrm{DM}}$}
\label{sec:2-dimensional mass recovery of upsilon and mdm}
\begin{figure}
    \centering
    \includegraphics[width=0.7\textwidth]{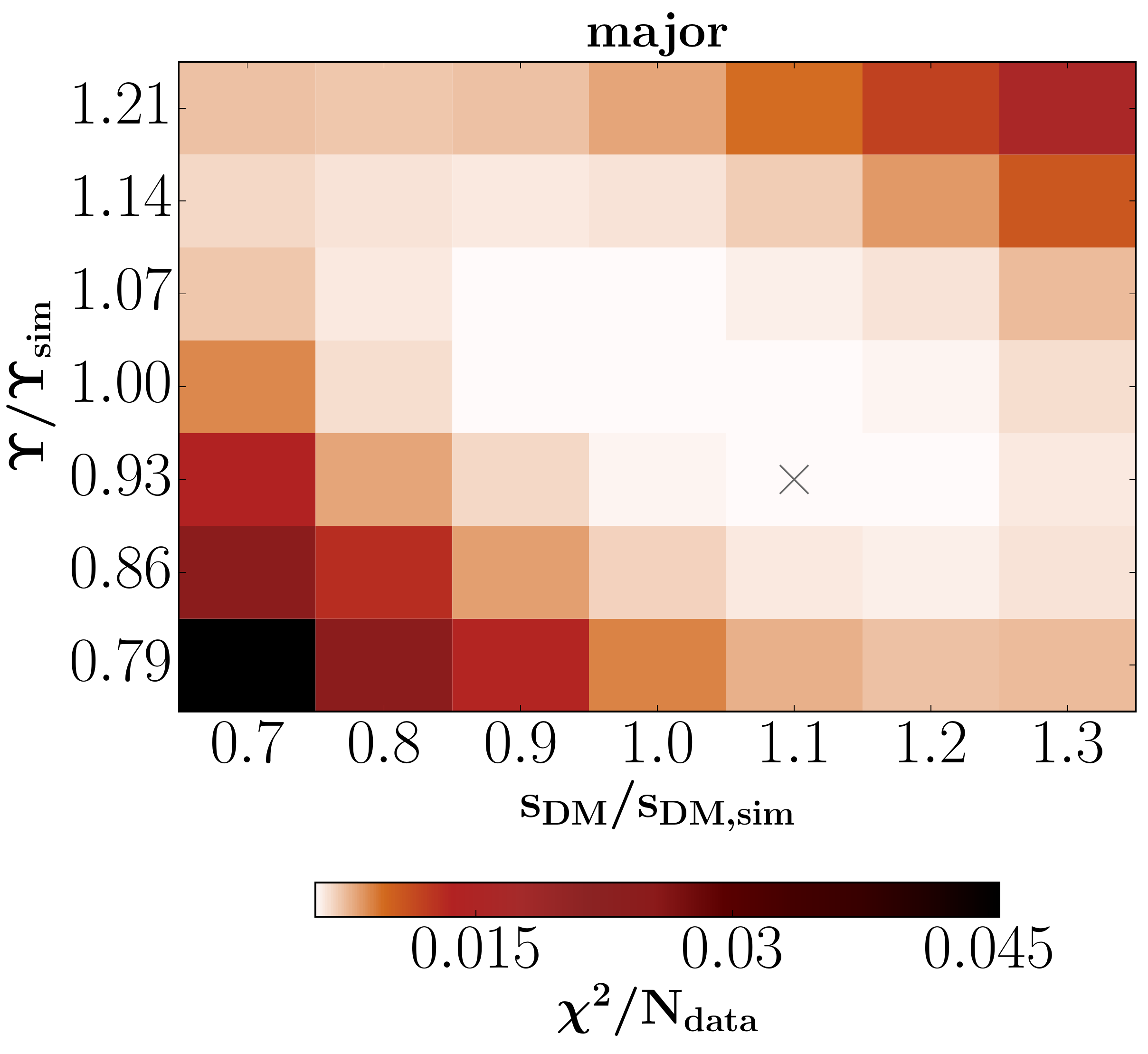}
    \caption{2-dimensional mass recovery results of $\Upsilon$ and $s_{\mathrm{DM}}$ for the positive half of the major axis projection. $s_{\mathrm{DM}}$ thereby is the mass multiplication scale factor of the predetermined halo shape of the simulation. We evaluate 49 models with different $\Upsilon$- and $s_{\mathrm{DM}}$-input-masses covering a 2-dimensional grid with a step size of 7\% for the stellar mass-to-light ratio and 10\% for the dark matter scaling multiplication factor. The favored model is marked with a grey cross. The model slightly overestimates the dark matter halo scale factor by 10\% and slightly underestimates the stellar mass-to-light ratio by 7\%.}
    \label{fig:Figure6}
\end{figure}

For recovering $\Upsilon$ and $s_{\mathrm{DM}}$ we sample a two-dimensional grid of input masses for 49 models along the major axis projection with $\Upsilon \in [0.79 \Upsilon_{\mathrm{sim}},1.21 \Upsilon_{\mathrm{sim}}]$ and 
$s_{\mathrm{DM}} \in [0.7 s_{\mathrm{DM},\mathrm{sim}},1.3 s_{\mathrm{DM},\mathrm{sim}}]$.
We here model the right half of the major axis projection.
The result is shown in Fig.~\ref{fig:Figure6}. Evaluated at $\alpha(min(rms_{\sigma}))$, the dark matter scale factor is slightly overestimated by 10\% and the stellar mass-to-light ratio is slightly underestimated by $7\%$ confirming the accurate mass reconstruction from the previous tests. \\ \\ 
In conclusion, these mass recovery results demonstrate that our orbit sampling and superposition algorithms allow for a very accurate reconstruction of the mass composition and orbital structure of triaxial systems with known density shapes. \\
\subsection{Beyond 2nd order velocity moments}
\begin{figure*}
    \centering
    \includegraphics[width=0.8\textwidth]{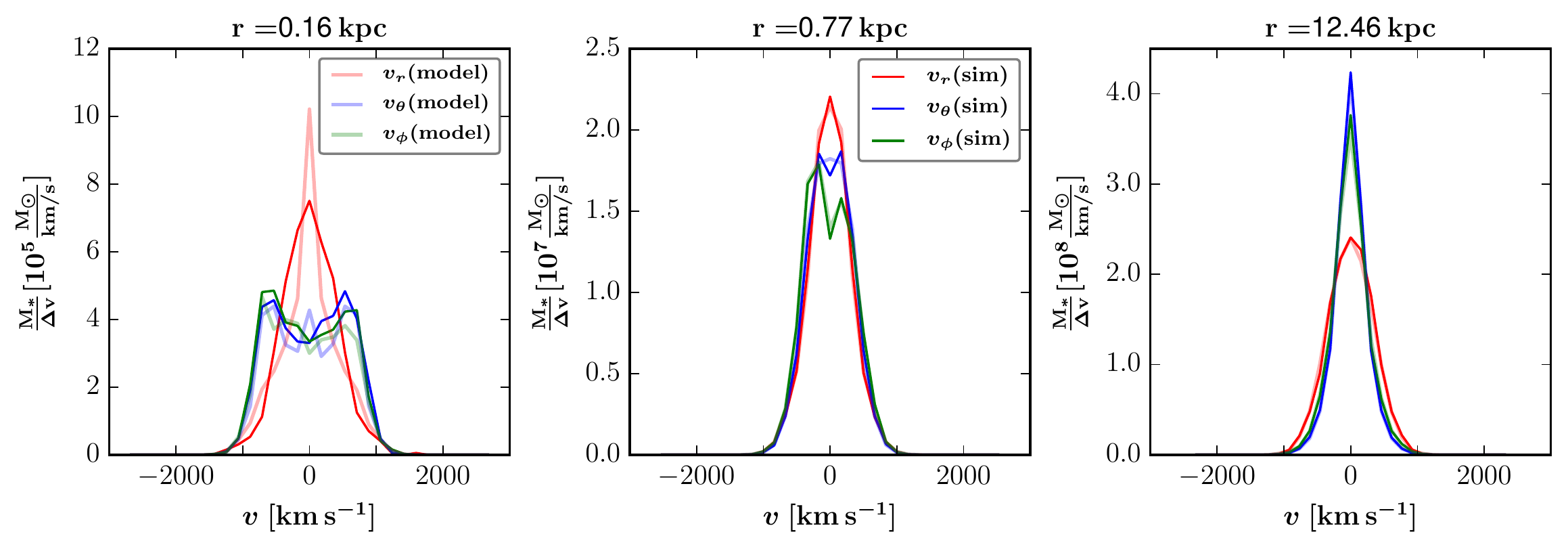}
    \caption{Internal velocity distributions in radial (red), elevation (blue) and azimuthal (green) direction when integrating over the respective spatial bin of the model and the other velocity components for three different radii of $r=0.16\mathrm{\,kpc}$ (left panel), $r=0.77\mathrm{\,kpc}$ (middle panel) and $r=12.46\mathrm{\,kpc}$ (right panel). The internal velocity distributions are calculated for 31 velocity bins within the positive and negative escape velocity evaluated at each radial and angular bin of the \texttt{SMART}-specific grid. We here show the mass-weighted stellar internal velocity distributions per velocity bin averaged within spherical shells. The non-shaded lines correspond to the simulation data and the shaded lines show the modeled results. In closer vicinity of the black hole, the elevation and azimuthal distribution show two maxima corresponding to the tangentially anisotropic orbit distribution. \texttt{SMART} is able to reproduce the internal velocity distributions in general and is also able to follow this specific behaviour in the central bins.}
    \label{fig:Figure7}
\end{figure*}
So far, we have focused on the reproduction of the first and second order internal velocity moments. The previous sections have shown that the full shape of the LOSVDs contains enough information to accurately reconstruct the mass and the anisotropy structure of the orbit distribution. In Fig.~\ref{fig:Figure7} we show the full mass-weighted stellar velocity distributions per velocity bin against the velocity in $\mathrm{\,km\,s^{-1}}$ in radial, longitudinal and azimuthal direction when integrating over the other velocity components and respective spatial bins of the model. 
We find that the central azimuthal and longitudinal velocity distributions within $r<r_{\mathrm{SOI}}$ have two maxima which become more pronounced closer to the center (see Fig.~\ref{fig:Figure7}). This likely reflects the strong tangential anisotropy produced during the formation of the core and is probably linked to the negative $h_4$ parameter at the center of the merger remnant (see Fig.~\ref{fig:Figure3} and~\ref{fig:Figure19}), which will be investigated in more detail in a separate paper.
The whole internal velocity distributions contain more information about the formation process than the velocity moments alone can do. So, we extended \texttt{SMART} to calculate the internal velocity distributions for 31 velocity bins within the positive and negative escape velocity, i.e., $v^{max}_{min}=\pm v_{esc}(r,\theta,\phi)$, evaluated at each radial and angular bin of the \texttt{SMART}-specific grid (see Section~\ref{sec:Coordinate systems and binning}). The internal velocity distributions averaged within spherical shells reproduce the ones from the simulation sufficiently well, with a deviation of $rms=0.07$ averaged over all velocity bins with $v<1000\mathrm{\,km\,s^{-1}}$ and radial bins within $r \in [0.25\mathrm{\,kpc},15\mathrm{\,kpc}]$ (see Fig.~\ref{fig:Figure7}). Including the outer wings of the internal velocity distributions with $v>1000\mathrm{\,km\,s^{-1}}$, the $rms$ increases, however, the number of simulation particles in these bins is very small. We will apply this ability of \texttt{SMART} to model the whole internal velocity distribution in future studies of real observational data. \\
We checked the behavior of the $rms$-profile in dependence of the regularization as deviation of the whole internal velocity distributions and compared it with the $rms_\sigma$- and $rms_{v,\sigma}$-profiles. It thereby showed the same form and its minimum appeared in the same regularization region and therefore does not provide additional information when determining the most suitable regularization value. \\ \\

\subsection{Robustness and Uniqueness checks}
In order to test the robustness of the results against modifications of our fiducial setup and in order to test the uniqueness of the results we make the following checks and considerations:\\
\subsubsection{PSF convolution and noise}
\label{sec:PSF convolution and noise}
\begin{figure}
    \centering
    \includegraphics[width=1.0\textwidth]{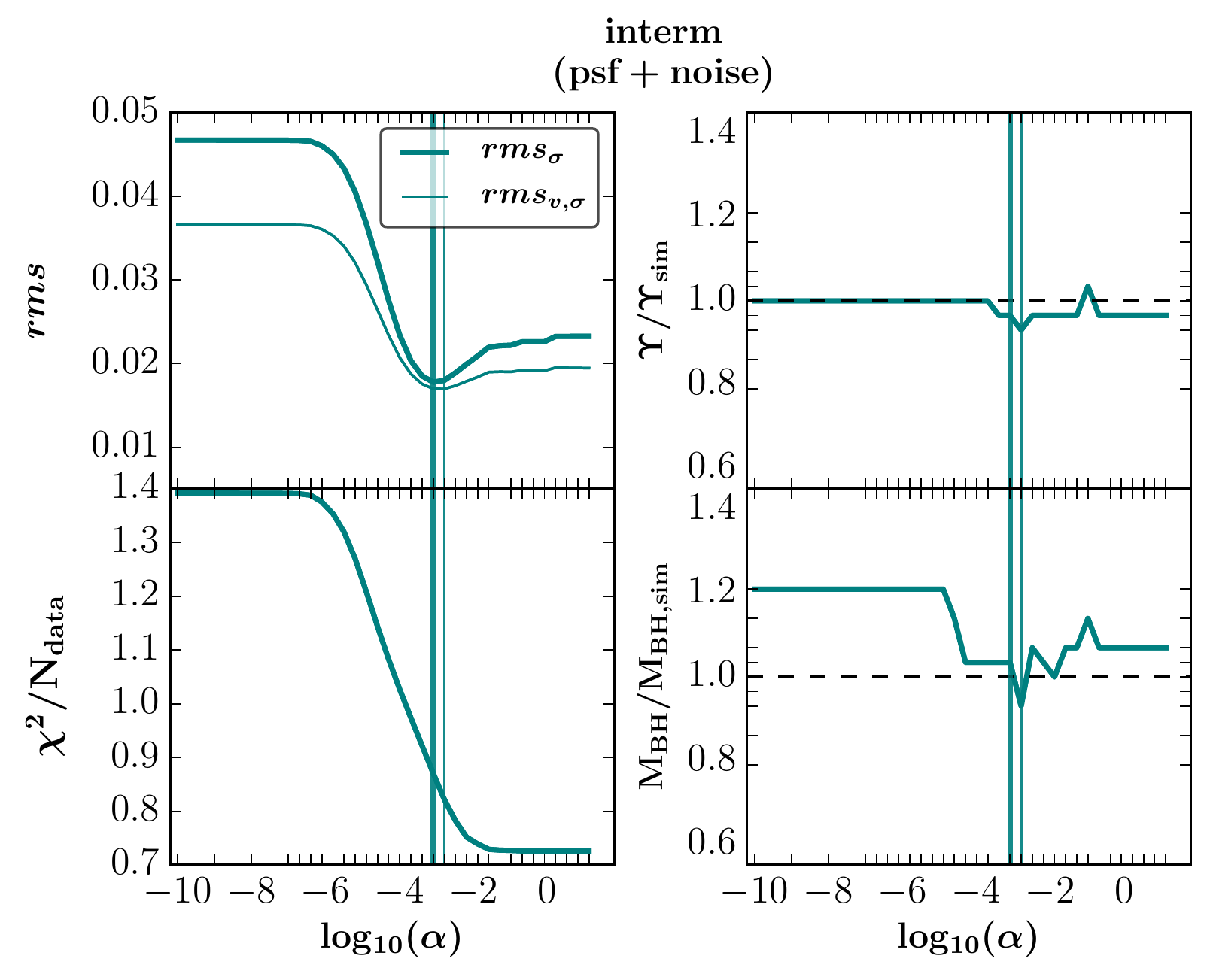}
    \caption{Choice of regularization and 1-dimensional mass recovery results for the positive half of a psf convolved noisy version of the intermediate axis projection. The $rms$- and $\chi^2/N_\mathrm{data}$-profile show the same shape and suitable regularization region than for the noiseless and non-psf-convolved case. Also, the $\Upsilon$- and $M_{BH}$-reproduction shows no remarkable change within the overall scatter.}
    \label{fig:Figure8}
\end{figure}
\begin{figure}
    \centering
    \includegraphics[width=0.53\textwidth]{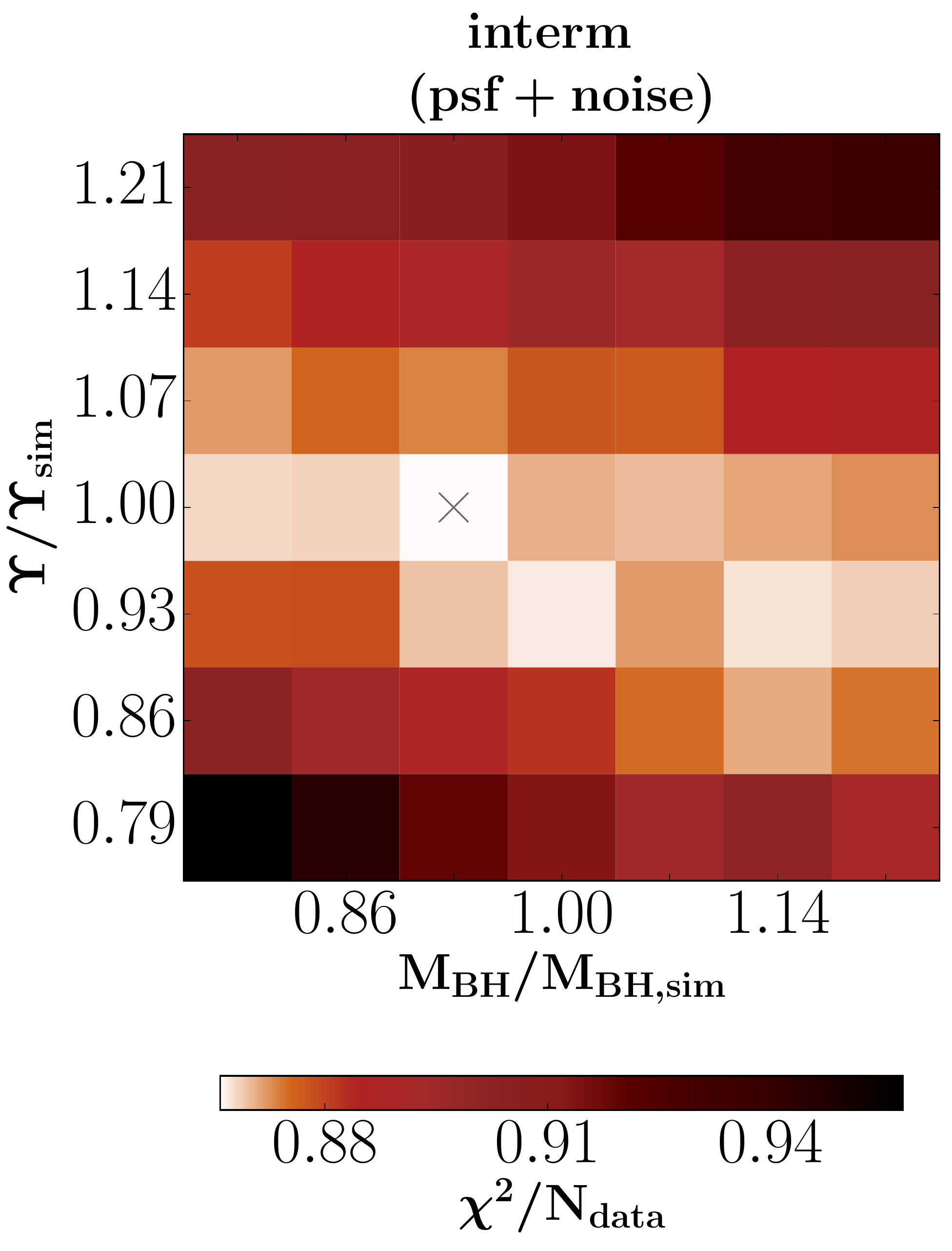}
    \caption{2-dimensional mass recovery results of $\Upsilon$ and $M_{BH}$ for the positive half of a psf convolved noisy version of the intermediate axis projection. We again evaluate 49 models with different $\Upsilon$- and $M_{BH}$-input-masses covering a 2-dimensional grid with a step size of 7\% around the correct parameters. The favored model is marked with a grey cross. The model finds the correct stellar mass-to-light ratio and slightly underestimates the black hole mass by $7\%$. Within the overall scatter this resembles the results of the models without noise and psf convolution. }
    \label{fig:Figure9}
\end{figure}
    While \citet{Valluri04} describe that three-integral, axisymmetric, orbit-based modeling algorithms in general show a flat-bottomed $\chi^2$ distribution, being unable to determine the black hole mass to better than a factor of $\sim 3.3$, 
    \citet{Magorrian06} demonstrates that this is only true for noiseless data.
    According to this, our model should not be able to precisely determine the correct black hole mass of the simulation due to the lack of an "error". However, the previous results have already shown that \texttt{SMART} achieves a well defined black hole mass due to a well defined minimum in the $\chi^2$-profile, which was probably supported by the intrinsic noise of the $N-$body simulation. Nevertheless, we check whether the minimum in the $\chi^2$-curve changes when simulating an "error" in the kinematic input data. For this, we model the positive half of the intermediate axis projection by adding Gaussian noise to the simulation chosen so that the velocity dispersion of the noisy kinematic simulation data results in an observationally realistic error of $\sim3\%$ ($\bar{v}=(0.14\pm7.64)\mathrm{\,km\,s^{-1}, } \text{ }\bar{\sigma}=(288.86\pm7.80)\mathrm{\,km\,s^{-1}, } \text{ }\bar{h_3}=0.00\pm0.02\text{, }\bar{h_4}=0.02\pm0.02$). To achieve even more realistic conditions we furthermore smooth the data by simulating a psf convolution with a FWHM of $2.43\mathrm{\,arcsec}$ which corresponds to $0.24\mathrm{\,kpc}$, i.e., about a fourth of the sphere of influence, when assuming the galaxy to be at a distance of $20\mathrm{\,Mpc}$. The so constructed velocity maps can be seen in Fig.~\ref{fig:Figure21}. We provide \texttt{SMART} with the information about the used FWHM-value for the two dimensional psf convolution and test the same $\Upsilon$ and $M_{BH}$ input masses as in Section~\ref{sec:Mass recovery}. The corresponding results for the 1-dimensional mass recoveries when modeling this modified input data are plotted as turquoise lines in Fig.~\ref{fig:Figure8}, and do not differ decisively from the ones without psf convolution and noise. The stellar mass-to-light ratio is slightly underestimated by $\Delta \Upsilon(\alpha(min(rms_\sigma))=3.5\%$ or $\Delta \Upsilon(\alpha(min(rms_{v,\sigma}))=7\%$ and the black hole mass is overestimated by $\Delta M_{BH}(\alpha(min(rms_\sigma))=3.5\%$ or underestimated by $\Delta M_{BH}(\alpha(min(rms_{v,\sigma}))=7\%$. The $rms_{\sigma}-, rms_{v,\sigma}-$ and $\chi^2/N_{\mathrm{data}}$-profiles are of course shifted upwards but follow the same form than the ones without noise and psf convolution (cf. Fig.~\ref{fig:Figure2}).
    Fig.~\ref{fig:Figure9} shows the 2-dimensional mass recovery result of $\Upsilon$ and $M_{BH}$ for this model. The stellar mass-to-light ratio is again correctly reproduced and the black hole mass is underestimated by only 7\%. With this, \texttt{SMART} is able to model the noisy and psf convolved kinematics as well as the default kinematics without any "error" equally well within the overall scatter.    

\subsubsection{Changing the orbital bias factors $\omega_i$}   
\label{sec:Changing the orbital bias factors omegai}
\begin{figure}
    \centering
    \includegraphics[width=1.0\textwidth]{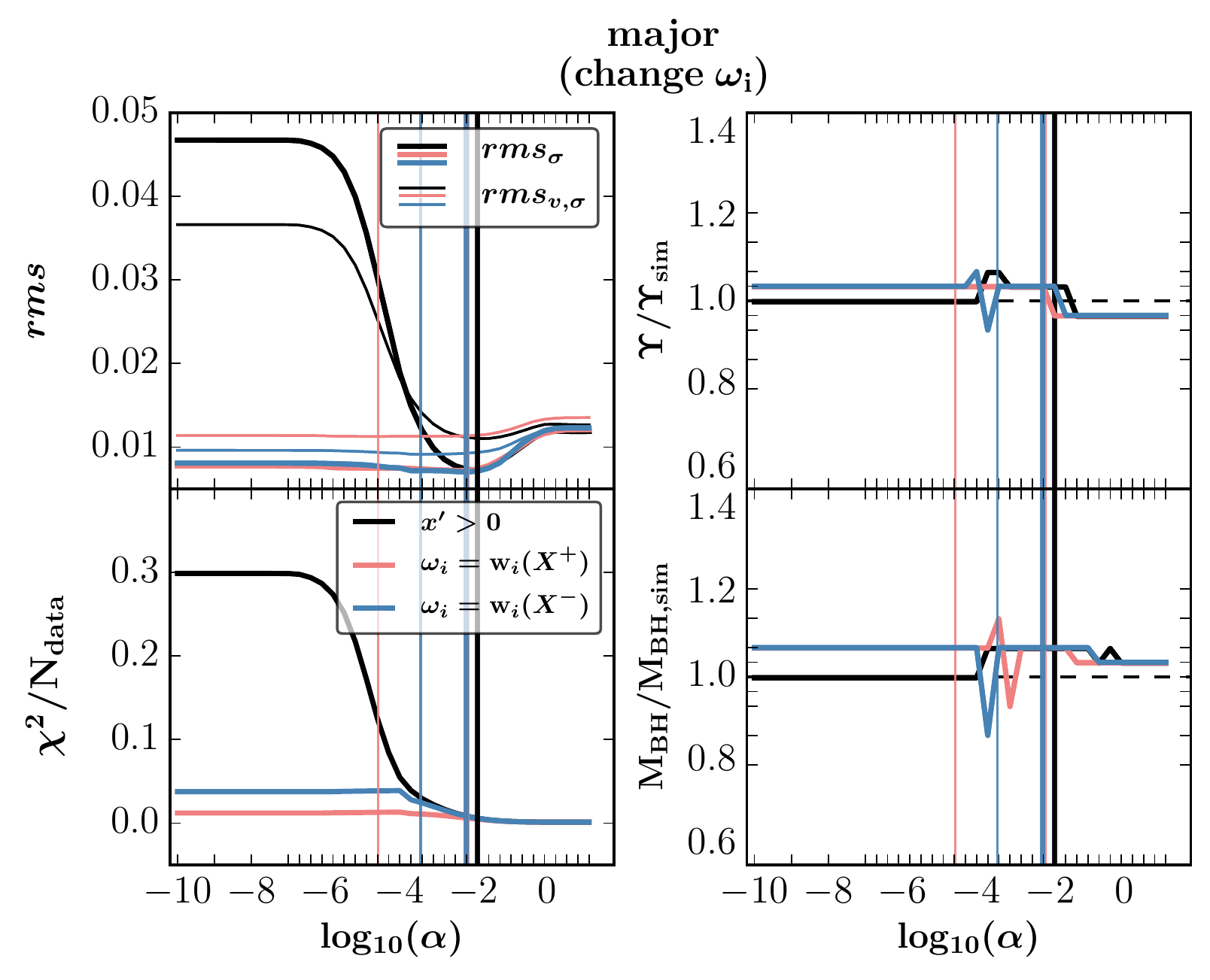}
    \caption{Choice of regularization and 1-dimensional mass recovery results for the positive half of the major axis projection when using the default constant orbital bias factors $\omega_i=$ const. (black line), when using the best-fit orbital weights $w_i(\boldsymbol{X}^+)$ of the positive half of the major axis as orbital bias factors, i.e., $\omega_i=w_i(\boldsymbol{X}^+)$ (pink line),  and when using $w_i(\boldsymbol{X}^-)$ of the negative half of the major axis as orbital bias factors, i.e., $\omega_i=w_i(\boldsymbol{X}^-)$ (blue line). The modified orbital bias factors do not improve the results indicating that the choice of $\omega_i=$ const. is sufficient. }
    \label{fig:Figure10}
\end{figure}
As discussed in Section~\ref{sec:Orbit superposition}, the orbital bias factors $\omega_i$ of Equation~\ref{eq:smoothing} can be used to control the orbital weights $w_i$. In the absence of other constraints $w_i \sim \omega_i$. Our choice of $\omega_i = 1$ is somewhat arbitrary. In fact, it biases the $w_i$s strongly away from the true solution. Thus, we want to test whether this affects our fits. Specifically, we test the opposite extreme: We remodel the positive half of the major axis, abbreviated below as $\boldsymbol{X}^+$, by setting $\omega_i=w_i(\boldsymbol{X}^+)$, where $w_i(\boldsymbol{X}^+)$ are the orbital weights of the best-fit model for $\boldsymbol{X}^+$. We also remodel $\boldsymbol{X}^+$ by setting the bias factors $\omega_i=w_i(\boldsymbol{X}^-)$ to the orbital weights of the best-fit model for the negative half of the major axis $\boldsymbol{X}^-$.
Fig.~\ref{fig:Figure10} shows the results for these completely independent model fits. As expected,
the $rms$-  and $\chi^2/N_{\mathrm{data}}$-profiles start with smaller values, since at $\alpha = 0$, the $\omega_i$s bias the orbital weights the strongest. As motivated above, in the specific case here, the weights are biased
towards a previous fit, which explains the better initial $\chi^2$ and $rms$. However, at our fiducial $\alpha$ range, the reproduction of the internal moments and the quality of the fit is the same as for the case with identical $\omega_i$s.
As a consequence, this implies that the simplifying assumption of constant $\omega_i$ does not change the modeling results significantly. 
    
\subsubsection{Degeneracy}   
\label{sec:Degeneracy}
\begin{figure}
    \centering
    \includegraphics[width=1\textwidth]{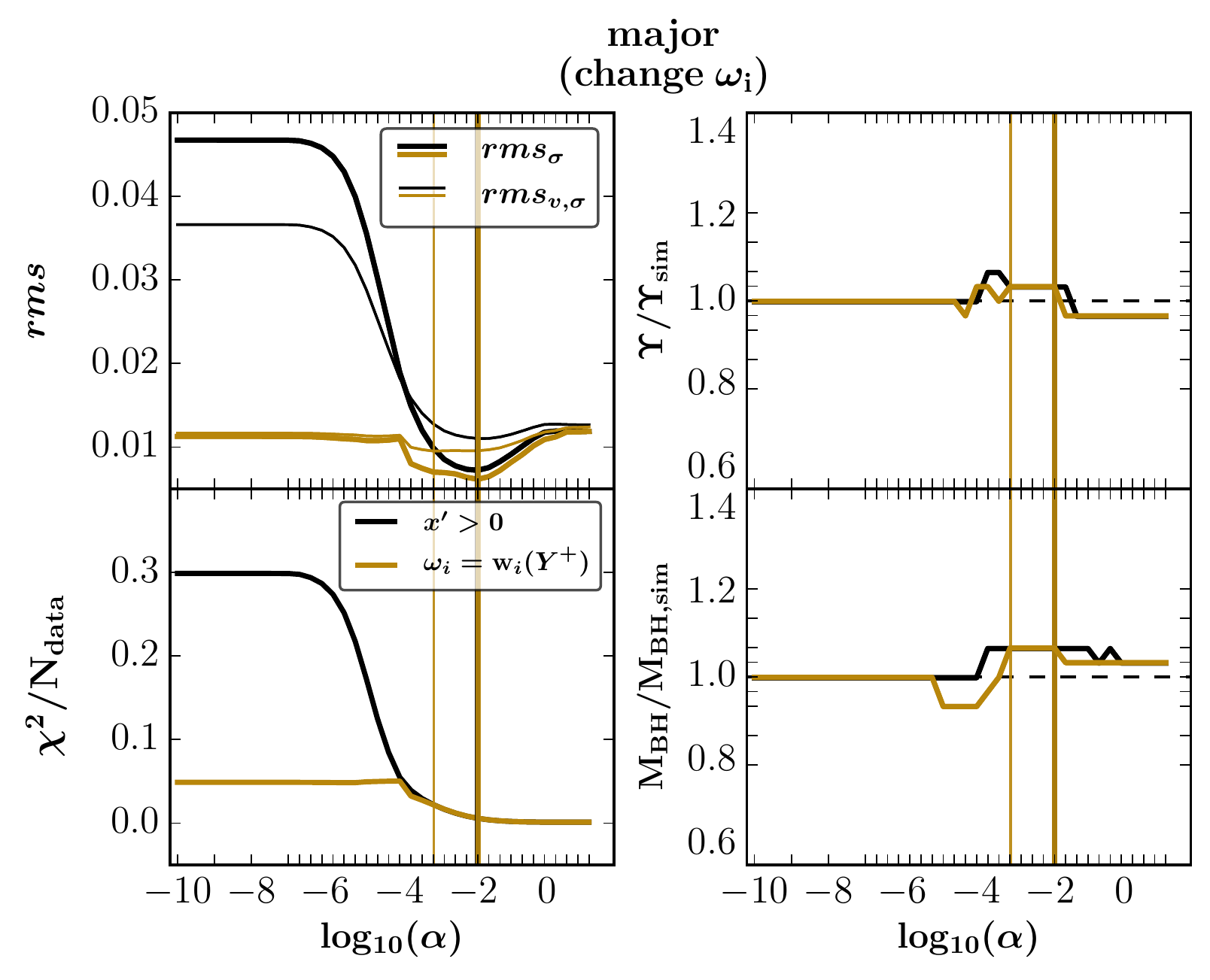}
    \caption{Choice of regularization and 1-dimensional mass recovery results for the positive half of the major axis projection when using the default constant initial orbital bias factors $\omega_i=$ const. (black line) and when using the best-fit orbital weights $w_i(\boldsymbol{Y}^+)$ of the positive half of the intermediate axis as orbital bias factors, i.e., $\omega_i=w_i(\boldsymbol{Y}^+)$ (brown line). The additional information from the second projection axis appears to reduce the degeneracy leading to a minor improvement in the internal moments reproduction as seen in the lower $rms_\sigma$- and $rms_{v,\sigma}$-values. }
    \label{fig:Figure11}
\end{figure}
    When redoing the just described analysis (Subsection~\ref{sec:Changing the orbital bias factors omegai}) but using the orbital bias factors of the right half of the \textit{intermediate} axis projection $\omega_i(\boldsymbol{Y}^+)$ (brown line in Fig.~\ref{fig:Figure11}) as initial values for remodeling $\boldsymbol{X}^+$ we gain a minor improvement in the reconstruction of the internal moments since the $rms_{\sigma}$- and $rms_{v,\sigma}$-values are a bit smaller compared to the default model (black line), though the mass recovery shows equal results. 
    The fact that, for the same quality of fit (i.e. same $\chi^2$), the $rms$ of the internal moments is smaller for the model with $\omega_i$ assumed to be equal to orbital weights from another modeled projection direction (in this case from the intermediate axis projection $\boldsymbol{Y}$) suggests that these $\omega_i$ contain some information in addition to the kinematics of the given projection (in this case the major axis projection $\boldsymbol{X}$). The fact that information from another line of sight can improve the model is not surprising. In fact, it shows that some of the already very small residual $rms$ in the internal velocity moments is due to remaining degeneracies in the recovery of the orbital weights. These degeneracies seem to be surprisingly small. However, our results imply that as long as the deprojected light profile and normalized DM halo are known and the orbit sampling is dense, masses and anisotropies can be recovered with very high accuracy, independent of the viewing angle.
    
\subsubsection{Change in orbit sampling technique}    
\label{sec:Change in orbit sampling technique}
\begin{figure}
    \centering
    \includegraphics[width=1.0\textwidth]{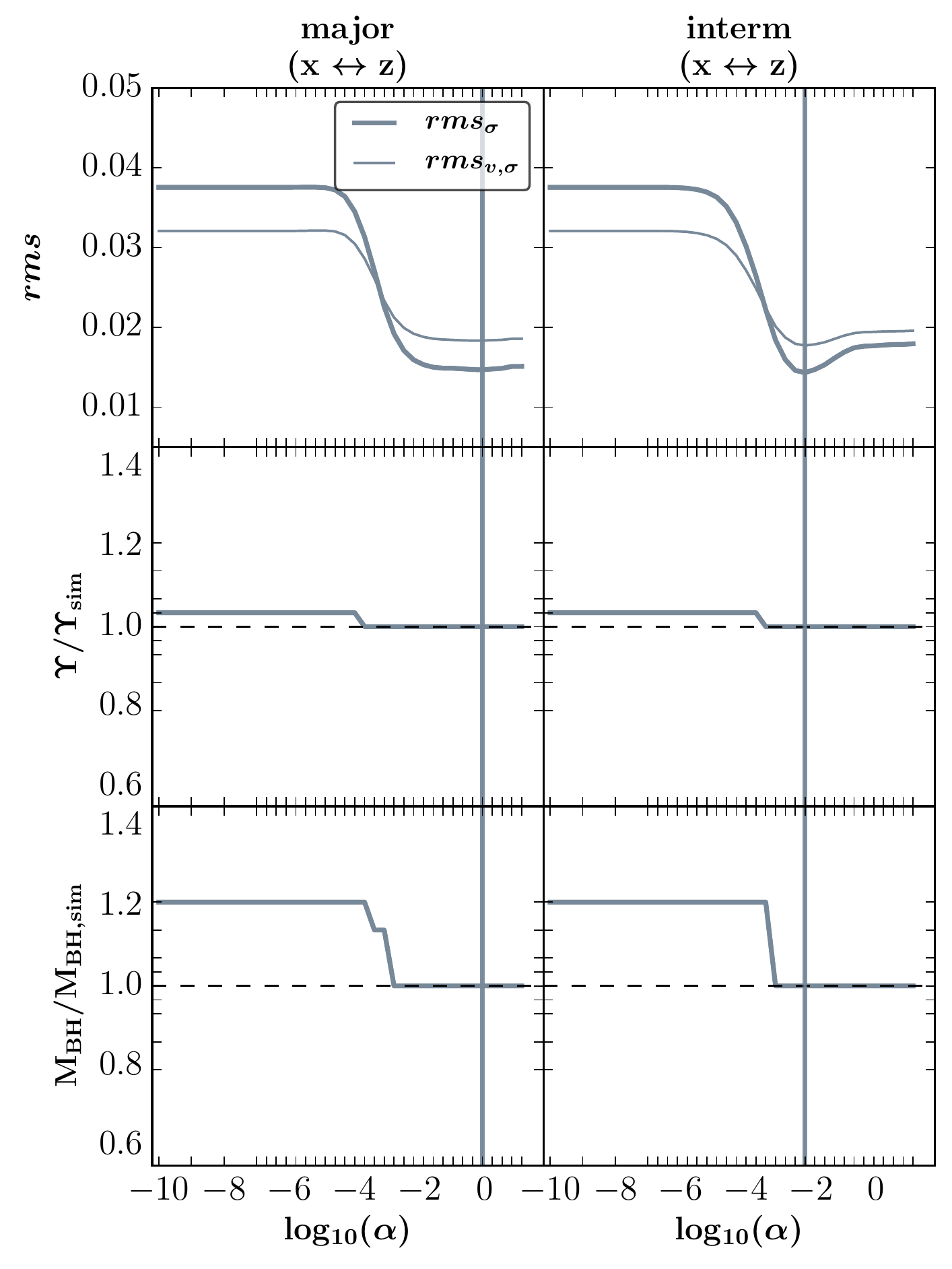}
    \caption{Choice of regularization and 1-dimensional mass recovery results for the positive half of the major axis projection (left column) and positive half of the intermediate axis projection (right column) when changing the $x$- and $z$-coordinates in the simulation so that the orbit sampling in \texttt{SMART} changes from a sampling of $v_\phi$ to $v_\theta$. Whereas the mass recovery gets slightly improved, the $rms_\sigma$- and $rms_{v,\sigma}$-profiles show slightly deteriorated values. Overall, both techniques basically lead to the same results.}
    \label{fig:Figure12}
\end{figure}
     Even though our results already prove the efficient ability of our orbit library to completely reproduce all necessary orbits in a triaxial potential, we want to check the robustness of our orbit sampling method (cf. App.~\ref{sec:Orbital Representation of the Phase Space}). We therefore change the angular momentum sampling direction from the minor axis to the major axis. This corresponds to a change of $v_{\phi}$- to $v_{\theta}$-sampling. 
    The basic idea is that a more homogeneous angular momentum sampling along the major axis instead of the minor axis might improve the mass recovery. Thus, we change the major and minor axis coordinates within the simulation and rerun the models for the new major- (left column in Fig.~\ref{fig:Figure12}) and intermediate-axis-kinematics (right column in Fig.~\ref{fig:Figure12}). With this, the angular momentum orbit sampling is proceeded along the major axis of the simulated remnant. The results are shown in Fig.~\ref{fig:Figure12}. 
    One can see that both orbit libraries produce the same results. In fact, the modified orbit sampling procedure reveals a slightly better mass recovery but slightly worse internal moments reproduction (higher $rms_{\sigma}$- and $rms_{v,\sigma}$-values) for equally good kinematic fits.  Nevertheless, regardless of the chosen angular momentum sampling axis, our orbit sampling technique (cf. App.~\ref{sec:Orbital Representation of the Phase Space}) of creating initial conditions, which belong to certain energy shells and angular momentum sequences, has proven to be a highly efficient technique. It produces a general and complete set of orbits for triaxial potentials being able to deal with changing structure in the integrals-of-motion space since it manages to reproduce all relevant properties and to radially adapt itself to the more spherical center as well as the more prolate outskirts of the simulated galaxy. \\ 
    
\subparagraph*{Summary of Robustness and Uniqueness Checks.}
In conclusion, these checks prove that \texttt{SMART} is robust against minor internal modifications as well as input data changes. \texttt{SMART} has proved its ability to handle with noisy and psf convolved data. Furthermore we have shown that constant orbital bias factors are a good approximation. 
Even though it is impossible to get observations from two viewing points, we have demonstrated that this would allow to reduce the minor degeneracies allowed by the kinematic data even further. We have verified this by using the orbital bias factors from a second projection direction. Moreover, the results are not affected by changing the orbit sampling technique from setting up $L_z$-sequences to setting up $L_x$-sequences. This shows that, as expected, the choice of sampling axis is not decisive and that the orbit sampling routine is universal. \\

\section{The quasi-uniqueness of the anisotropy reconstruction when fitting full LOSVDs}
\label{sec:The quasi-uniqueness of the anisotropy reconstruction when fitting full LOSVDs}

The results of the previous Section have shown that the anisotropy of the $N$-body merger remnant can be reconstructed with very high accuracy from the Schwarzschild models that we fitted to the full LOSVDs. This not only demonstrates the high accuracy of our orbit superposition model, but it also implies that when models can exploit the full information contained in the entire LOSVDs, the remaining degeneracy in the recovery of the distribution function can not affect the anisotropy or mass recovery significantly. In this Section we want to use the maximum entropy technique to explore this in more depth.

\subsection{The Maximum Entropy Technique and the Mathematical Structure of the Solution Space} 
\label{sec:Maximum Entropy Technique and Structure of Solution Space} 
As already described in Section~\ref{sec:Orbit superposition}, we solve for the orbital weights $w_i$ by using a maximum entropy 
technique (cf. eqs. \ref{eq:costfunc} and \ref{eq:smoothing}). 
The $\chi^2$ term in Equation~\ref{eq:costfunc} contains the kinematical
constraints and is the deviation between observed LOSVDs $\ldat$ and the model prediction $\lmod$, i.e. the weighted sum over the contributions of all orbits to LOSVD $j^\prime$ and line-of-sight velocity bin $k$, 
\begin{equation}
{\lmod}^{j^\prime,k} \equiv \sum_{i=1}^{\norb} w_i \, {\lorb}^{j^\prime,k,i}.
\end{equation}

It is convenient to think of the observed $\kdat$ and the model
predictions $\kmod$ as vectors with $\nkin = \nbin \times \nvel$
elements. Then,
\begin{equation}
\label{eq:kinmat}
\kmod = \mathrm{\korb} \cdot \vect{w},
\end{equation}
where $\mathrm{\korb}$ is a matrix with $\nkin$ rows and $\norb$ columns. \\
In addition to the kinematical observations $\kdat$, 
the orbital weights $\vect{w}$ are subject to photometric constraints.
In analogy to Equation~\ref{eq:kinmat}:
\begin{equation}
\phmod = \mathrm{\phorb} \cdot \vect{w},
\end{equation}
where $\phmod$ is a vector with the model predictions for the 3D
luminosity density at spatial position $j_3$ in the galaxy ($j_3 =
1,\ldots,\nphot$). To guarantee the self-consistency of our model, the
respective observed $\phdat$ are not included via a $\chi^2$ term.
Instead, we treat them as boundary conditions for the fit:
\begin{equation}
\label{eq:constraints}
\phdat \stackrel{!}{=} \phmod.
\end{equation}
Hence, we seek for the maximum of Equation~\ref{eq:costfunc} subject to
the linear equality constraints 
\begin{equation}
\label{eq:densityconstraints}
\phdat - \mathrm{\phorb} \cdot \vect{w} = 0.
\end{equation}
For convenience, we normalise the $\phdat$ such that
$\sum \phdat = 1$. Since we are only interested in positive orbital
weights (see below), the orbital weights obey
$0 \leq w_i \leq 1$ and we can restrict the maximization to the
respective $\norb$-dimensional convex quader. \\
To show that Equation~\ref{eq:costfunc} has a unique global maximum that
can be controlled through the bias factors $\omega_i$ it is convenient
to consider the equivalent minimisation problem for
$f \equiv - \hat{S}$ given by multiplying Equation~\ref{eq:costfunc} with
$-1$. \\
Let us first consider Equation~\ref{eq:costfunc} without the entropy term
$S$. The $\chi^2$ term can be written as
\begin{equation}
\chi^2 = -\kdat^{T} \mathrm{C_v} \kdat + 2 \kdat^{T} \mathrm{C_v}
\mathrm{\korb}  \vect{w} - \vect{w}^{T} \mathrm{\korb}^{T} \mathrm{C_v} \mathrm{\korb} \vect{w},
\end{equation}
where $\mathrm{C_v}$ is the covariance matrix of the observed
LOSVDs and is positive definite. The Hesse matrix of $\chi^2$
reads
\begin{equation}
\mathrm{\nabla^2} \chi^2 = 2  \mathrm{\korb}^{T} \mathrm{C_v} \mathrm{\korb} .
\end{equation}
Because $\mathrm{\korb}$ is positive by construction, the symmetric
matrix $\mathrm{\nabla^2} \chi^2$ is at least positive semi-definite
and $\chi^2$ is convex. \\
The minimisation of $\chi^2$ alone, subject to the linear equality
constraints Equation~\ref{eq:densityconstraints}, is therefore a convex
optimisation problem with affine equality constraints in standard
form. As such, it only has a global minimum \citep[e.g.][]{convex}. In
general
we cannot assume that $\chi^2$ is {\it strictly} convex (i.e. that
$\mathrm{\nabla^2} \chi^2$ is positive definite). The set of orbital
weights that solve $\chi^2(\vect{w}) = \chi^2_\mathrm{min}$ may 
therefore be non-unique. This is not surprising given that the
linear equation
\begin{equation}
\vect{\nabla} \chi^2 = 2 \vect{w}^{T} \mathrm{\korb}^{T} \mathrm{C_v}
\mathrm{\korb} + 2 \kdat^{T} \mathrm{C_v}
\mathrm{\korb} \equiv 0
\end{equation}
will in general be underconstrained if
$\norb > \nkin$.  As already mentioned above, the reconstruction of the
distribution function is not unique even if the deprojected density
and the potential are known. We give an example in Appendix~\ref{sec:Optimization Algorithm and Testing}. \\
Because the Hesse matrix of $-S$ is diagonal with the $i$-th element
equal to $1/w_i$, the entropy $-S$ is {\it strictly} convex. In
contrast to the case of $\chi^2$, the set of orbital weights that
minimise $-S$ (or, equivalently, that maximise $S$) is {\it unique}.
For the entropy alone this is easy to see since
\begin{equation}
\nabla S = - \log \frac{w_i}{\omega_i} - 1 \equiv 0
\end{equation}
can be solved analytically: $w_i = \exp(-1) \cdot \omega_i$. The constraints
from Equation~\ref{eq:densityconstraints} will shift the solution, but
the strict convexity still guarantees it to remain unique \citep[e.g.][]{convex}. \\
For general $\alpha>0$, the Hesse matrix of $f$ is the sum
$\mathrm{\nabla^2} (-S) + \alpha \mathrm{\nabla^2} \chi^2$ and $f$ is
always strictly convex. Equation~\ref{eq:costfunc} subject to the
constraints (\ref{eq:densityconstraints}) hence always has a unique
solution. \\
As already mentioned above, 
from an algorithmic point of view it is advantageous to have a unique
solution. From the physical point of view this is not desirable
because any algorithm that picks up only one of the potentially many
solutions that minimise $\chi^2$ in Equation~\ref{chi squared} may lead to
a {\it bias}. However, suppose $\vect{s_1}$ and $\vect{s_2}$ are two
solutions which lead to the same $\chi^2_\mathrm{min}$. By setting
$\omega_i = s_{1,i} \cdot \exp(1)$ we can make $\vect{s_1}$ the global
solution of Equation~\ref{eq:costfunc}. Likewise, by setting
$\omega_i = s_{2,i} \cdot \exp(1)$ we can make $\vect{s_2}$ the global
solution of Equation~\ref{eq:costfunc}. This shows that the
maximum-entropy formulation of the problem allows in principle the
reconstruction of the entire solution space (via variation of the
$\omega_i$).

\subsection{Testing the Uniqueness of the Anisotropy Recovery}
\label{sec:Experiment to testing the Uniqueness of the Anisotropy Recovery}

\begin{figure*}
    \centering
    \includegraphics[width=0.9\textwidth]{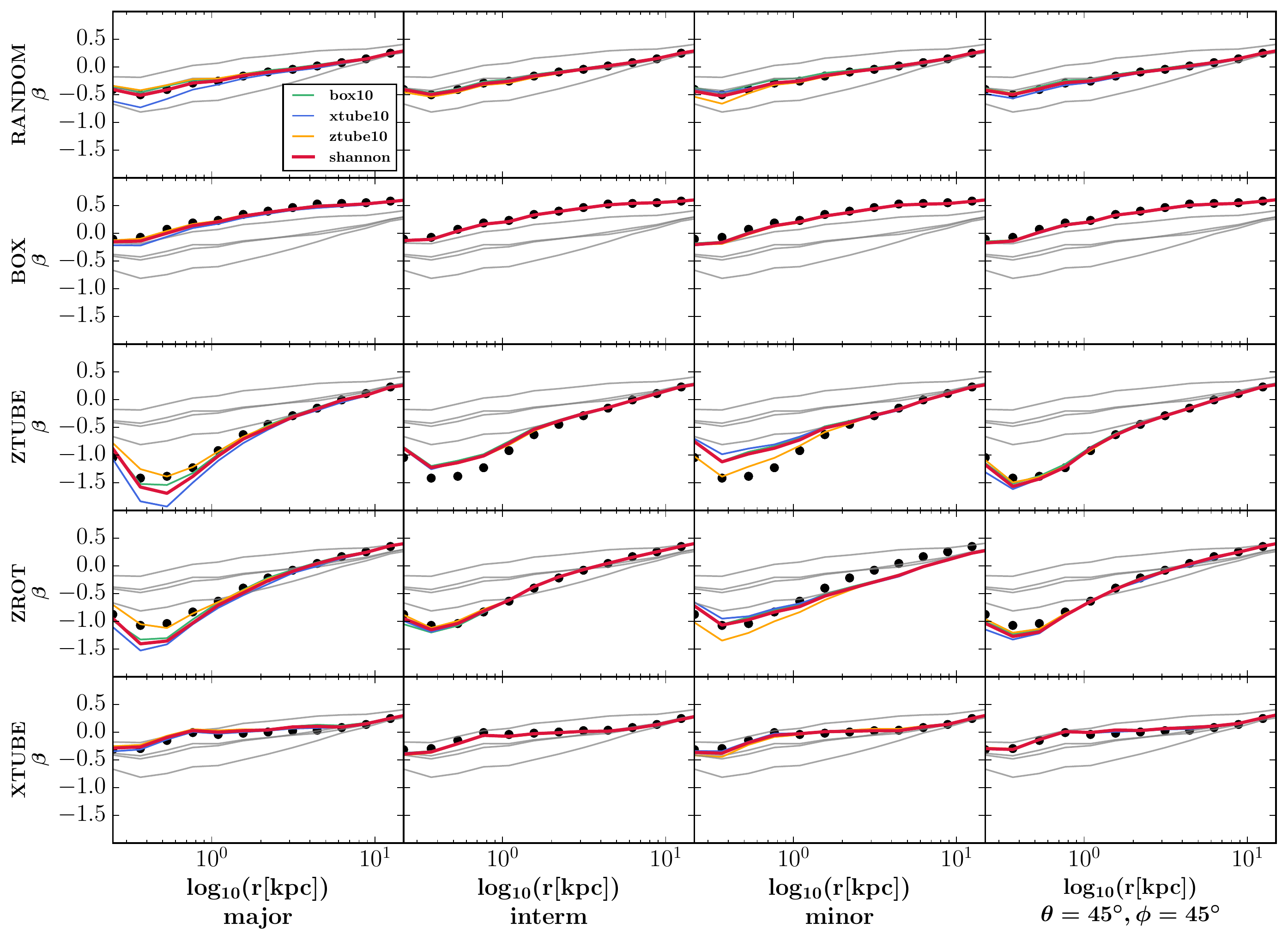}
    \caption{Recovery of the anisotropy profiles of different input toy models by SMART fits with different entropy functions. The input models (black dots) are constructed to have specific orbit classes overrated, resulting in different anisotropy profiles (the overrated orbit type is labelled on the y-axis of each individual row, details in the text). All the five toy models can be well recovered. This is independent of the choice of the projection axis (different columns; from left to right major, intermediate, minor and a diagonal axis) and from the assumed entropy function in the fit (colored lines). For example, the anisotropy profile of the toy model with an overpopulation of box/chaotic orbits (called BOX; second row) is well reproduced by models maximising the Shannon-entropy (red lines, labelled shannon) but also with other entropy functions that use a ten times higher bias factor for box-orbits (green, labelled box10), for $x$-tubes (blue, xtube10) or for $z$-tubes (orange, ztube10). The grey lines correspond to the anisotropies implied by maximising these four different entropy functions without fitting the kinematic data. They symbolise the variety of anisotropy profiles which are in principle possible for different choices of $\omega_i$. After fitting the kinematic data, the average deviation (averaged over all radii and toy models fitted with different entropy techniques) between recovered and input anisotropy is very small, $|\Delta \beta| = 0.05$.}
    \label{fig:Figure13}
\end{figure*}

The previous Section has shown that the recovery of the distribution function (or, equivalently, of the orbital weights) is in general not unique, even when the deprojection and the potential are known. The maximum entropy technique recovers one of the many possible solutions, which is unique for every given set of $\omega_i$. Variation of the $\omega_i$ allows to sample the full solution space. On the other hand, the fits of the $N$-body merger remnant have shown that the anisotropy recovery is very accurate and stable even to variations of the $\omega_i$. Moreover, in Appendix~\ref{sec:Optimization Algorithm and Testing} we explicitly construct two different phase-space distribution functions that fit a given set of kinematics equally well. Even though they have different orbital weights, they reveal very similar anisotropies in the second-order velocity moments. This suggests that while the recovery of the full distribution function is non-unique, the anisotropy of the second-order moments is actually very well constrained by the information contained in the full LOSVDs. \\
To investigate this further, we create kinematic data of several toy models with different intrinsic properties: with randomised orbit weights (called RANDOM in the following), with an overpopulation of box/chaotic orbits (called BOX), with an overpopulation of $z$-tubes (called ZTUBE), with an overpopulation of only the prograde $z$-tubes (called ZROT) and with an overpopulation of $x$-tubes (called XTUBE).
We then fit the mock kinematic data of these toy models under different choices for the $\omega_i$, trying to push 
the fitted orbit model towards extreme shapes. The goal is to test how accurate and stable the recovery of the internal velocity anisotropy is, when we fit the entire information contained in the LOSVDs. \\
The toy models are constructed as maximum-entropy models (i.e. through maximisation of eq.~\ref{eq:costfunc} with $\alpha = 0$; cf.~\citealt{Thomas07}). For the BOX model,
we increase the $\omega_i$ of all box orbits by a factor of 1000, for the XTUBE and ZTUBE models we increase the $\omega_i$ of the respective tube orbits by the same amount. For the ZROT toy model we increase the $\omega_i$ only for prograde z-tubes and for the RANDOM model we use randomised $\omega_i$ (cf. Appendix~\ref{sec:Optimization Algorithm and Testing}). The weights are then still forced to satisfy the density constraints of the N-body simulation. For each toy model, we create kinematic mock data for  
four different projections (major-, intermediate-, minor-axis projection and the $\theta=45^\circ, \phi=45^\circ$-projection).
We then model every projection of all toy models (in total 20 different input data) and test four different methods for the fits: We use (i) our default constant orbital bias factors, i.e. $\omega_i=1$ corresponding to the Shannon-entropy (abbreviated as 'shannon' in Fig.~\ref{fig:Figure13}), (ii) increased orbital bias factors by a factor of 10 for the box/chaotic orbits (box10), (iii) increased orbital bias factors by a factor of 10 for the $z$-tubes (ztube10) and (iv) increased orbital bias factors by a factor of 10 for the $x$-tubes (xtube10). \\
Figure~\ref{fig:Figure13} shows the resulting anisotropy-profiles (when averaging over shells) for these models. The different rows correspond to the different input toy models and the different columns show the individual projection directions. The dotted data points symbolise the anisotropy profiles of the input toy models (which are of course the same for the different projections). The colored lines (green, blue, orange, red) show the recovered anisotropies of \texttt{SMART} fits using entropy functions with enhanced bias factors for specific orbit types as described under (i)-(iv) above.  The grey lines, for comparison, show the anisotropies that result when we maximise the above entropy functions (i)-(iv) without fitting the mock LOSVDs of the toy models. The grey lines therefore illustrate the variety of different anisotropy profiles that can be constructed by varying the orbital bias factors $\omega_i$. They also indicate the range of different anisotropy profiles that are consistent with the given density distribution. \\
As one can see, even though the entropy functions tend to push the fits into extreme directions, the range of anisotropy profiles recovered after the fit to the LOVSDs is very narrow. When averaging over all radii and toy models fitted with different entropies, the mean deviation to the input models (dots) is $|\Delta \beta| = 0.05$. The average spread in beta inside the sphere of influence  is slightly larger $|\Delta \beta(r<r_{\mathrm{SOI}})| = 0.09$ than $|\Delta \beta(r_{\mathrm{SOI}}<r<r_{\mathrm{FOV}})| = 0.03$ outside $r_{\mathrm{SOI}}$. One possible explanation for this might be the increase of degrees of freedom of spherical orbits near the center. In addition, because $\beta$ involves the ratio of the intrinsic dispersions, the same fractional error in the intrinsic dispersions results in a 4 times larger $|\Delta \beta|$ when the anisotropy is as tangential as $\beta = -1.5$  compared to the isotropic case. Overall, the Shannon entropy (red line) is able to recover the beta anisotropy best. This is our default entropy used in \texttt{SMART}.\\
These results together with Section~\ref{sec:Results} strongly suggest that the information contained in the full LOSVDs constrains the anisotropy in the second-order velocity moments very well. In turn, this is the reason why our models can reproduce the mass of the black hole and of the stars in the N-body simulation very well.  \\
At larger radii, solely the reconstruction of the intrinsic anisotropy of the toy model with enhanced prograde $z$-tubes (i.e. ZROT) turns out to be difficult when viewed along the minor axis. This, however, is expected since any rotation around the minor-axis can not be observed and, thus, not be reconstructed from this viewing direction. Since we use equal $\omega_i$ for prograde and retrograde orbits in our {\it fitted} models, these models do not have intrinsic rotation in the $z$-tubes for this projection. Consequently, the tangential velocity {\it dispersion} is larger than in the toy model and the fitted $\beta$ becomes too negative. We checked that if we use the true second-order velocity moment rather than the velocity dispersion in the tangential direction, then the differences between the outer profiles of the ZROT model and the fits along the minor axis disappear.

\begin{figure*}
    \centering
    \includegraphics[width=0.9\textwidth]{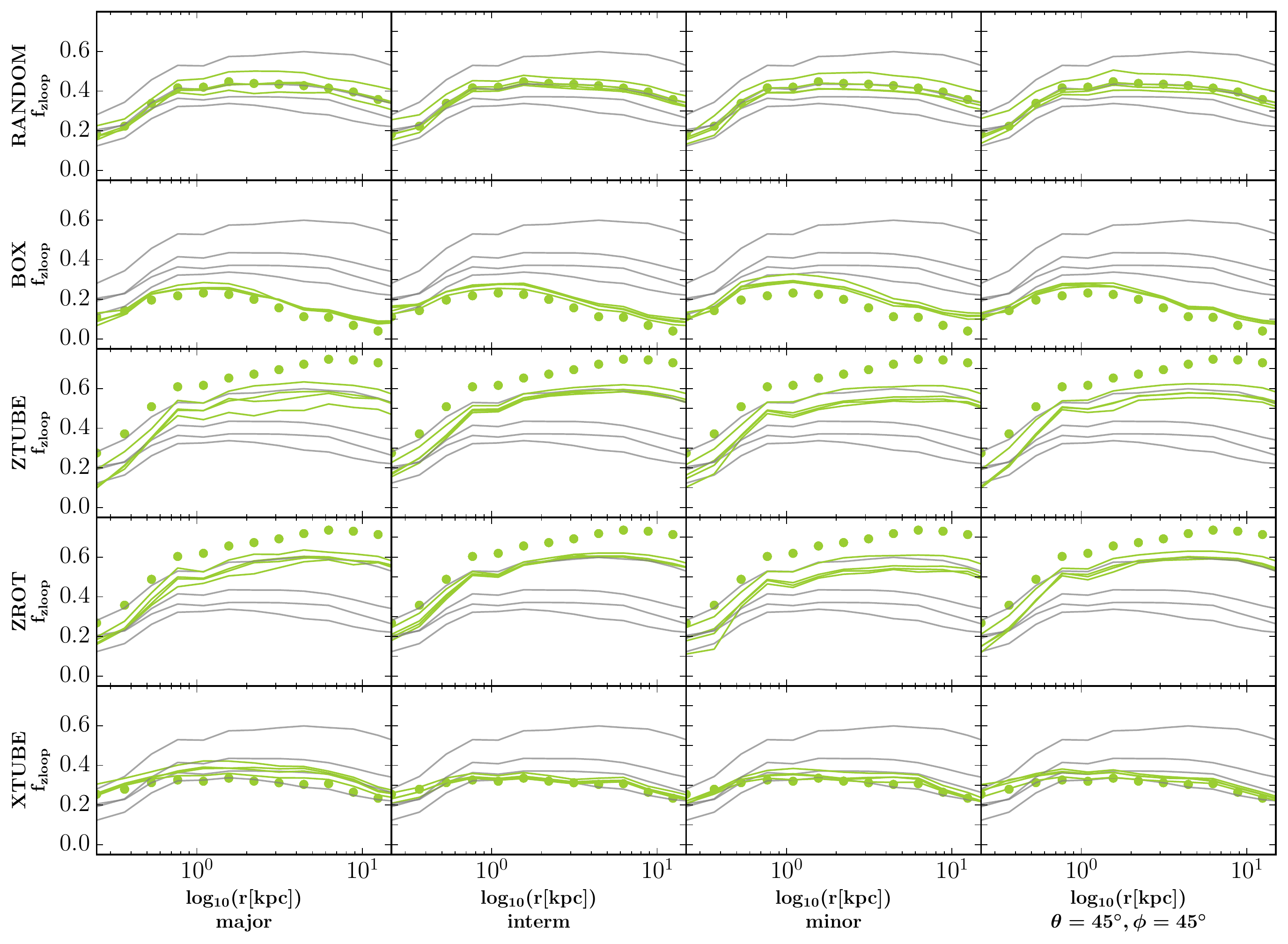}
    \caption{Recovery of the $z$-loop orbit fractions of different input toy models by models with different entropy methods. We here show the same analysis as in Figure~\ref{fig:Figure13} but now for the reconstruction of the fraction of orbits classified as $z$-tubes. The color coding is adapted to Fig.~\ref{fig:Figure4} and~\ref{fig:Figure23}. Independent of the tested projection, the $z$-loop fractions of the individual input toy models are well recovered by the models using different  entropy methods. The same is true for the other orbit class fractions (see Fig.~\ref{fig:Figure28}-~\ref{fig:Figure30}).}
    \label{fig:Figure14}
\end{figure*}

For dynamical models which aim for a full phase-space reconstruction (like Schwarzschild models) and which use the full information encoded in the LOSVDs (see also \citealt{Vasiliev19}) the anisotropy should be recoverable with a typical error of $|\Delta \beta| \approx 0.05$. We solely found larger anisotropy discrepancies (up to $|\Delta \beta| = 0.5$) in extremely tangentially biased regions inside the sphere of influence. \\
As an example of the corresponding accordance of the reconstructed orbit fractions, Figure~\ref{fig:Figure14} shows the case of $z$-tubes. The intended overpopulation of the $z$-tubes in the ZTUBE toy model (third row) and the ZROT toy model (fourth row) can be clearly seen in comparison to the other toy models. Independent of the chosen line of sight and entropy method (i.e. the bias factors $\omega_i$), the fraction of $z$-tubes is qualitatively recovered and follows the enhancement tendencies. The orbit fractions however are less well determined than the anistropy by the data and show a stronger dependence on the entropy. The same is true for the $x$-tubes, box/chaotic and spherical/Kepler orbits shown in Figures~\ref{fig:Figure28} to~\ref{fig:Figure30}.  


\section{Intermediate axis rotation}
\begin{figure}
    \centering
    \includegraphics[width=0.95\textwidth]{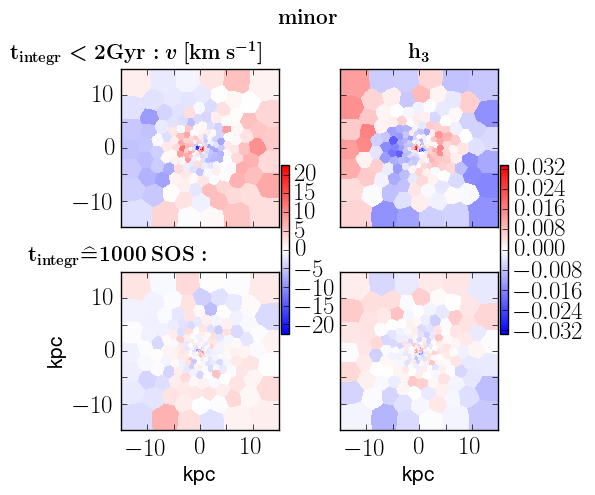}
    \caption{Fitting mock kinematic input data containing intermediate axis rotation. When modeling artificial LOSVDs showing a net rotation along the intermediate axis, \texttt{SMART} is able to reproduce a y-rotation signal when integrating the orbits for a limited integration time of 2Gyrs (first row). The y-rotation becomes visible in the $v-$ (left panel) and $h_3$ (right panel)-maps. When integrating for 1000 SOS-crossings (second row), the y-rotation signal cannot be fitted by the model any more.}
    \label{fig:Figure15}
\end{figure}
One of the first investigations to check whether tube orbits with net rotation around the intermediate axis are stable in a triaxial ellipsoid was done by  \citet{Heiligman79}, who studied a triaxial model with fixed axis ratios of 1:1.25:2 by numerical methods and stated that “Y-tube orbits nearly certainly do not exist in the adopted model”. Also \citet[p. 263]{Binney08} assert that “tube orbits around the intermediate axis are unstable” in a triaxial potential. \citet{Adams07} analysed the orbit instability of orbits in a triaxial cusp potential, which are initially confined to one of the three principal axes, under a perturbation along the perpendicular direction. They found that orbits around any of the principal axes are unstable to perpendicular motions. However, according to previous results, they again state that orbits around the intermediate axis are more likely to be unstable. This instability is strongest for original box orbits lying in the x-z plane when the axis ratio of these two axes in the original plane is largest. \\
Our orbit classification routine in \verb'SMART' finds no y-tubes in the sense that there is no sign conservation of the angular momentum along the intermediate axis over our default integration period of 100 SOS-crossings, agreeing with the aforementioned works done by other groups. However, some of the orbits integrated for the N-body models do show y-rotation for a limited time-span. When providing \verb'SMART' with artificial projected input line-of-sight velocity distributions that mimick a net rotation along the intermediate axis, the model is able to produce a y-rotation signal of the order of $10\mathrm{\,km\,s^{-1}}$ in the fit. A rotation of this magnitude is small but in principle detectable with today's telescopes' resolution. Fig.~\ref{fig:Figure15} shows the \verb'SMART' fit to the major-axis projection of the simulation (cf. Fig.~\ref{fig:Figure3}) when assuming the viewing angles of the minor-axis projection. With this, we simulate a hypothetical rotation along the intermediate axis. The top row shows the velocity- and $h_3$-map when stopping the orbit integration after 2 Gyrs if this is shorter than the time needed for 100 SOS-crossings. Indeed, the model reproduces a y-rotation signal.  The amplitude of this residual y-rotation becomes smaller and smaller when the orbital integration time is increased, and vanishes when all orbits are integrated for 1000 SOS-crossings (bottom row). \\
This analysis indicates that the model's triaxial potential, which is constructed based on the 3D density from the realistic $N$-body merger simulation, --contrary to expectations-- contains orbits with y-rotation for a physically relevant time span. What remains unclear at this moment is whether such a y-rotation indeed appears in real elliptical galaxies. 
If so, then the connection between kinematic misalignments and photometric twists is less constrained than often assumed when only rotation around the intrinsic long and short axes is considered.

\section{Discussion}
\label{sec:Discussion}
\subsection{Remaining sources of systematics}
All relevant properties of the simulated merger remnant galaxy were proven to be recovered with a convincing precision and the deviations we found are almost negligible.
The remaining deviations (in the $\sim 5-10\%$ level) can be either originated by \texttt{SMART} or the simulation. 
One remaining contribution to the scatter in the final mass recovery certainly comes from the finite binning resolution of the simulation data and of the \verb'SMART' models (see Sections~\ref{sec:Triaxial Schwarzschild code SMART} and~\ref{sec:The N-body simulation}). Especially the need of extrapolating the density towards the center due to limited resolution of the simulation holds uncertainties. \\
One  more inaccuracy is potentially induced by the softening length in the simulation used to avoid unrealistic 2-body encounters between massive particles. Force calculations for radii smaller than the softening length are consequently modified by the softening. In the close vicinity of the black hole, the stellar particles in the simulation are modeled using a nonsoftened algorithmic chain regularization technique (ARCHAIN; \citealt{Mikkola06, Mikkola08}) including post-Newtonian corrections (e.g., \citealt{Will06}). The particles outside this chain radius are treated by using softened gravitational force calculations based on the GADGET-3 \citep{Springel05} leapfrog integrator. The chain radius $r_{\mathrm{chain}}$ is chosen to be at least 2.8 times larger than the GADGET-3 softening length $\epsilon$ ($r_{\mathrm{chain}}>2.8\epsilon$) to ensure that the particles within the chain remain nonsoftened. The  softening length in the simulation for the stellar particles is $\epsilon_*=3.5\mathrm{\,pc}$ and the softening length for the dark matter particles is $\epsilon_{\mathrm{DM}}=100\mathrm{\,pc}$. The mass recovery of the global stellar mass-to-light ratio and dark matter scale factor will be unaffected by these relatively small values. However, there might be a remaining influence on the black hole mass recovery within $r_{\mathrm{SOI}}$. Nevertheless, this effect is expected to be small. 
    
\subsection{Comparison to other triaxial Schwarzschild Models} 
We found two other dynamical modeling codes in the literature using Schwarzschild's orbit superposition technique dealing with triaxiality by \citet{vandenBosch08} and \citet{Vasiliev19}.
The code \verb'FORSTAND' by \citet{Vasiliev19} is applicable to galaxies of all morphological types. When assuming that the deprojection and dark matter halo is known and provided to the code, the models of noise-free axisymmetric disc mock datasets taken from $N$-body simulations showed very weak constraints on $M_{BH}$: any value between zero and 5-10 times the true black hole mass was equally consistent with the data. Read from Fig. 2 in their paper, the stellar mass-to-light ratio showed a variation of around 20\%. \\
\citet{Jin19} tested the triaxial Schwarzschild code by \citet{vandenBosch08} by applying it to nine triaxial galaxies from the large scale, high resolution Illustris-1 simulation \citep{Vogelsberger14}, which provides a stellar and dark matter resolution of $\sim10^6M_{\odot}$. When fixing the black hole mass and allowing the model to deproject the mock dataset, the stellar mass within an average effective radius is underestimated by $\sim$24\% and the dark matter is overestimated by $\sim$38\%. Their averaged model results obtained from mock data with different viewing angles tend to be too radial in the outer regions with better anisotropy matches in the inner region.  \\
Of course, these results cannot be used for direct comparison due to a widely varying resolution and in case of the analysis by \citet{Jin19} the deprojection probably causes the major deviation. However, \texttt{SMART} for sure is able to add further progress in modeling triaxial galaxies and convinces with its proved precision. 

\section{Summary and Conclusion}
\label{sec:Summary and Conclusion}
We have developed a new triaxial dynamical Schwarzschild code called \texttt{SMART} and tested its efficiency and reliability by applying it to an $N$-body merger  simulation including supermassive black holes. 
The simulation was deliberately selected due to its high accuracy, reasonable formation process, realistic internal structure and ability to precisely calculate the dynamics close to the central black hole. This ensured the possibility to check whether \texttt{SMART} is able to recover all relevant properties including the mass of a supermassive black hole of a realistic triaxial galaxy when providing the deprojected light profile and normalized DM halo.\\
\texttt{SMART} is assembled with the feature to compute the potential and force by expansion into spherical harmonics allowing to deal with non-parametric densities and halos. Its orbit library contains 50000 integrated orbits which are set up by creating random initial radial and velocity values within given energy shells and angular momentum sequences and by filling the surfaces of section. This ensures the ability to adapt itself to a radially changing number of integrals of motion. The orbit superposition is executed by maximizing an entropy-like quantity and by using the full line-of-sight velocity distributions instead of only Gauss-Hermite parameters alone. \\
These benefits enable \texttt{SMART} to reconstruct all relevant properties and features of the merger remnant with an excellent precision. We will now recap the requirements which were set on \texttt{SMART} and proved to be fulfilled in this analysis:
\begin{itemize}
    \item[-] \texttt{SMART} is able to reproduce the anisotropy profile and internal velocity dispersions with an $rms_\sigma$ of only $1.2\%$. 
    \item[-] \texttt{SMART} reproduces the stellar mass-to-light ratio,
    \item[-] black hole mass,
    \item[-] and mass scale factor of the dark matter density profile with a precision on the $5$--$10\%$ level. To our knowledge this is the first time that the intrinsic precision at given deprojected light profile has been quantified to be so high. This sets the basis for further investigations of the whole modeling procedure.
    \item[-] \texttt{SMART} well fits the line-of-sight velocity distributions with mean values and deviations of only $\bar{v}=(0.11\pm2.09)\mathrm{\,km\,s^{-1}, } \text{ } \bar{\sigma}=(309.75\pm2.54)\mathrm{\,km\,s^{-1}, } \text{ } \bar{h_3}=0.00\pm0.01\text{ and }\bar{h_4}=0.01\pm0.01$.
\end{itemize}

For the determination of these accuracy values, the simulation is modeled from up to five different projections. The mass recovery precision can be achieved for noiseless as well as noisy and psf convolved kinematic input data.

We extensively discuss that the maximum-entropy technique provides an elegant technique to study the range of possible orbit distributions consistent with a given set of data. Our tests with the $N$-body data and with additional toy models strongly suggest that when the full information contained in the entire LOSVDs is used to constrain the model, then the remaining degeneracies in the recovery of the exact phase-space distribution function do not affect 'macroscopic' properties of the galaxy models, like the anisotropy in the second-order velocity moments. This is the basis for the very good reconstruction of the orbital structure and mass of the black hole and stars with \texttt{SMART}. 

It was shown that the orbit library is robust against axis changes and generates a complete set of well superpositioned orbits necessary to model a triaxial galaxy with all corresponding internal structures. \\
Also the accurate mass parameter recovery accomplished by \texttt{SMART} suggests only minor degeneracies contained in the projected kinematic data, provided that the deprojection is known. We showed that these remaining minor degeneracies could in principle be narrowed even more if information about the orbital bias factors from a second projection direction were provided. 

When analysing the elevation and azimuthal internal velocity distributions of the simulation we find that the central radial bins show two maxima. This corresponds to the negative $h_4$-parameter in the center and the strong tangential anisotropy produced during the core formation. \texttt{SMART} is able to reconstruct this phenomenon with an accuracy of $\sim 7\%$. 

One more discovery of scientific interest is intermediate axis rotation which is produced by orbits contained in the model's orbit library representing the simulation's triaxial potential. Independent of the question whether such intermediate axis rotation really appears in the real universe, it was shown that our model contains orbits with y-rotation whose stability was empirically found to be maintained up to at least 2 Gyrs. \\

\section*{Acknowledgements}
We acknowledge the support by the DFG Cluster of Excellence "Origin and Structure of the Universe". The dynamical models have been done on the computing facilities of the Computational Center for Particle and Astrophysics (C2PAP) and we are grateful for the support by F. Beaujean through the C2PAP.

\section*{Data Availability Statement}
The data underlying this article will be shared on reasonable request to the corresponding author.

\clearpage
\bibliographystyle{mnras}
\bibliography{bib}

\begin{thebibliography}{}
\makeatletter
\relax
\def\mn@urlcharsother{\let\do\@makeother \do\$\do\&\do\#\do\^\do\_\do\%\do\~}
\def\mn@doi{\begingroup\mn@urlcharsother \@ifnextchar [ {\mn@doi@}
  {\mn@doi@[]}}
\def\mn@doi@[#1]#2{\def\@tempa{#1}\ifx\@tempa\@empty \href
  {http://dx.doi.org/#2} {doi:#2}\else \href {http://dx.doi.org/#2} {#1}\fi
  \endgroup}
\def\mn@eprint#1#2{\mn@eprint@#1:#2::\@nil}
\def\mn@eprint@arXiv#1{\href {http://arxiv.org/abs/#1} {{\tt arXiv:#1}}}
\def\mn@eprint@dblp#1{\href {http://dblp.uni-trier.de/rec/bibtex/#1.xml}
  {dblp:#1}}
\def\mn@eprint@#1:#2:#3:#4\@nil{\def\@tempa {#1}\def\@tempb {#2}\def\@tempc
  {#3}\ifx \@tempc \@empty \let \@tempc \@tempb \let \@tempb \@tempa \fi \ifx
  \@tempb \@empty \def\@tempb {arXiv}\fi \@ifundefined
  {mn@eprint@\@tempb}{\@tempb:\@tempc}{\expandafter \expandafter \csname
  mn@eprint@\@tempb\endcsname \expandafter{\@tempc}}}

\bibitem[\protect\citeauthoryear{{Aarseth} \& {Binney}}{{Aarseth} \&
  {Binney}}{1978}]{Aarseth78}
{Aarseth} S.~J.,  {Binney} J.,  1978, \mn@doi [\mnras]
  {10.1093/mnras/185.2.227}, \href
  {https://ui.adsabs.harvard.edu/abs/1978MNRAS.185..227A} {185, 227}

\bibitem[\protect\citeauthoryear{{Adams}, {Bloch}, {Butler}, {Druce}  \&
  {Ketchum}}{{Adams} et~al.}{2007}]{Adams07}
{Adams} F.~C.,  {Bloch} A.~M.,  {Butler} S.~C.,  {Druce} J.~M.,   {Ketchum}
  J.~A.,  2007, \mn@doi [\apj] {10.1086/522581}, \href
  {https://ui.adsabs.harvard.edu/abs/2007ApJ...670.1027A} {670, 1027}

\bibitem[\protect\citeauthoryear{Anderson et~al.,}{Anderson
  et~al.}{1999}]{Anderson99}
Anderson E.,  et~al., 1999, {LAPACK} Users' Guide, third edn.
Society for Industrial and Applied Mathematics, Philadelphia, PA

\bibitem[\protect\citeauthoryear{Audet \& {Dennis, Jr.}}{Audet \& {Dennis,
  Jr.}}{2006}]{Audet06}
Audet C.,  {Dennis, Jr.} J.,  2006, \mn@doi [SIAM Journal on Optimization]
  {doi:10.1137/040603371}, 17, 188

\bibitem[\protect\citeauthoryear{{Bailin} \& {Steinmetz}}{{Bailin} \&
  {Steinmetz}}{2005}]{Bailin05}
{Bailin} J.,  {Steinmetz} M.,  2005, \mn@doi [\apj] {10.1086/430397}, \href
  {https://ui.adsabs.harvard.edu/abs/2005ApJ...627..647B} {627, 647}

\bibitem[\protect\citeauthoryear{{Barnes}}{{Barnes}}{1992}]{Barnes92}
{Barnes} J.~E.,  1992, \mn@doi [\apj] {10.1086/171522}, \href
  {https://ui.adsabs.harvard.edu/abs/1992ApJ...393..484B} {393, 484}

\bibitem[\protect\citeauthoryear{{Barnes} \& {Hernquist}}{{Barnes} \&
  {Hernquist}}{1996}]{Barnes96}
{Barnes} J.~E.,  {Hernquist} L.,  1996, \mn@doi [\apj] {10.1086/177957}, \href
  {https://ui.adsabs.harvard.edu/abs/1996ApJ...471..115B} {471, 115}

\bibitem[\protect\citeauthoryear{{Begelman}, {Blandford}  \& {Rees}}{{Begelman}
  et~al.}{1980}]{Begelman80}
{Begelman} M.~C.,  {Blandford} R.~D.,   {Rees} M.~J.,  1980, \mn@doi [\nat]
  {10.1038/287307a0}, \href
  {https://ui.adsabs.harvard.edu/abs/1980Natur.287..307B} {287, 307}

\bibitem[\protect\citeauthoryear{{Bender}}{{Bender}}{1988a}]{Bender88_a}
{Bender} R.,  1988a, \aap, \href
  {https://ui.adsabs.harvard.edu/abs/1988A&A...193L...7B} {193, L7}

\bibitem[\protect\citeauthoryear{{Bender}}{{Bender}}{1988b}]{Bender88_b}
{Bender} R.,  1988b, \aap, \href
  {https://ui.adsabs.harvard.edu/abs/1988A&A...202L...5B} {202, L5}

\bibitem[\protect\citeauthoryear{{Bender} \& {Moellenhoff}}{{Bender} \&
  {Moellenhoff}}{1987}]{Bender87}
{Bender} R.,  {Moellenhoff} C.,  1987, \aap, \href
  {https://ui.adsabs.harvard.edu/abs/1987A&A...177...71B} {177, 71}

\bibitem[\protect\citeauthoryear{{Bender}, {Surma}, {Doebereiner},
  {Moellenhoff}  \& {Madejsky}}{{Bender} et~al.}{1989}]{Bender89}
{Bender} R.,  {Surma} P.,  {Doebereiner} S.,  {Moellenhoff} C.,   {Madejsky}
  R.,  1989, \aap, \href
  {https://ui.adsabs.harvard.edu/abs/1989A&A...217...35B} {217, 35}

\bibitem[\protect\citeauthoryear{{Bender}, {Burstein}  \& {Faber}}{{Bender}
  et~al.}{1992}]{Bender92}
{Bender} R.,  {Burstein} D.,   {Faber} S.~M.,  1992, \mn@doi [\apj]
  {10.1086/171940}, \href
  {https://ui.adsabs.harvard.edu/abs/1992ApJ...399..462B} {399, 462}

\bibitem[\protect\citeauthoryear{{Bender}, {Saglia}  \& {Gerhard}}{{Bender}
  et~al.}{1994}]{Bender94}
{Bender} R.,  {Saglia} R.~P.,   {Gerhard} O.~E.,  1994, \mn@doi [\mnras]
  {10.1093/mnras/269.3.785}, \href
  {https://ui.adsabs.harvard.edu/abs/1994MNRAS.269..785B} {269, 785}

\bibitem[\protect\citeauthoryear{{Bender} et~al.,}{{Bender}
  et~al.}{2005}]{Bender05}
{Bender} R.,  et~al., 2005, \mn@doi [\apj] {10.1086/432434}, \href
  {https://ui.adsabs.harvard.edu/abs/2005ApJ...631..280B} {631, 280}

\bibitem[\protect\citeauthoryear{{Bertola} \& {Galletta}}{{Bertola} \&
  {Galletta}}{1978}]{Bertola78}
{Bertola} F.,  {Galletta} G.,  1978, \mn@doi [\apjl] {10.1086/182844}, \href
  {https://ui.adsabs.harvard.edu/abs/1978ApJ...226L.115B} {226, L115}

\bibitem[\protect\citeauthoryear{{Bertola} \& {Galletta}}{{Bertola} \&
  {Galletta}}{1979}]{Bertola79}
{Bertola} F.,  {Galletta} G.,  1979, \aap, \href
  {https://ui.adsabs.harvard.edu/abs/1979A&A....77..363B} {77, 363}

\bibitem[\protect\citeauthoryear{{Binney} \& {Tremaine}}{{Binney} \&
  {Tremaine}}{2008}]{Binney08}
{Binney} J.,  {Tremaine} S.,  2008, {Galactic Dynamics}, {2nd} edn.
{Princeton Univers. Press}, p.~263

\bibitem[\protect\citeauthoryear{{Bissantz}, {Debattista}  \&
  {Gerhard}}{{Bissantz} et~al.}{2004}]{Bissantz04}
{Bissantz} N.,  {Debattista} V.~P.,   {Gerhard} O.,  2004, \mn@doi [\apjl]
  {10.1086/382043}, \href
  {https://ui.adsabs.harvard.edu/abs/2004ApJ...601L.155B} {601, L155}

\bibitem[\protect\citeauthoryear{{Boyd} \& {Vandenberghe}}{{Boyd} \&
  {Vandenberghe}}{2004}]{convex}
{Boyd} S.,  {Vandenberghe} L.,  2004, {Convex Optimization}.
{Cambridge University Press}

\bibitem[\protect\citeauthoryear{{Brainerd}, {Goldberg}, {Adams}, {Garcia},
  {McKay}  \& {Christian}}{{Brainerd} et~al.}{1996}]{Brainerd96}
{Brainerd} W.~S.,  {Goldberg} C.~H.,  {Adams} J.~C.,  {Garcia} A.,  {McKay} S.,
    {Christian} W.,  1996, Computers in Physics, \href
  {https://ui.adsabs.harvard.edu/abs/1996ComPh..10..135B} {10, 135}

\bibitem[\protect\citeauthoryear{{Brown} \& {Magorrian}}{{Brown} \&
  {Magorrian}}{2013}]{Brown13}
{Brown} C.~K.,  {Magorrian} J.,  2013, \mn@doi [\mnras] {10.1093/mnras/stt104},
  \href {https://ui.adsabs.harvard.edu/abs/2013MNRAS.431...80B} {431, 80}

\bibitem[\protect\citeauthoryear{{Cappellari}}{{Cappellari}}{2002}]{Cappellari02}
{Cappellari} M.,  2002, \mn@doi [\mnras] {10.1046/j.1365-8711.2002.05412.x},
  \href {https://ui.adsabs.harvard.edu/abs/2002MNRAS.333..400C} {333, 400}

\bibitem[\protect\citeauthoryear{{Cappellari}}{{Cappellari}}{2016}]{Cappellari16}
{Cappellari} M.,  2016, \mn@doi [\araa] {10.1146/annurev-astro-082214-122432},
  \href {https://ui.adsabs.harvard.edu/abs/2016ARA&A..54..597C} {54, 597}

\bibitem[\protect\citeauthoryear{{Cappellari} \& {Copin}}{{Cappellari} \&
  {Copin}}{2003}]{Cappellari03}
{Cappellari} M.,  {Copin} Y.,  2003, \mn@doi [\mnras]
  {10.1046/j.1365-8711.2003.06541.x}, \href
  {https://ui.adsabs.harvard.edu/abs/2003MNRAS.342..345C} {342, 345}

\bibitem[\protect\citeauthoryear{{Cappellari} et~al.,}{{Cappellari}
  et~al.}{2007}]{Cappellari07}
{Cappellari} M.,  et~al., 2007, \mn@doi [\mnras]
  {10.1111/j.1365-2966.2007.11963.x}, \href
  {https://ui.adsabs.harvard.edu/abs/2007MNRAS.379..418C} {379, 418}

\bibitem[\protect\citeauthoryear{{Cappellari} et~al.,}{{Cappellari}
  et~al.}{2012}]{Cappellari12}
{Cappellari} M.,  et~al., 2012, \mn@doi [\nat] {10.1038/nature10972}, \href
  {https://ui.adsabs.harvard.edu/abs/2012Natur.484..485C} {484, 485}

\bibitem[\protect\citeauthoryear{Cash \& Karp}{Cash \& Karp}{1990}]{Cash90}
Cash J.,  Karp A.,  1990, \mn@doi [ACM Trans. Math. Softw.]
  {10.1145/79505.79507}, 16, 201

\bibitem[\protect\citeauthoryear{{Chabrier}}{{Chabrier}}{2003}]{Chabrier03}
{Chabrier} G.,  2003, \mn@doi [\pasp] {10.1086/376392}, \href
  {https://ui.adsabs.harvard.edu/abs/2003PASP..115..763C} {115, 763}

\bibitem[\protect\citeauthoryear{{Cretton}, {de Zeeuw}, {van der Marel}  \&
  {Rix}}{{Cretton} et~al.}{1999}]{Cretton99}
{Cretton} N.,  {de Zeeuw} P.~T.,  {van der Marel} R.~P.,   {Rix} H.-W.,  1999,
  \mn@doi [\apjs] {10.1086/313264}, \href
  {https://ui.adsabs.harvard.edu/abs/1999ApJS..124..383C} {124, 383}

\bibitem[\protect\citeauthoryear{{Dehnen}}{{Dehnen}}{1993}]{Dehnen93}
{Dehnen} W.,  1993, \mn@doi [\mnras] {10.1093/mnras/265.1.250}, \href
  {https://ui.adsabs.harvard.edu/abs/1993MNRAS.265..250D} {265, 250}

\bibitem[\protect\citeauthoryear{{Ebisuzaki}, {Makino}  \&
  {Okumura}}{{Ebisuzaki} et~al.}{1991}]{Ebisuzaki91}
{Ebisuzaki} T.,  {Makino} J.,   {Okumura} S.~K.,  1991, \mn@doi [\nat]
  {10.1038/354212a0}, \href
  {https://ui.adsabs.harvard.edu/abs/1991Natur.354..212E} {354, 212}

\bibitem[\protect\citeauthoryear{{Emsellem}, {Monnet}  \& {Bacon}}{{Emsellem}
  et~al.}{1994}]{Emsellem94}
{Emsellem} E.,  {Monnet} G.,   {Bacon} R.,  1994, \aap, \href
  {https://ui.adsabs.harvard.edu/abs/1994A&A...285..723E} {285, 723}

\bibitem[\protect\citeauthoryear{{Emsellem} et~al.,}{{Emsellem}
  et~al.}{2007}]{Emsellem07}
{Emsellem} E.,  et~al., 2007, \mn@doi [\mnras]
  {10.1111/j.1365-2966.2007.11752.x}, \href
  {https://ui.adsabs.harvard.edu/abs/2007MNRAS.379..401E} {379, 401}

\bibitem[\protect\citeauthoryear{{Faber}, {Dressler}, {Davies}, {Burstein},
  {Lynden Bell}, {Terlevich}  \& {Wegner}}{{Faber} et~al.}{1987}]{Faber87}
{Faber} S.~M.,  {Dressler} A.,  {Davies} R.~L.,  {Burstein} D.,  {Lynden Bell}
  D.,  {Terlevich} R.,   {Wegner} G.,  1987, in {Faber} S.~M.,  ed., Nearly
  Normal Galaxies. From the Planck Time to the Present. p.~175

\bibitem[\protect\citeauthoryear{{Finozzi}}{{Finozzi}}{2018}]{Finozzi18}
{Finozzi} F.,  2018, {Triaxial models of massive elliptical galaxies}.
PhD thesis, Ludwig-Maximilians-University Munich

\bibitem[\protect\citeauthoryear{{Franx} \& {Illingworth}}{{Franx} \&
  {Illingworth}}{1988}]{Franx88}
{Franx} M.,  {Illingworth} G.~D.,  1988, \mn@doi [\apjl] {10.1086/185139},
  \href {https://ui.adsabs.harvard.edu/abs/1988ApJ...327L..55F} {327, L55}

\bibitem[\protect\citeauthoryear{{Gebhardt} et~al.,}{{Gebhardt}
  et~al.}{2000}]{Gebhardt00}
{Gebhardt} K.,  et~al., 2000, \mn@doi [\aj] {10.1086/301240}, \href
  {https://ui.adsabs.harvard.edu/abs/2000AJ....119.1157G} {119, 1157}

\bibitem[\protect\citeauthoryear{{Genzel}, {Tacconi}, {Rigopoulou}, {Lutz}  \&
  {Tecza}}{{Genzel} et~al.}{2001}]{Genzel01}
{Genzel} R.,  {Tacconi} L.~J.,  {Rigopoulou} D.,  {Lutz} D.,   {Tecza} M.,
  2001, \mn@doi [\apj] {10.1086/323772}, \href
  {https://ui.adsabs.harvard.edu/abs/2001ApJ...563..527G} {563, 527}

\bibitem[\protect\citeauthoryear{{Gerhard}}{{Gerhard}}{1993}]{Gerhard93}
{Gerhard} O.~E.,  1993, \mn@doi [\mnras] {10.1093/mnras/265.1.213}, \href
  {https://ui.adsabs.harvard.edu/abs/1993MNRAS.265..213G} {265, 213}

\bibitem[\protect\citeauthoryear{{Heiligman} \& {Schwarzschild}}{{Heiligman} \&
  {Schwarzschild}}{1979}]{Heiligman79}
{Heiligman} G.,  {Schwarzschild} M.,  1979, \mn@doi [\apj] {10.1086/157449},
  \href {https://ui.adsabs.harvard.edu/abs/1979ApJ...233..872H} {233, 872}

\bibitem[\protect\citeauthoryear{{Henon} \& {Heiles}}{{Henon} \&
  {Heiles}}{1964}]{Henon64}
{Henon} M.,  {Heiles} C.,  1964, \mn@doi [\aj] {10.1086/109234}, \href
  {https://ui.adsabs.harvard.edu/abs/1964AJ.....69...73H} {69, 73}

\bibitem[\protect\citeauthoryear{{Hills} \& {Fullerton}}{{Hills} \&
  {Fullerton}}{1980}]{Hills80}
{Hills} J.~G.,  {Fullerton} L.~W.,  1980, \mn@doi [\aj] {10.1086/112798}, \href
  {https://ui.adsabs.harvard.edu/abs/1980AJ.....85.1281H} {85, 1281}

\bibitem[\protect\citeauthoryear{{Hohl} \& {Zang}}{{Hohl} \&
  {Zang}}{1979}]{Hohl79}
{Hohl} F.,  {Zang} T.~A.,  1979, \mn@doi [\aj] {10.1086/112454}, \href
  {https://ui.adsabs.harvard.edu/abs/1979AJ.....84..585H} {84, 585}

\bibitem[\protect\citeauthoryear{{Hopkins}, {Cox}  \& {Hernquist}}{{Hopkins}
  et~al.}{2008}]{Hopkins08}
{Hopkins} P.~F.,  {Cox} T.~J.,   {Hernquist} L.,  2008, \mn@doi [\apj]
  {10.1086/592105}, \href
  {https://ui.adsabs.harvard.edu/abs/2008ApJ...689...17H} {689, 17}

\bibitem[\protect\citeauthoryear{{Illingworth}}{{Illingworth}}{1977}]{Illingworth77}
{Illingworth} G.,  1977, \mn@doi [\apjl] {10.1086/182572}, \href
  {https://ui.adsabs.harvard.edu/abs/1977ApJ...218L..43I} {218, L43}

\bibitem[\protect\citeauthoryear{{Jahnke} \& {Macci{\`o}}}{{Jahnke} \&
  {Macci{\`o}}}{2011}]{Jahnke11}
{Jahnke} K.,  {Macci{\`o}} A.~V.,  2011, \mn@doi [\apj]
  {10.1088/0004-637X/734/2/92}, \href
  {https://ui.adsabs.harvard.edu/abs/2011ApJ...734...92J} {734, 92}

\bibitem[\protect\citeauthoryear{{Jin}, {Zhu}, {Long}, {Mao}, {Xu}, {Li}  \&
  {van de Ven}}{{Jin} et~al.}{2019}]{Jin19}
{Jin} Y.,  {Zhu} L.,  {Long} R.~J.,  {Mao} S.,  {Xu} D.,  {Li} H.,   {van de
  Ven} G.,  2019, \mn@doi [\mnras] {10.1093/mnras/stz1170}, \href
  {https://ui.adsabs.harvard.edu/abs/2019MNRAS.486.4753J} {486, 4753}

\bibitem[\protect\citeauthoryear{{Jing} \& {Suto}}{{Jing} \&
  {Suto}}{2002}]{Jing02}
{Jing} Y.~P.,  {Suto} Y.,  2002, \mn@doi [\apj] {10.1086/341065}, \href
  {https://ui.adsabs.harvard.edu/abs/2002ApJ...574..538J} {574, 538}

\bibitem[\protect\citeauthoryear{{Johansson}, {Naab}  \& {Burkert}}{{Johansson}
  et~al.}{2009}]{Johansson09}
{Johansson} P.~H.,  {Naab} T.,   {Burkert} A.,  2009, \mn@doi [\apj]
  {10.1088/0004-637X/690/1/802}, \href
  {https://ui.adsabs.harvard.edu/abs/2009ApJ...690..802J} {690, 802}

\bibitem[\protect\citeauthoryear{{Karl}, {Aarseth}, {Naab}, {Haehnelt}  \&
  {Spurzem}}{{Karl} et~al.}{2015}]{Karl15}
{Karl} S.~J.,  {Aarseth} S.~J.,  {Naab} T.,  {Haehnelt} M.~G.,   {Spurzem} R.,
  2015, \mn@doi [\mnras] {10.1093/mnras/stv1453}, \href
  {https://ui.adsabs.harvard.edu/abs/2015MNRAS.452.2337K} {452, 2337}

\bibitem[\protect\citeauthoryear{{Kormendy} \& {Bender}}{{Kormendy} \&
  {Bender}}{1996}]{Kormendy96}
{Kormendy} J.,  {Bender} R.,  1996, \mn@doi [\apjl] {10.1086/310095}, \href
  {https://ui.adsabs.harvard.edu/abs/1996ApJ...464L.119K} {464, L119}

\bibitem[\protect\citeauthoryear{{Kroupa}}{{Kroupa}}{2001}]{Kroupa01}
{Kroupa} P.,  2001, \mn@doi [\mnras] {10.1046/j.1365-8711.2001.04022.x}, \href
  {https://ui.adsabs.harvard.edu/abs/2001MNRAS.322..231K} {322, 231}

\bibitem[\protect\citeauthoryear{{Le~Digabel}}{{Le~Digabel}}{2011}]{LeDigabel11}
{Le~Digabel} S.,  2011, {ACM} Transactions on Mathematical Software, 37, 1

\bibitem[\protect\citeauthoryear{{Magorrian}}{{Magorrian}}{2006}]{Magorrian06}
{Magorrian} J.,  2006, \mn@doi [\mnras] {10.1111/j.1365-2966.2006.11054.x},
  \href {https://ui.adsabs.harvard.edu/abs/2006MNRAS.373..425M} {373, 425}

\bibitem[\protect\citeauthoryear{{Mehrgan}, {Thomas}, {Saglia}, {Mazzalay},
  {Erwin}, {Bender}, {Kluge}  \& {Fabricius}}{{Mehrgan}
  et~al.}{2019}]{Mehrgan19}
{Mehrgan} K.,  {Thomas} J.,  {Saglia} R.,  {Mazzalay} X.,  {Erwin} P.,
  {Bender} R.,  {Kluge} M.,   {Fabricius} M.,  2019, \mn@doi [\apj]
  {10.3847/1538-4357/ab5856}, \href
  {https://ui.adsabs.harvard.edu/abs/2019ApJ...887..195M} {887, 195}

\bibitem[\protect\citeauthoryear{{Merritt}}{{Merritt}}{2006}]{Merritt06}
{Merritt} D.,  2006, \mn@doi [\apj] {10.1086/506139}, \href
  {https://ui.adsabs.harvard.edu/abs/2006ApJ...648..976M} {648, 976}

\bibitem[\protect\citeauthoryear{{Mikkola} \& {Merritt}}{{Mikkola} \&
  {Merritt}}{2006}]{Mikkola06}
{Mikkola} S.,  {Merritt} D.,  2006, \mn@doi [\mnras]
  {10.1111/j.1365-2966.2006.10854.x}, \href
  {https://ui.adsabs.harvard.edu/abs/2006MNRAS.372..219M} {372, 219}

\bibitem[\protect\citeauthoryear{{Mikkola} \& {Merritt}}{{Mikkola} \&
  {Merritt}}{2008}]{Mikkola08}
{Mikkola} S.,  {Merritt} D.,  2008, \mn@doi [\aj]
  {10.1088/0004-6256/135/6/2398}, \href
  {https://ui.adsabs.harvard.edu/abs/2008AJ....135.2398M} {135, 2398}

\bibitem[\protect\citeauthoryear{{Mikkola} \& {Valtonen}}{{Mikkola} \&
  {Valtonen}}{1992}]{Mikkola92}
{Mikkola} S.,  {Valtonen} M.~J.,  1992, \mn@doi [\mnras]
  {10.1093/mnras/259.1.115}, \href
  {https://ui.adsabs.harvard.edu/abs/1992MNRAS.259..115M} {259, 115}

\bibitem[\protect\citeauthoryear{{Miller} \& {Smith}}{{Miller} \&
  {Smith}}{1979}]{Miller79}
{Miller} R.~H.,  {Smith} B.~F.,  1979, \mn@doi [\apj] {10.1086/156745}, \href
  {https://ui.adsabs.harvard.edu/abs/1979ApJ...227..407M} {227, 407}

\bibitem[\protect\citeauthoryear{{Milosavljevi{\'c}} \&
  {Merritt}}{{Milosavljevi{\'c}} \& {Merritt}}{2001}]{Milosavljevic01}
{Milosavljevi{\'c}} M.,  {Merritt} D.,  2001, \mn@doi [\apj] {10.1086/323830},
  \href {https://ui.adsabs.harvard.edu/abs/2001ApJ...563...34M} {563, 34}

\bibitem[\protect\citeauthoryear{{Monnet}, {Bacon}  \& {Emsellem}}{{Monnet}
  et~al.}{1992}]{Monnet92}
{Monnet} G.,  {Bacon} R.,   {Emsellem} E.,  1992, \aap, \href
  {https://ui.adsabs.harvard.edu/abs/1992A&A...253..366M} {253, 366}

\bibitem[\protect\citeauthoryear{{Moster}, {Naab}  \& {White}}{{Moster}
  et~al.}{2019}]{Moster19}
{Moster} B.~P.,  {Naab} T.,   {White} S. D.~M.,  2019, arXiv e-prints, \href
  {https://ui.adsabs.harvard.edu/abs/2019arXiv191009552M} {p. arXiv:1910.09552}

\bibitem[\protect\citeauthoryear{{Naab} \& {Burkert}}{{Naab} \&
  {Burkert}}{2003}]{Naab03}
{Naab} T.,  {Burkert} A.,  2003, \mn@doi [\apj] {10.1086/378581}, \href
  {https://ui.adsabs.harvard.edu/abs/2003ApJ...597..893N} {597, 893}

\bibitem[\protect\citeauthoryear{{Naab} \& {Ostriker}}{{Naab} \&
  {Ostriker}}{2017}]{Naab17}
{Naab} T.,  {Ostriker} J.~P.,  2017, \mn@doi [\araa]
  {10.1146/annurev-astro-081913-040019}, \href
  {https://ui.adsabs.harvard.edu/abs/2017ARA&A..55...59N} {55, 59}

\bibitem[\protect\citeauthoryear{{Naab}, {Jesseit}  \& {Burkert}}{{Naab}
  et~al.}{2006}]{Naab06}
{Naab} T.,  {Jesseit} R.,   {Burkert} A.,  2006, \mn@doi [\mnras]
  {10.1111/j.1365-2966.2006.10902.x}, \href
  {https://ui.adsabs.harvard.edu/abs/2006MNRAS.372..839N} {372, 839}

\bibitem[\protect\citeauthoryear{Navarro, Frenk  \& White}{Navarro
  et~al.}{1996}]{Navarro96}
Navarro J.~F.,  Frenk C.~S.,   White S. D.~M.,  1996, \mn@doi [The
  Astrophysical Journal] {10.1086/177173}, 462, 563

\bibitem[\protect\citeauthoryear{{Novak}, {Cox}, {Primack}, {Jonsson}  \&
  {Dekel}}{{Novak} et~al.}{2006}]{Novak06}
{Novak} G.~S.,  {Cox} T.~J.,  {Primack} J.~R.,  {Jonsson} P.,   {Dekel} A.,
  2006, \mn@doi [\apjl] {10.1086/506605}, \href
  {https://ui.adsabs.harvard.edu/abs/2006ApJ...646L...9N} {646, L9}

\bibitem[\protect\citeauthoryear{{Oser}, {Ostriker}, {Naab}, {Johansson}  \&
  {Burkert}}{{Oser} et~al.}{2010}]{Oser10}
{Oser} L.,  {Ostriker} J.~P.,  {Naab} T.,  {Johansson} P.~H.,   {Burkert} A.,
  2010, \mn@doi [\apj] {10.1088/0004-637X/725/2/2312}, \href
  {https://ui.adsabs.harvard.edu/abs/2010ApJ...725.2312O} {725, 2312}

\bibitem[\protect\citeauthoryear{{Parikh} et~al.,}{{Parikh}
  et~al.}{2018}]{Parikh18}
{Parikh} T.,  et~al., 2018, \mn@doi [\mnras] {10.1093/mnras/sty785}, \href
  {https://ui.adsabs.harvard.edu/abs/2018MNRAS.477.3954P} {477, 3954}

\bibitem[\protect\citeauthoryear{{Peng}}{{Peng}}{2007}]{Peng07}
{Peng} C.~Y.,  2007, \mn@doi [\apj] {10.1086/522774}, \href
  {https://ui.adsabs.harvard.edu/abs/2007ApJ...671.1098P} {671, 1098}

\bibitem[\protect\citeauthoryear{{Posacki}, {Cappellari}, {Treu}, {Pellegrini}
  \& {Ciotti}}{{Posacki} et~al.}{2015}]{Posacki15}
{Posacki} S.,  {Cappellari} M.,  {Treu} T.,  {Pellegrini} S.,   {Ciotti} L.,
  2015, \mn@doi [\mnras] {10.1093/mnras/stu2098}, \href
  {https://ui.adsabs.harvard.edu/abs/2015MNRAS.446..493P} {446, 493}

\bibitem[\protect\citeauthoryear{{Press}, {Vetterling}, {Metcalf}, {Flannery}
  \& {Teukolsky}}{{Press} et~al.}{1996}]{Press07}
{Press} W.,  {Vetterling} W.,  {Metcalf} M.,  {Flannery} B.,   {Teukolsky} S.,
  1996, Cambridge University Press, Cambridge

\bibitem[\protect\citeauthoryear{{Rantala}, {Pihajoki}, {Johansson}, {Naab},
  {Lah{\'e}n}  \& {Sawala}}{{Rantala} et~al.}{2017}]{Rantala17}
{Rantala} A.,  {Pihajoki} P.,  {Johansson} P.~H.,  {Naab} T.,  {Lah{\'e}n} N.,
   {Sawala} T.,  2017, \mn@doi [\apj] {10.3847/1538-4357/aa6d65}, \href
  {https://ui.adsabs.harvard.edu/abs/2017ApJ...840...53R} {840, 53}

\bibitem[\protect\citeauthoryear{{Rantala}, {Johansson}, {Naab}, {Thomas}  \&
  {Frigo}}{{Rantala} et~al.}{2018}]{Rantala18}
{Rantala} A.,  {Johansson} P.~H.,  {Naab} T.,  {Thomas} J.,   {Frigo} M.,
  2018, \mn@doi [\apj] {10.3847/1538-4357/aada47}, \href
  {https://ui.adsabs.harvard.edu/abs/2018ApJ...864..113R} {864, 113}

\bibitem[\protect\citeauthoryear{{Rantala}, {Johansson}, {Naab}, {Thomas}  \&
  {Frigo}}{{Rantala} et~al.}{2019}]{Rantala19}
{Rantala} A.,  {Johansson} P.~H.,  {Naab} T.,  {Thomas} J.,   {Frigo} M.,
  2019, \mn@doi [\apjl] {10.3847/2041-8213/ab04b1}, \href
  {https://ui.adsabs.harvard.edu/abs/2019ApJ...872L..17R} {872, L17}

\bibitem[\protect\citeauthoryear{{Richstone}}{{Richstone}}{1982}]{Richstone82}
{Richstone} D.~O.,  1982, \mn@doi [\apj] {10.1086/159578}, \href
  {https://ui.adsabs.harvard.edu/abs/1982ApJ...252..496R} {252, 496}

\bibitem[\protect\citeauthoryear{{Richstone} \& {Tremaine}}{{Richstone} \&
  {Tremaine}}{1984}]{Richstone84}
{Richstone} D.~O.,  {Tremaine} S.,  1984, \mn@doi [\apj] {10.1086/162572},
  \href {https://ui.adsabs.harvard.edu/abs/1984ApJ...286...27R} {286, 27}

\bibitem[\protect\citeauthoryear{{Richstone} \& {Tremaine}}{{Richstone} \&
  {Tremaine}}{1985}]{Richstone85}
{Richstone} D.~O.,  {Tremaine} S.,  1985, \mn@doi [\apj] {10.1086/163455},
  \href {https://ui.adsabs.harvard.edu/abs/1985ApJ...296..370R} {296, 370}

\bibitem[\protect\citeauthoryear{{Richstone} \& {Tremaine}}{{Richstone} \&
  {Tremaine}}{1988}]{Richstone88}
{Richstone} D.~O.,  {Tremaine} S.,  1988, \mn@doi [\apj] {10.1086/166171},
  \href {https://ui.adsabs.harvard.edu/abs/1988ApJ...327...82R} {327, 82}

\bibitem[\protect\citeauthoryear{{Rix}, {de Zeeuw}, {Cretton}, {van der Marel}
  \& {Carollo}}{{Rix} et~al.}{1997}]{Rix97}
{Rix} H.-W.,  {de Zeeuw} P.~T.,  {Cretton} N.,  {van der Marel} R.~P.,
  {Carollo} C.~M.,  1997, \mn@doi [\apj] {10.1086/304733}, \href
  {https://ui.adsabs.harvard.edu/abs/1997ApJ...488..702R} {488, 702}

\bibitem[\protect\citeauthoryear{{R{\"o}ttgers}, {Naab}  \&
  {Oser}}{{R{\"o}ttgers} et~al.}{2014}]{Roettgers14}
{R{\"o}ttgers} B.,  {Naab} T.,   {Oser} L.,  2014, \mn@doi [\mnras]
  {10.1093/mnras/stu1762}, \href
  {https://ui.adsabs.harvard.edu/abs/2014MNRAS.445.1065R} {445, 1065}

\bibitem[\protect\citeauthoryear{{Schechter} \& {Gunn}}{{Schechter} \&
  {Gunn}}{1978}]{Schechter78}
{Schechter} P.~L.,  {Gunn} J.~E.,  1978, \mn@doi [\aj] {10.1086/112324}, \href
  {https://ui.adsabs.harvard.edu/abs/1978AJ.....83.1360S} {83, 1360}

\bibitem[\protect\citeauthoryear{{Schwarzschild}}{{Schwarzschild}}{1979}]{Schwarzschild79}
{Schwarzschild} M.,  1979, \mn@doi [\apj] {10.1086/157282}, \href
  {https://ui.adsabs.harvard.edu/abs/1979ApJ...232..236S} {232, 236}

\bibitem[\protect\citeauthoryear{{Schwarzschild}}{{Schwarzschild}}{1993}]{Schwarzschild93}
{Schwarzschild} M.,  1993, \mn@doi [\apj] {10.1086/172687}, \href
  {https://ui.adsabs.harvard.edu/abs/1993ApJ...409..563S} {409, 563}

\bibitem[\protect\citeauthoryear{{S{\'e}rsic}}{{S{\'e}rsic}}{1963}]{Sersic63}
{S{\'e}rsic} J.~L.,  1963, Boletin de la Asociacion Argentina de Astronomia La
  Plata Argentina, \href
  {https://ui.adsabs.harvard.edu/abs/1963BAAA....6...99S} {6, 99}

\bibitem[\protect\citeauthoryear{{Siopis} et~al.,}{{Siopis}
  et~al.}{2009}]{Siopis09}
{Siopis} C.,  et~al., 2009, \mn@doi [\apj] {10.1088/0004-637X/693/1/946}, \href
  {https://ui.adsabs.harvard.edu/abs/2009ApJ...693..946S} {693, 946}

\bibitem[\protect\citeauthoryear{{Smith}, {Lucey}  \& {Conroy}}{{Smith}
  et~al.}{2015}]{Smith15}
{Smith} R.~J.,  {Lucey} J.~R.,   {Conroy} C.,  2015, \mn@doi [\mnras]
  {10.1093/mnras/stv518}, \href
  {https://ui.adsabs.harvard.edu/abs/2015MNRAS.449.3441S} {449, 3441}

\bibitem[\protect\citeauthoryear{{Somerville} \& {Dav{\'e}}}{{Somerville} \&
  {Dav{\'e}}}{2015}]{Somerville15}
{Somerville} R.~S.,  {Dav{\'e}} R.,  2015, \mn@doi [\araa]
  {10.1146/annurev-astro-082812-140951}, \href
  {https://ui.adsabs.harvard.edu/abs/2015ARA&A..53...51S} {53, 51}

\bibitem[\protect\citeauthoryear{{Springel}}{{Springel}}{2005}]{Springel05}
{Springel} V.,  2005, \mn@doi [\mnras] {10.1111/j.1365-2966.2005.09655.x},
  \href {https://ui.adsabs.harvard.edu/abs/2005MNRAS.364.1105S} {364, 1105}

\bibitem[\protect\citeauthoryear{{Syer} \& {Tremaine}}{{Syer} \&
  {Tremaine}}{1996}]{Syer96}
{Syer} D.,  {Tremaine} S.,  1996, \mn@doi [\mnras] {10.1093/mnras/282.1.223},
  \href {https://ui.adsabs.harvard.edu/abs/1996MNRAS.282..223S} {282, 223}

\bibitem[\protect\citeauthoryear{{Tacconi} et~al.,}{{Tacconi}
  et~al.}{2005}]{Tacconi05}
{Tacconi} L.~J.,  et~al., 2005, The Messenger, \href
  {https://ui.adsabs.harvard.edu/abs/2005Msngr.122...28T} {122, 28}

\bibitem[\protect\citeauthoryear{{Thomas}, {Saglia}, {Bender}, {Thomas},
  {Gebhardt}, {Magorrian}  \& {Richstone}}{{Thomas} et~al.}{2004}]{Thomas04}
{Thomas} J.,  {Saglia} R.~P.,  {Bender} R.,  {Thomas} D.,  {Gebhardt} K.,
  {Magorrian} J.,   {Richstone} D.,  2004, \mn@doi [\mnras]
  {10.1111/j.1365-2966.2004.08072.x}, \href
  {https://ui.adsabs.harvard.edu/abs/2004MNRAS.353..391T} {353, 391}

\bibitem[\protect\citeauthoryear{{Thomas}, {Saglia}, {Bender}, {Thomas},
  {Gebhardt}, {Magorrian}, {Corsini}  \& {Wegner}}{{Thomas}
  et~al.}{2005}]{Thomas05}
{Thomas} J.,  {Saglia} R.~P.,  {Bender} R.,  {Thomas} D.,  {Gebhardt} K.,
  {Magorrian} J.,  {Corsini} E.~M.,   {Wegner} G.,  2005, \mn@doi [\mnras]
  {10.1111/j.1365-2966.2005.09139.x}, \href
  {https://ui.adsabs.harvard.edu/abs/2005MNRAS.360.1355T} {360, 1355}

\bibitem[\protect\citeauthoryear{{Thomas}, {Jesseit}, {Naab}, {Saglia},
  {Burkert}  \& {Bender}}{{Thomas} et~al.}{2007}]{Thomas07}
{Thomas} J.,  {Jesseit} R.,  {Naab} T.,  {Saglia} R.~P.,  {Burkert} A.,
  {Bender} R.,  2007, \mn@doi [\mnras] {10.1111/j.1365-2966.2007.12343.x},
  \href {https://ui.adsabs.harvard.edu/abs/2007MNRAS.381.1672T} {381, 1672}

\bibitem[\protect\citeauthoryear{{Thomas} et~al.,}{{Thomas}
  et~al.}{2009}]{Thomas09}
{Thomas} J.,  et~al., 2009, \mn@doi [\mnras]
  {10.1111/j.1365-2966.2008.14238.x}, \href
  {https://ui.adsabs.harvard.edu/abs/2009MNRAS.393..641T} {393, 641}

\bibitem[\protect\citeauthoryear{{Thomas} et~al.,}{{Thomas}
  et~al.}{2011}]{Thomas11}
{Thomas} J.,  et~al., 2011, \mn@doi [\mnras]
  {10.1111/j.1365-2966.2011.18725.x}, \href
  {https://ui.adsabs.harvard.edu/abs/2011MNRAS.415..545T} {415, 545}

\bibitem[\protect\citeauthoryear{{Thomas}, {Ma}, {McConnell}, {Greene},
  {Blakeslee}  \& {Janish}}{{Thomas} et~al.}{2016}]{Thomas16}
{Thomas} J.,  {Ma} C.-P.,  {McConnell} N.~J.,  {Greene} J.~E.,  {Blakeslee}
  J.~P.,   {Janish} R.,  2016, \mn@doi [\nat] {10.1038/nature17197}, \href
  {https://ui.adsabs.harvard.edu/abs/2016Natur.532..340T} {532, 340}

\bibitem[\protect\citeauthoryear{{Toomre} \& {Toomre}}{{Toomre} \&
  {Toomre}}{1972}]{Toomre72}
{Toomre} A.,  {Toomre} J.,  1972, \mn@doi [\apj] {10.1086/151823}, \href
  {https://ui.adsabs.harvard.edu/abs/1972ApJ...178..623T} {178, 623}

\bibitem[\protect\citeauthoryear{{Tremaine}, {Henon}  \&
  {Lynden-Bell}}{{Tremaine} et~al.}{1986}]{Tremaine86}
{Tremaine} S.,  {Henon} M.,   {Lynden-Bell} D.,  1986, \mn@doi [\mnras]
  {10.1093/mnras/219.2.285}, \href
  {https://ui.adsabs.harvard.edu/abs/1986MNRAS.219..285T} {219, 285}

\bibitem[\protect\citeauthoryear{{Treu}, {Auger}, {Koopmans}, {Gavazzi},
  {Marshall}  \& {Bolton}}{{Treu} et~al.}{2010}]{Treu10}
{Treu} T.,  {Auger} M.~W.,  {Koopmans} L. V.~E.,  {Gavazzi} R.,  {Marshall}
  P.~J.,   {Bolton} A.~S.,  2010, \mn@doi [\apj]
  {10.1088/0004-637X/709/2/1195}, \href
  {https://ui.adsabs.harvard.edu/abs/2010ApJ...709.1195T} {709, 1195}

\bibitem[\protect\citeauthoryear{{Tsallis}}{{Tsallis}}{1988}]{Tsallis88}
{Tsallis} C.,  1988, \mn@doi [Journal of Statistical Physics]
  {10.1007/BF01016429}, \href
  {https://ui.adsabs.harvard.edu/abs/1988JSP....52..479T} {52, 479}

\bibitem[\protect\citeauthoryear{{Valluri}, {Merritt}  \& {Emsellem}}{{Valluri}
  et~al.}{2004}]{Valluri04}
{Valluri} M.,  {Merritt} D.,   {Emsellem} E.,  2004, \mn@doi [\apj]
  {10.1086/380896}, \href
  {https://ui.adsabs.harvard.edu/abs/2004ApJ...602...66V} {602, 66}

\bibitem[\protect\citeauthoryear{{Vasiliev} \& {Valluri}}{{Vasiliev} \&
  {Valluri}}{2019}]{Vasiliev19}
{Vasiliev} E.,  {Valluri} M.,  2019, arXiv e-prints, \href
  {https://ui.adsabs.harvard.edu/abs/2019arXiv191204288V} {p. arXiv:1912.04288}

\bibitem[\protect\citeauthoryear{{Vazdekis} et~al.,}{{Vazdekis}
  et~al.}{2015}]{Vazdekis15}
{Vazdekis} A.,  et~al., 2015, \mn@doi [\mnras] {10.1093/mnras/stv151}, \href
  {https://ui.adsabs.harvard.edu/abs/2015MNRAS.449.1177V} {449, 1177}

\bibitem[\protect\citeauthoryear{{Vincent} \& {Ryden}}{{Vincent} \&
  {Ryden}}{2005}]{Vincent05}
{Vincent} R.~A.,  {Ryden} B.~S.,  2005, \mn@doi [\apj] {10.1086/428765}, \href
  {https://ui.adsabs.harvard.edu/abs/2005ApJ...623..137V} {623, 137}

\bibitem[\protect\citeauthoryear{{Virtanen} et~al.,}{{Virtanen}
  et~al.}{2019}]{Virtanen19}
{Virtanen} P.,  et~al., 2019, arXiv e-prints, \href
  {https://ui.adsabs.harvard.edu/abs/2019arXiv190710121V} {p. arXiv:1907.10121}

\bibitem[\protect\citeauthoryear{{Vogelsberger} et~al.,}{{Vogelsberger}
  et~al.}{2014}]{Vogelsberger14}
{Vogelsberger} M.,  et~al., 2014, \mn@doi [\nat] {10.1038/nature13316}, \href
  {https://ui.adsabs.harvard.edu/abs/2014Natur.509..177V} {509, 177}

\bibitem[\protect\citeauthoryear{{White} \& {Narayan}}{{White} \&
  {Narayan}}{1987}]{White87}
{White} S. D.~M.,  {Narayan} R.,  1987, \mn@doi [\mnras]
  {10.1093/mnras/229.1.103}, \href
  {https://ui.adsabs.harvard.edu/abs/1987MNRAS.229..103W} {229, 103}

\bibitem[\protect\citeauthoryear{{Will}}{{Will}}{2006}]{Will06}
{Will} C.~M.,  2006, \mn@doi [Living Reviews in Relativity]
  {10.12942/lrr-2006-3}, \href
  {https://ui.adsabs.harvard.edu/abs/2006LRR.....9....3W} {9, 3}

\bibitem[\protect\citeauthoryear{{Williams} \& {Schwarzschild}}{{Williams} \&
  {Schwarzschild}}{1979}]{Williams79}
{Williams} T.~B.,  {Schwarzschild} M.,  1979, \mn@doi [\apjs] {10.1086/190616},
  \href {https://ui.adsabs.harvard.edu/abs/1979ApJS...41..209W} {41, 209}

\bibitem[\protect\citeauthoryear{{de Lorenzi}, {Debattista}, {Gerhard}  \&
  {Sambhus}}{{de Lorenzi} et~al.}{2007}]{deLorenzi07}
{de Lorenzi} F.,  {Debattista} V.~P.,  {Gerhard} O.,   {Sambhus} N.,  2007,
  \mn@doi [\mnras] {10.1111/j.1365-2966.2007.11434.x}, \href
  {https://ui.adsabs.harvard.edu/abs/2007MNRAS.376...71D} {376, 71}

\bibitem[\protect\citeauthoryear{{de Nicola}, {Saglia}, {Thomas}, {Dehnen}  \&
  {Bender}}{{de Nicola} et~al.}{2020}]{deNicola20}
{de Nicola} S.,  {Saglia} R.~P.,  {Thomas} J.,  {Dehnen} W.,   {Bender} R.,
  2020, \mn@doi [\mnras] {10.1093/mnras/staa1703}, \href
  {https://ui.adsabs.harvard.edu/abs/2020MNRAS.496.3076D} {496, 3076}

\bibitem[\protect\citeauthoryear{{de Zeeuw}}{{de Zeeuw}}{1985}]{deZeeuw85}
{de Zeeuw} T.,  1985, \mn@doi [\mnras] {10.1093/mnras/216.2.273}, \href
  {https://ui.adsabs.harvard.edu/abs/1985MNRAS.216..273D} {216, 273}

\bibitem[\protect\citeauthoryear{{van Dokkum} \& {Conroy}}{{van Dokkum} \&
  {Conroy}}{2010}]{vanDokkum10}
{van Dokkum} P.~G.,  {Conroy} C.,  2010, \mn@doi [\nat] {10.1038/nature09578},
  \href {https://ui.adsabs.harvard.edu/abs/2010Natur.468..940V} {468, 940}

\bibitem[\protect\citeauthoryear{{van den Bosch} \& {de Zeeuw}}{{van den Bosch}
  \& {de Zeeuw}}{2010}]{vandenBosch10}
{van den Bosch} R. C.~E.,  {de Zeeuw} P.~T.,  2010, \mn@doi [\mnras]
  {10.1111/j.1365-2966.2009.15832.x}, \href
  {https://ui.adsabs.harvard.edu/abs/2010MNRAS.401.1770V} {401, 1770}

\bibitem[\protect\citeauthoryear{{van den Bosch}, {van de Ven}, {Verolme},
  {Cappellari}  \& {de Zeeuw}}{{van den Bosch} et~al.}{2008}]{vandenBosch08}
{van den Bosch} R.~C.~E.,  {van de Ven} G.,  {Verolme} E.~K.,  {Cappellari} M.,
    {de Zeeuw} P.~T.,  2008, \mn@doi [\mnras]
  {10.1111/j.1365-2966.2008.12874.x}, \href
  {https://ui.adsabs.harvard.edu/abs/2008MNRAS.385..647V} {385, 647}

\bibitem[\protect\citeauthoryear{{van der Marel} \& {Franx}}{{van der Marel} \&
  {Franx}}{1993}]{vanderMarel93}
{van der Marel} R.~P.,  {Franx} M.,  1993, \mn@doi [\apj] {10.1086/172534},
  \href {https://ui.adsabs.harvard.edu/abs/1993ApJ...407..525V} {407, 525}

\bibitem[\protect\citeauthoryear{{van der Marel}, {Cretton}, {de Zeeuw}  \&
  {Rix}}{{van der Marel} et~al.}{1998}]{vanderMarel98}
{van der Marel} R.~P.,  {Cretton} N.,  {de Zeeuw} P.~T.,   {Rix} H.-W.,  1998,
  \mn@doi [\apj] {10.1086/305147}, \href
  {https://ui.adsabs.harvard.edu/abs/1998ApJ...493..613V} {493, 613}

\makeatother
\end{thebibliography}

\appendix

\section{Surface Brightness-, ellipticity- and Position-Angle-profile}
\begin{figure}
    \centering
    \includegraphics[width=0.9\textwidth]{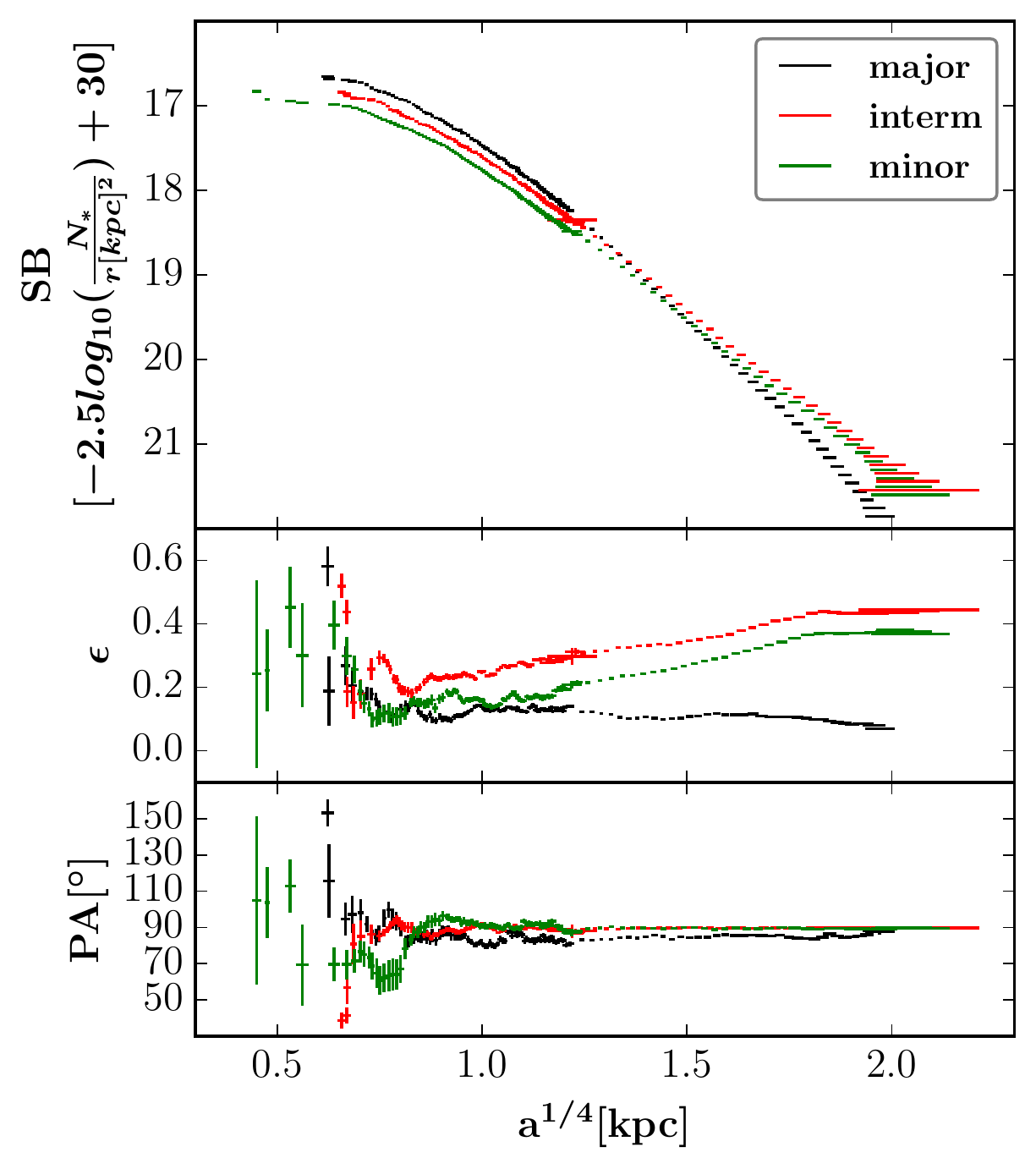}
    \caption{Surface brightness- (first row), ellipticity- (second row) and position angle-profile (third row) for the major- (black), intermediate- (red) and minor-axis (green) projection of the simulation plotted against the semi-major axis $a$ of the elliptical isophotes. The simulation demonstrates to be generic triaxial.}
    \label{fig:Figure16}
\end{figure}

When fitting elliptical isophotes \citep{Bender87} to the surface brightness maps  for the principal axis directions of the simulation in units of stellar simulation particles one can see that all three projections differ in their surface-brightness- (SB) and ellipticity-profile ($\epsilon=1-b/a$ with $a$ being the semi-major and $b$ being the semi-minor axis of the elliptical isophotes). The position-angle-profile (PA) shows some isophotal twists. Overall, the simulation appears to show typical triaxial behavior (see Fig.~\ref{fig:Figure16}). 

\section{\texorpdfstring{Velocity-, Surface Brightness- and $\chi^2$-maps}{Velocity-, Surface Brightness- and chi-maps}}
Fig.~\ref{fig:Figure19} and~\ref{fig:Figure21} show the velocity maps of the kinematic input data and their fits by \texttt{SMART} for four different projection axes and an additional psf convolved version of the intermediate axis. The plots also contain surface brightness maps in form of stellar simulation particles and $\chi^2$-maps as deviation between the simulation data and modeled fit. As already described in Section~\ref{sec:Choice of Regularization} one can see that \texttt{SMART} manages to fit the kinematic input data independent of the chosen projection.

\begin{figure*}[ht]
  \centering
\begin{subfigure}[c]{1.0\textwidth}
    \includegraphics[width=1.0\textwidth]{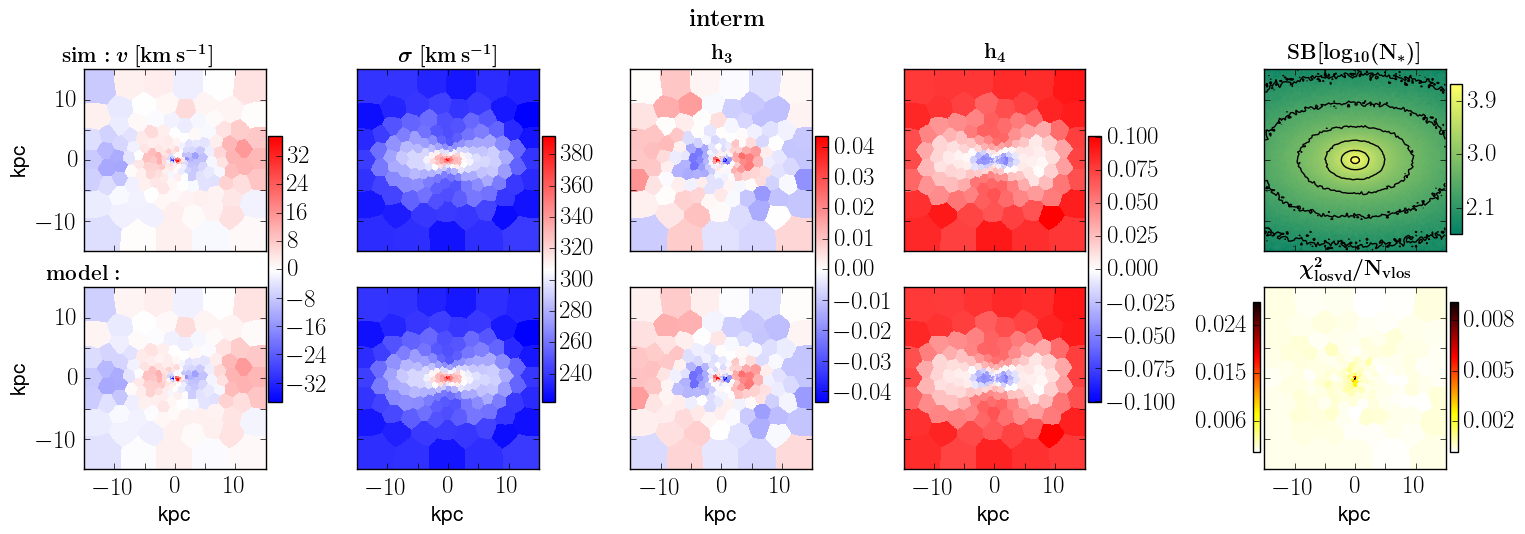} \\[\abovecaptionskip]
  \end{subfigure}
\begin{subfigure}[c]{1.0\textwidth}
    \includegraphics[width=1.0\textwidth]{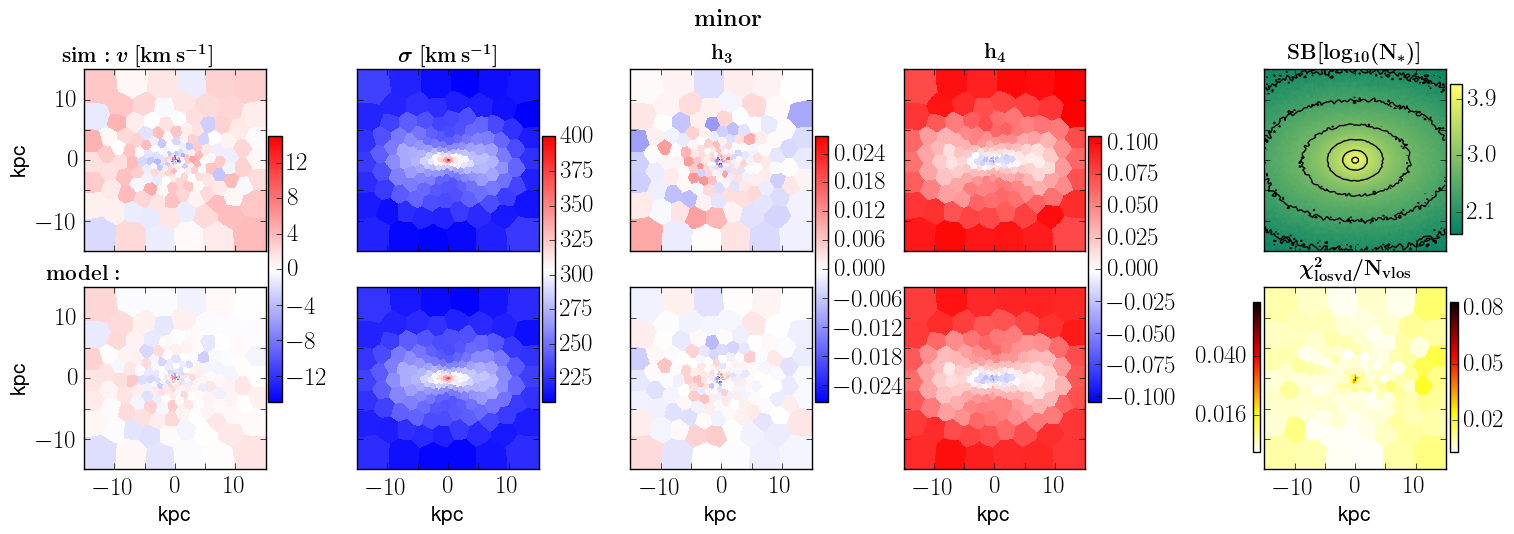} \\[\abovecaptionskip]
  \end{subfigure}
  \begin{subfigure}[c]{1.0\textwidth}
    \includegraphics[width=1.0\textwidth]{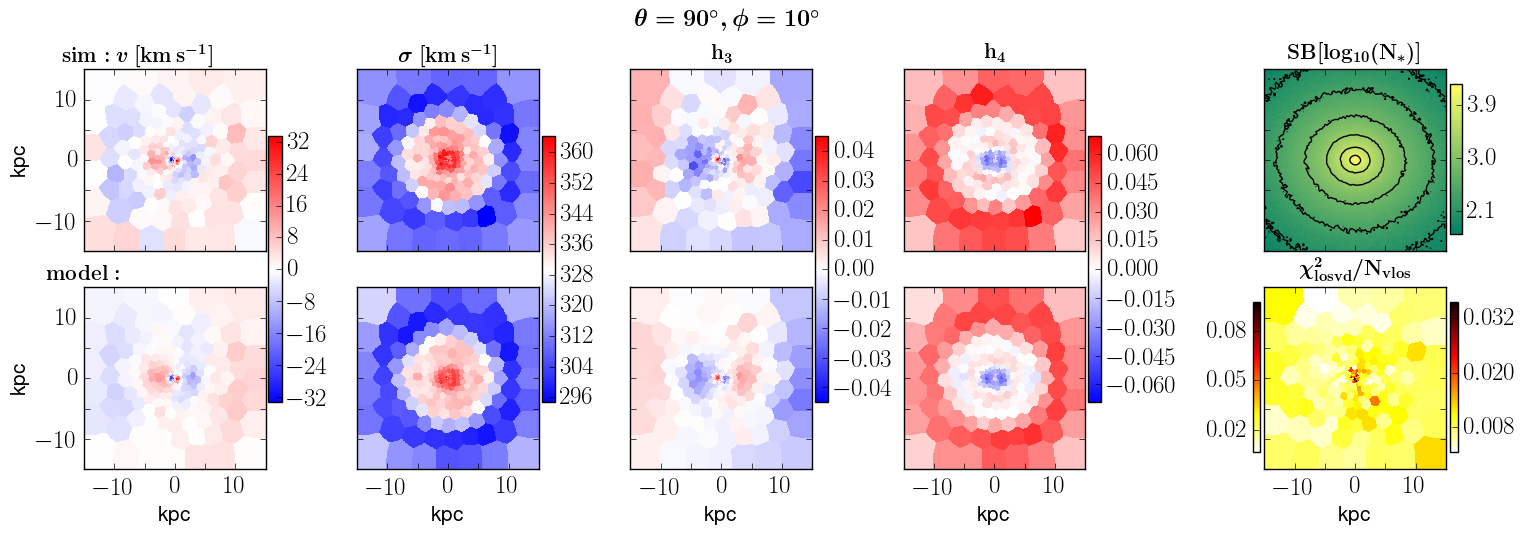} \\[\abovecaptionskip]
  \end{subfigure}
\caption{Velocity and surface brightness maps of the simulation (top row) as well as velocity maps of the model and $\chi^2$-map as deviation from the kinematic input data with the modeled fit (bottom row) for different projections. The individual projection axis can be read from the title (for the major axis projection and a more detailed caption description see Fig.~\ref{fig:Figure3}).}\label{fig:Figure19}
\end{figure*}
\begin{figure*}[htbp]\ContinuedFloat
\begin{subfigure}[c]{1.0\textwidth}
    \includegraphics[width=1.0\textwidth]{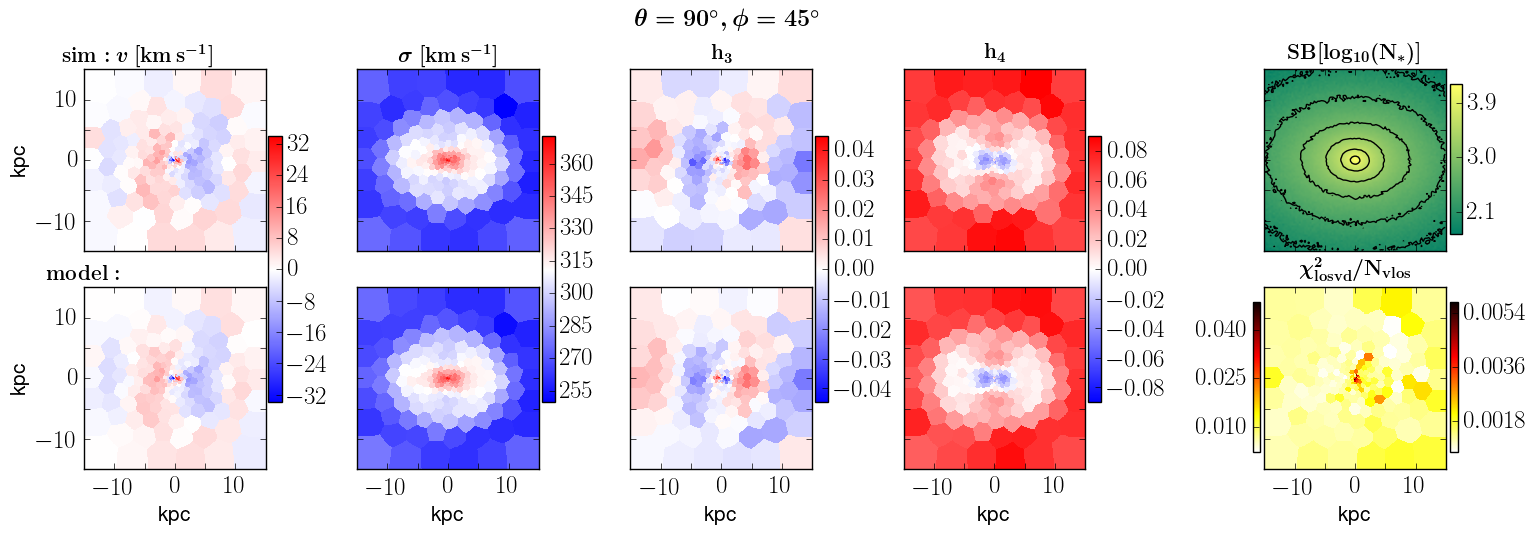} \\[\abovecaptionskip]
  \end{subfigure}
  \contcaption{(continued)}
\end{figure*}

\begin{figure*}
    \centering
    \includegraphics[width=1.0\textwidth]{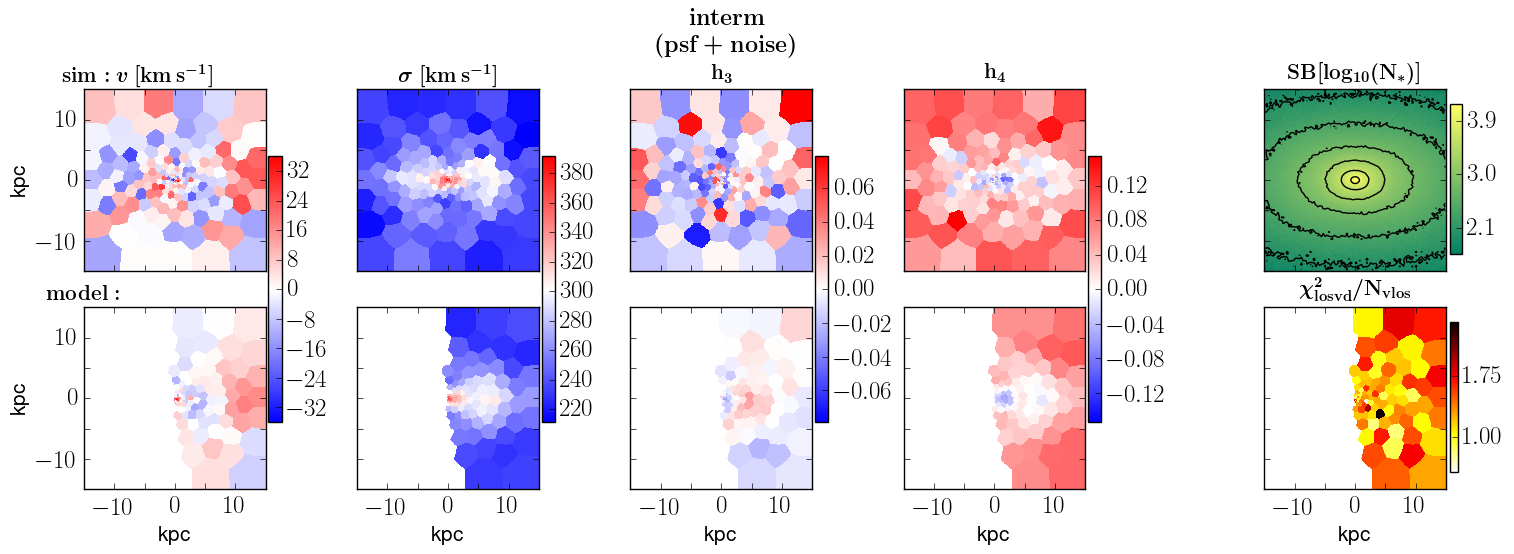}
    \caption{Velocity and surface brightness maps of the simulation (top row) as well as velocity maps of the model and $\chi^2$-map as deviation from the kinematic input data with the modeled fit (bottom row) for the noisy and psf convolved intermediate axis projection (for a more detailed description see caption in Fig.~\ref{fig:Figure3} and Section~\ref{sec:PSF convolution and noise}).}
    \label{fig:Figure21}
\end{figure*}

\section{Orbital Representation of the Phase Space}
\label{sec:Orbital Representation of the Phase Space}

For generating our orbital initial conditions we set up energy shells by calculating the maximum allowed potential energy for $N_E$ radii between the minimum allowed pericenter chosen to be $r_{\mathrm{peri,min}}=r_{\mathrm{min}}$ and the maximum allowed apocenter of $r_{\mathrm{apo,max}}=r_{\mathrm{max}}$. The radial bins are sampled by using formula~\ref{eq:radbin_cval_celz}. Each energy shell $i_{E} \in [1,N_E]$ is then subdivided in $i_E(L_z)$ sequences. We empirically find $i_E(L_z)=i_E^2$ as appropriate choice for the number of sequences per energy shell. The orbits for a specific sequence share the same initial z-component of the angular momentum $L_{z,i_E}$. These $L_{z,i_E}$-sequences are logarithmically sampled between the minimum and maximum allowed $L_z$-value in dependence of the radius $r(i_E)$ of the specific bin. \\
The final orbital initial spatial conditions $(r_i,\theta_i,\phi_i)$ are then given 
\begin{compactenum} 
 \item[-] by randomly sampling azimuthal angles $\phi_i \in [0,2 \pi]$,
 \item[-] by setting the elevation angles $\theta_i=0$ because every orbit crosses the equatorial plane and 
 \item[-] by randomly sampling possible radial values $r_i$ between the minimum and maximum allowed apsis, i.e., $r_i \in\left[r_{\mathrm{peri}}(L_{z,i_E}), r_{\mathrm{apo}}(L_{z,i_E})\right]$.
\end{compactenum}
While sampling the energy as an integral of motion is straightforward, the sampling of $L_z$ is an arbitrary way to qualify the azimuthal velocity $v_\phi$ as the parameter we actually want to sample. In analogy, one could also sample the elevation velocity $v_\theta$ in terms of the x-component of the angular momentum $L_x$, which is discussed in more detail in Section~\ref{sec:Change in orbit sampling technique}. 
The computation of $r_{\mathrm{peri}}(L_{z,i_E})$ and $r_{\mathrm{apo}}(L_{z,i_E})$ is thereby done under the wrong assumption that $L_{z,i_E}$ was conserved. Strictly speaking, this is only true for an axisymmetric potential as limiting case. However, our tests above show that this does not bias the results obtained with our orbit library. 
The remaining initial velocity conditions $(v_{r,i},v_{\theta,i},v_{\phi,i})$ can then be subsequently computed as 
\begin{compactenum}
 \item[-] $v_{\phi, i}=\frac{L_{z,i_E}}{r_{i}}$,
 \item[-] $v_{r,i} \in \left[0, v_{r, \max }\right]$ 
 \item[ ] with $v_{r, \max }=\sqrt{2\left(E-\Phi\left(r_{i}, \theta_{i}=0, \phi_{i}\right)\right)-v_{\phi, i}^{2}} \textit{  }$, where $E$ is 
 \item[ ] the total energy and $\Phi$ the potential and
 \item[-] $v_{\theta, i}=\sqrt{2\left(E-\Phi\left(r_{i}, \theta_{i}=0, \phi_{i}\right)\right)-v_{\phi, i}^{2}-v_{r, i}^{2}}$.
\end{compactenum}

More initial conditions are generated by filling the surfaces of section (SOS; \citealt{Henon64,Richstone82}). 
The surface of section, also called Poincaré section, consists of the $r_i$- and $v_{r,i}$-values of the orbits crossing the equatorial plane, i.e., $\theta_i=0$, in upwards direction. The topology of the SOS depends on the integrals of motion. The orbit library representatively fills the phase-space if the SOS is properly filled. For this, we randomly generate azimuthal angles $\phi_i$ which get divided in $N_{\mathrm{sector}}$ azimuthal sectors. For each azimuthal sector we then sample ($r_i, v_{r,i}$)-tuples which are chosen so that their distance to the nearest imprint-tuple is maximised. The remaining initial conditions $v_{\phi,i}$ and $v_{\theta,i}$ are generated as before. 
To create a homogeneous distribution of pro- and retrograde orbits, the orbits contained in the library which are later on classified as regular orbits (see Section~\ref{sec:Orbit integration and classification}) get duplicated by changing the sign of their direction of rotation. \\
For an illustration of our 5D starting space of the stellar orbits in Fig.~\ref{fig:Figure23} we show a series of 2D plots with velocity spheres on annulus sectors (chosen radial and azimuthal sections) in the equatorial plane (i.e. $\theta=0$).
Since the magnitude of the velocity vector $\sqrt{v_x^2+v_y^2+v_z^2}$ is determined by the energy $E$, it is sufficient to plot $v_x, v_y$-imprints for a certain energy shell (the remaining $v_z$-component is then given as velocity shell in dependence of the value of $E$). 
We show both, the velocity components $v_x, v_y$ of the orbital initial starting point (marked as thick dots) as well as the orbital imprints for every crossing event in the subsequent time evolution (thin dots). The velocity components are normalized by $v_{\mathrm{max}}(r,\theta,\phi)=\sqrt{2(E-\Phi(r,\theta,\phi))}$, with $\Phi$ being the gravitational potential. The different colors correspond to the different orbit types. 
Fig.~\ref{fig:Figure23} shows that the orbits of our orbit library homogeneously represent the accessible phase space. A sufficient coverage of the phase space is crucial for a reliable model. In addition, one can see that the actual starting points of the orbits can be chosen in very different ways. Our random choice is only one possibility that leads to an appropriate phase-space representation. Most of the space gets automatically filled by the orbital imprints over time.

\begin{figure*}[ht]
  \centering
\begin{subfigure}[c]{1.0\textwidth}
    \includegraphics[width=0.9\textwidth]{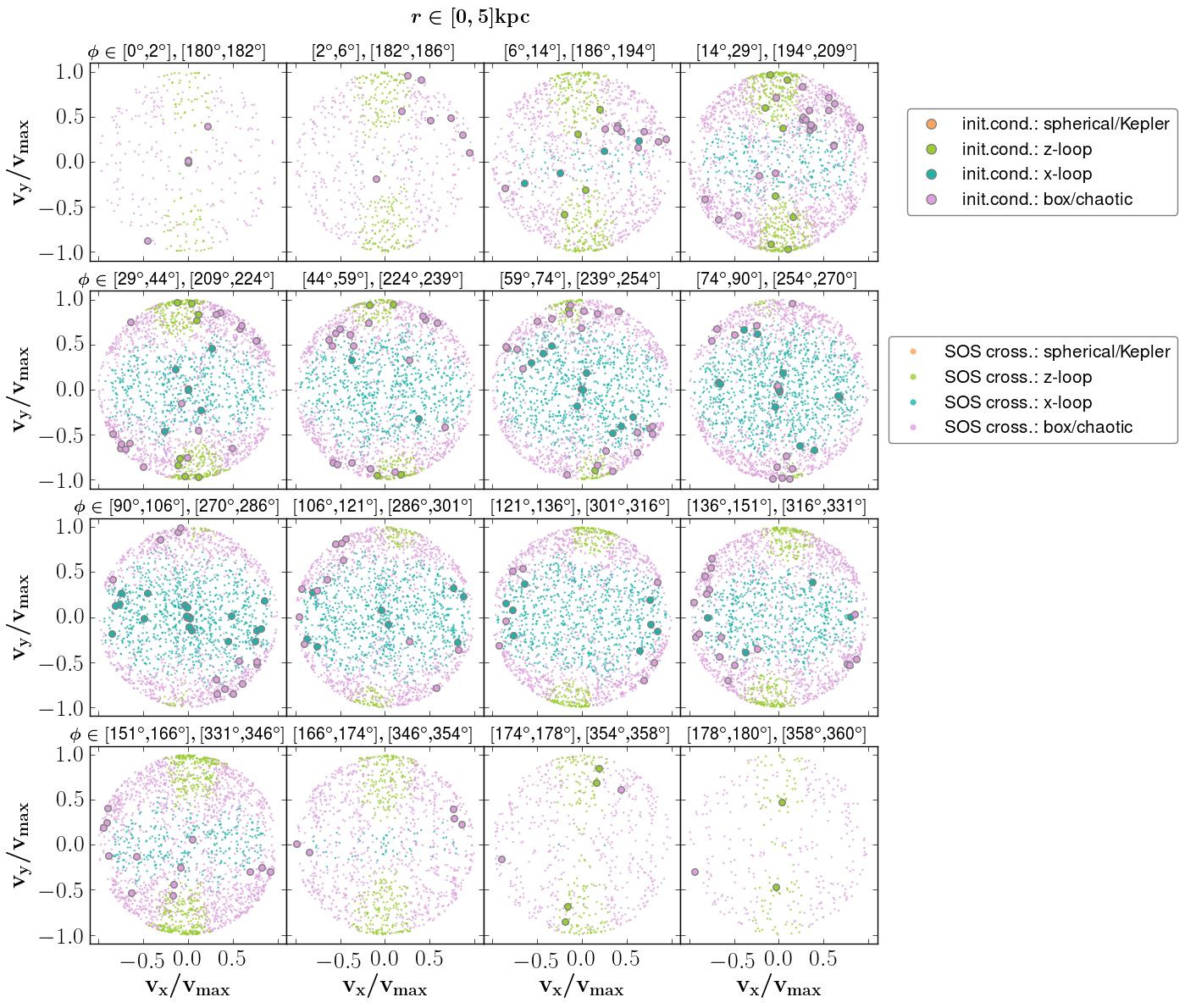} \\[\abovecaptionskip]
  \end{subfigure}
\caption{Orbital Phase-space coverage in the $x-y$ plane. We here plot the normalised velocity components $v_x, v_y$ of all orbits in a single energy shell. This includes the orbital starting points (thick dots) as well as SOS-crossings during the time evolution of the orbits (thin dots). To consider all five dimensions of our starting space we distinguish between several radial and azimuthal annulus sectors of the plane (the chosen radial and azimuthal intervals are labelled above the individual panels). The orbit library produces a representative coverage of the phase-space because the orbital imprints homogeneously fill the area enclosed by the unit circle. }\label{fig:Figure23}
\end{figure*}

\begin{figure*}[htbp]\ContinuedFloat
\begin{subfigure}[c]{0.9\textwidth}
    \includegraphics[width=1.0\textwidth]{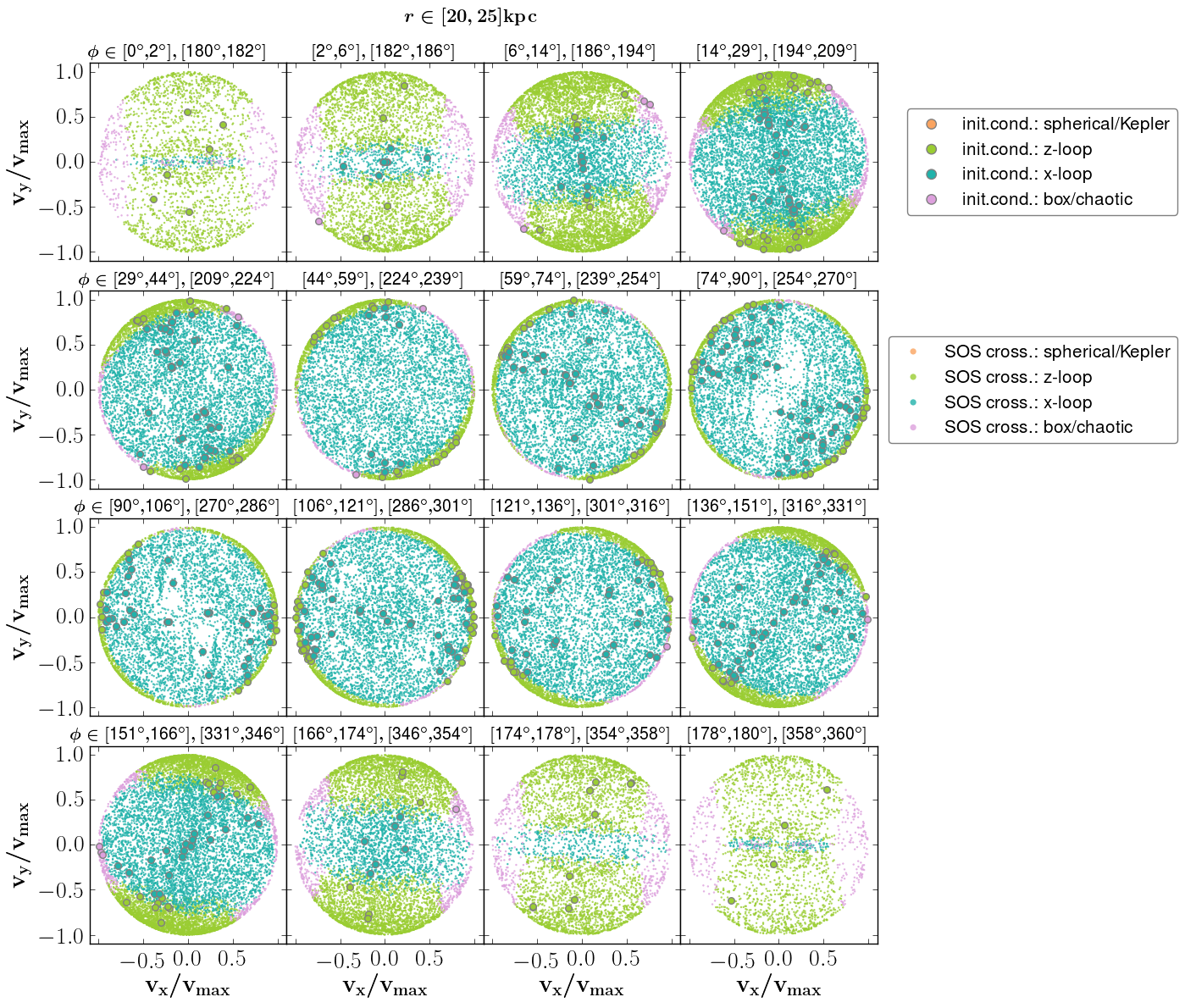} \\[\abovecaptionskip]
  \end{subfigure}
  \contcaption{(continued)}
\end{figure*}

\begin{figure*}[htbp]\ContinuedFloat
\begin{subfigure}[c]{0.9\textwidth}
    \includegraphics[width=1.0\textwidth]{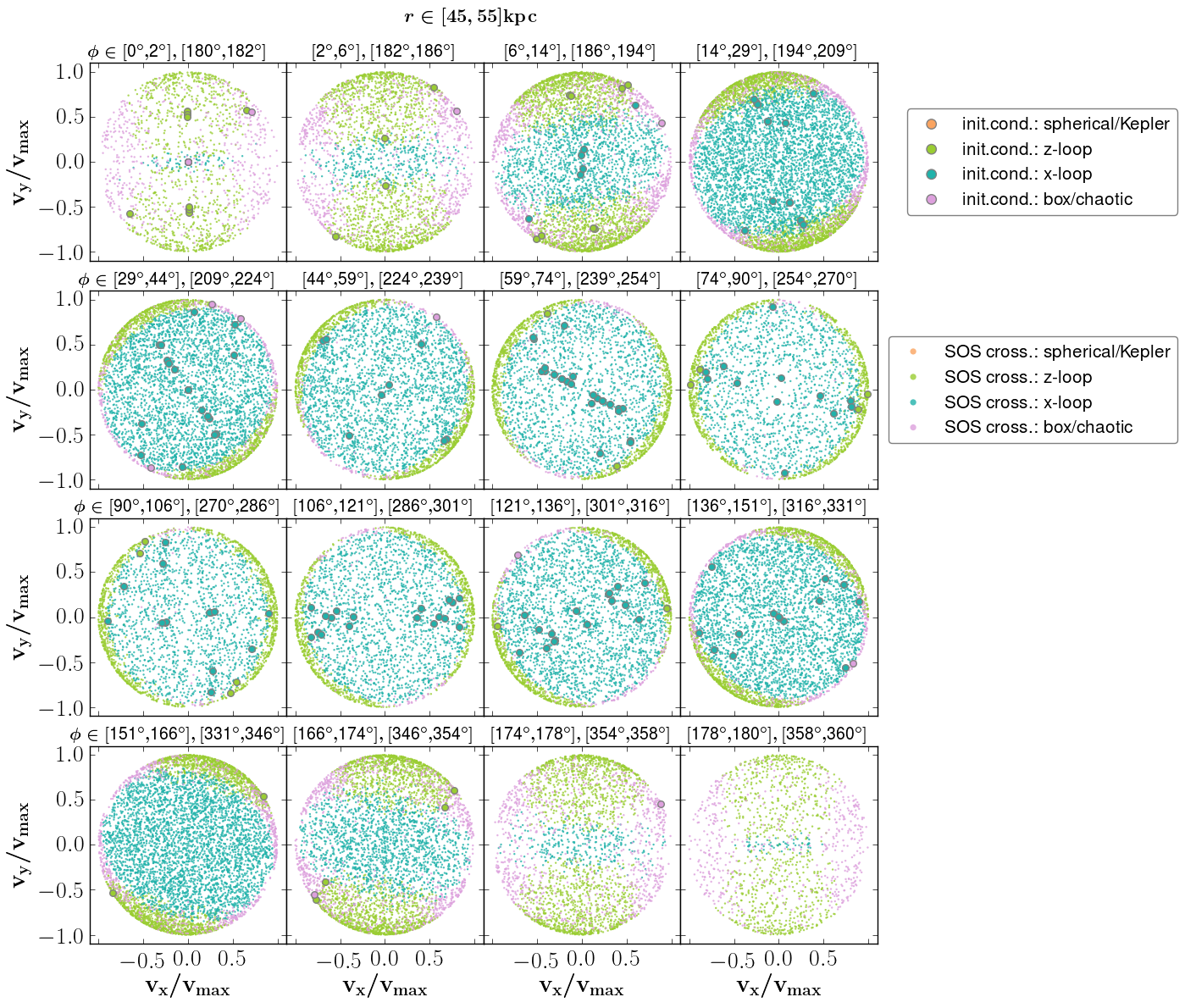} \\[\abovecaptionskip]
  \end{subfigure}
  \contcaption{(continued)}
\end{figure*}

\section{Optimization Algorithm and Test}
\label{sec:Optimization Algorithm and Testing}
To find the unique set of orbital weights that maximises $\hat{S}$
our code conceptionally seeks for a solution
\begin{equation}
\nabla_w \hat{S} = 0,
\end{equation}
subject to the constraints (\ref{eq:densityconstraints}).

Technically, we follow \citet{Richstone88} and treat the
orbital weights $\vect{w}$ and the kinematical model predictions
\begin{equation}
\vect{y} \equiv \kdat - \kmod
\end{equation}
as independent variables. The actual dependency
$\vect{y} = \vect{y}(\vect{w})$ (Equation~\ref{eq:kinmat}) is added to
the constraints.

Thus, let
\begin{equation}
\vect{x} \equiv 
\begin{pmatrix}
\vect{w} \\
\vect{y} \\
\end{pmatrix}
\end{equation}
be the $\norb+\nkin$ new variables, then the model solves
\begin{equation}
\label{eq:theequation}
\nabla_x \hat{S} - \mathrm{C}^\mathrm{T} \cdot \lambda = 0
\end{equation}
for the $x_i$ and for $\nkin+\nphot$ Lagrange multipliers $\lambda_j$
(combined in the vector $\vect{\lambda}$). The matrix
\begin{equation}
\mathrm{C} \equiv 
\begin{pmatrix}
\mathrm{\phorb} & \mathrm{0} \\
\mathrm{\korb} & \mathrm{1} \\
\end{pmatrix}
\end{equation}
contains the constraint equations
\begin{equation}
\label{eq:eigentlich}
\begin{pmatrix}
\phdat\\
\kdat\\
\end{pmatrix}
= \mathrm{C} \cdot \vect{x}.
\end{equation}

Equation~\ref{eq:theequation} is nonlinear and solved with a Newton
method:
\begin{equation}
\vect{x}^{(n+1)} = \vect{x}^{(n)} - \left(
\mathrm{\frac{\mathrm{d}g}{\mathrm{d}x}}
\right)^{-1} \vect{g},
\end{equation}
or,
\begin{equation}
\vect{\Delta x}^{(n)} = - \left(
\mathrm{\frac{\mathrm{d}g}{\mathrm{d}x}}
\right)^{-1} \vect{g},
\end{equation}
where 
\begin{equation}
g \equiv \nabla_x \hat{S} - \mathrm{C}^\mathrm{T} \cdot \lambda.
\end{equation}
At each iteration, the Lagrange multipliers are updated via
(\ref{eq:eigentlich}),
\begin{equation}
\begin{pmatrix}
\Delta \phdat\\
\Delta \kdat\\
\end{pmatrix}
= \mathrm{C} \cdot \Delta \vect{x},
\end{equation}
which leads to
\begin{equation}
\mathrm{A} \cdot \vect{\lambda} = 
\begin{pmatrix}
\Delta \phdat\\
\Delta \kdat\\
\end{pmatrix}
+ \mathrm{C} \cdot \left(
\mathrm{\frac{\mathrm{d}g}{\mathrm{d}x}}
\right)^{-1} \cdot \nabla_x \hat{S},
\end{equation}
\begin{equation}
\mathrm{A} \equiv 
\mathrm{C} \left(
\mathrm{\frac{\mathrm{d}g}{\mathrm{d}x}}
\right)^{-1} \mathrm{C}^{\mathrm{T}}.
\end{equation}
The Jacobian $\mathrm{\frac{\mathrm{d}g}{\mathrm{d}x}}$ is diagonal and
its inverse is easy to compute.
The only matrix that needs to be inverted numerically is
$\mathrm{A}$ and its dimension is $\nkin + \nphot < \norb$. 

Modulo the scaling by $(\mathrm{\frac{\mathrm{d}g}{\mathrm{d}x}})^{-1}$, $\mathrm{A}$ reads
\begin{equation}
\mathrm{A} = 
\begin{pmatrix}
\mathrm{\phorb} \, \mathrm{\phorb}^{\mathrm{T}} & \mathrm{\phorb} \, \mathrm{\korb}^{\mathrm{T}} \\
\mathrm{\korb} \, \mathrm{\phorb}^{\mathrm{T}} & \mathrm{\korb} \, \mathrm{\korb}^{\mathrm{T}} + \mathrm{1} \\
\end{pmatrix}.
\end{equation}
The required linear algebra operations are implemented through the LAPACK library \citep{Anderson99}. 

In practice, to find the best-fitting orbit model, we start with a
very small value $\alpha_0$ and then iteratively increase $\alpha$
until $\chi^2$ does not change anymore. Everytime we change
$\alpha \to \alpha + \mathrm{d}\alpha$ the unique solution to
Equation~\ref{eq:costfunc} will shift slightly from $\vect{w}$ to
$\vect{w} + \mathrm{d}\vect{w}$ in the space of the orbital
weights. Because $f \equiv - \hat{S}$ is differentiable with respect to $\alpha$ we can
keep the new solution $\vect{w} + \mathrm{d} \vect{w}$ in an
arbitrarily small neighborhood around $\vect{w}$ by limiting
$\mathrm{d}\alpha$ accordingly. For strictly convex $f$, the Newton
method converges locally (and the convergence is quadratic). This
means that once we have found a solution for some value of $\alpha$,
the convergence of the Newton steps for neighbouring
$\alpha + \Delta \alpha$ is guaranteed (and fast) if we use the solution at
$\alpha$ as starting point for the search at
$\alpha + \Delta \alpha$: it is only a question of limiting
$\Delta \alpha$, if necessary. In this way, we can iteratively solve
for any value of $\alpha$ starting from $\alpha_0$. In practice, we increase $\alpha$ by a
factor of 2 in each iteration. Intermediate steps with a smaller
$\alpha$ increment are inserted in case
an iteration has not converged (which almost never happens).

At the initial $\alpha_0$ we start from a homogeneous
distribution of orbital weights. Methods with guaranteed
convergence even from starting points outside of the constraint
conditions have been developed. We do not use such methods, because in
practice we always find the solution of Equation~\ref{eq:theequation}
even from an initially homogeneous distribution of orbital
weights. Since the solution is global and unique for each $\alpha$, 
convergence at any step of the $\alpha$ sequence can be tested  based on Equation~\ref{eq:theequation} itself. \\

To test our implementation we reconstruct the kinematical data of a toy model with randomised orbital weights. The toy model is constructed as a maximum-entropy model (i.e. through maximisation of $\hat{S}$ in eq.~\ref{eq:costfunc} with $\alpha = 0$). The bias factors $\omega_i$ for the orbital weights are set to $\omega_i = 10^{r_i}$, where $r_i \in [0,3]$ are random numbers. After maximisation of $\hat{S}$ subject to the density constraints, the resulting orbital weights $w_i$ satisfy the density constraints of the N-body simulation. Due to these density constraints, the $w_i$ are not directly proportional to the random  $ \omega_i$, yet they are still very strongly affected by them and thus we call this toy model with randomised orbital weights RANDOM in the following. We adapt $r_{\mathrm{peri,min}}=r_{\mathrm{min}}=0.05\mathrm{\,kpc}$ and $r_{\mathrm{apo,max}}=r_{\mathrm{max}}=21\mathrm{\,kpc}$. 
If an orbit conserves the sign of at least one angular-momentum component, then this orbit has a duplicate companion orbit with opposite direction of rotation in the library (this concerns tube orbits and spherical orbits; see  also Section~\ref{sec:Orbital Representation of the Phase Space}). Let $i^+$ and $i^-$ denote the orbital indices of such a pair of orbits, where $i^+$ refers to the orbit with initial velocity components $v_r, v_\theta, v_\phi >0$. To increase the degree of difficulty for reconstructing the RANDOM model, we modify the weights of orbits in such a pair: 
\begin{equation}
\label{eq:setting weights}
\begin{aligned}
    w_{i^+,\mathrm{ran}} & =(1-c_w) \cdot (w_{i^+} + w_{i^-}), \\
    w_{i^-,\mathrm{ran}} & = c_w \cdot (w_{i^+} + w_{i^-}).
\end{aligned}
\end{equation}
By using $c_w=10^{-3}$, Equation~\ref{eq:setting weights} leads to increased weights $w_{i^+,\mathrm{ran}}$ and almost eliminated weights $w_{i^-,\mathrm{ran}}$, corresponding to an internal rotation of the model (see also \citealt{Thomas07}). We then modelled the LOSVDs of the major axis projection of this RANDOM toy model with \texttt{SMART} in order to test our solver algorithm for the orbital weights.

Because the $\chi^2$ minimisation is non-unique (cf. Section~\ref{sec:The quasi-uniqueness of the anisotropy reconstruction when fitting full LOSVDs}), the solution of eq.~\ref{eq:costfunc} will in general not occur at the true weights $w_{i,\mathrm{ran}}$ of the RANDOM model. We therefore tried two different configurations: (i) we used the default orbital bias factors $\omega_i=1$ (Shannon entropy) and (ii) we set $\omega_i=\mathrm{exp(1) \cdot w_{i,\mathrm{ran}}}$. In the second case, we push the maximum entropy solution towards the original, randomly generated weights. Only in this case, we can predict the location of the global solution analytically and test our implementation rigorously. For ensuring fair conditions we set the initial weights to $w_{i,\mathrm{ini}}=1$ in both cases when we start the minimisation. An efficient optimization algorithm should be able to find two equivalent degenerate solutions, one for case (i) and a different one for case (ii). The true orbital weights of the RANDOM model should be recovered (only) in the latter case.  

\begin{figure}
    \centering
    \includegraphics[width=0.7\textwidth]{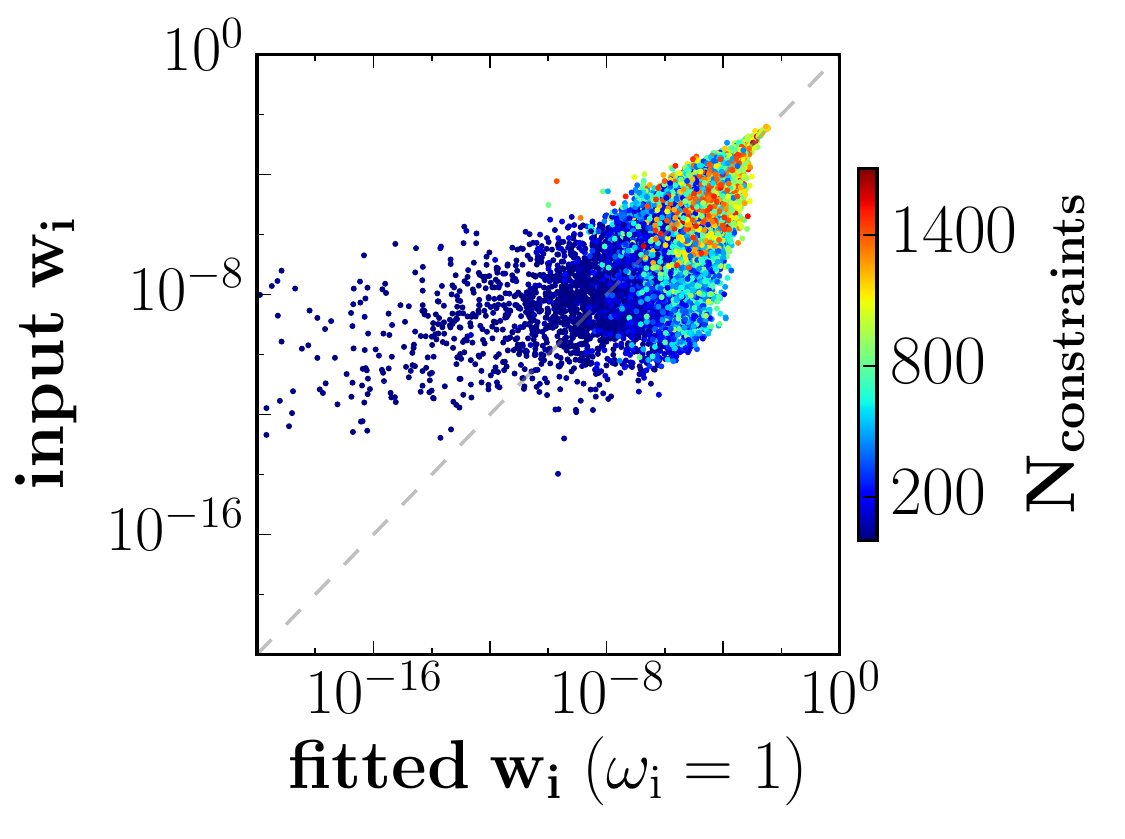}
    \caption{Recovery of the orbital weights of the RANDOM test model described in Section~\ref{sec:Optimization Algorithm and Testing}. Each dot represents one orbit with the color indicating the number of LOSVD data points $N_\mathrm{constraints}$ that the orbit contributes to. For the recovery of the weights we fitted the kinematics of the RANDOM model using our fiducial constant bias factors $\omega_i=\mathrm{const.}$ (Shannon entropy). Due to the degeneracy in the $\chi^2$ minimisation, the recovered orbital weights differ from the original weights. Larger differences are observed for orbits which are less constrained by data (blue points). 
    }
    \label{fig:Figure25}
\end{figure}

\begin{figure}
    \centering
    \includegraphics[width=0.7\textwidth]{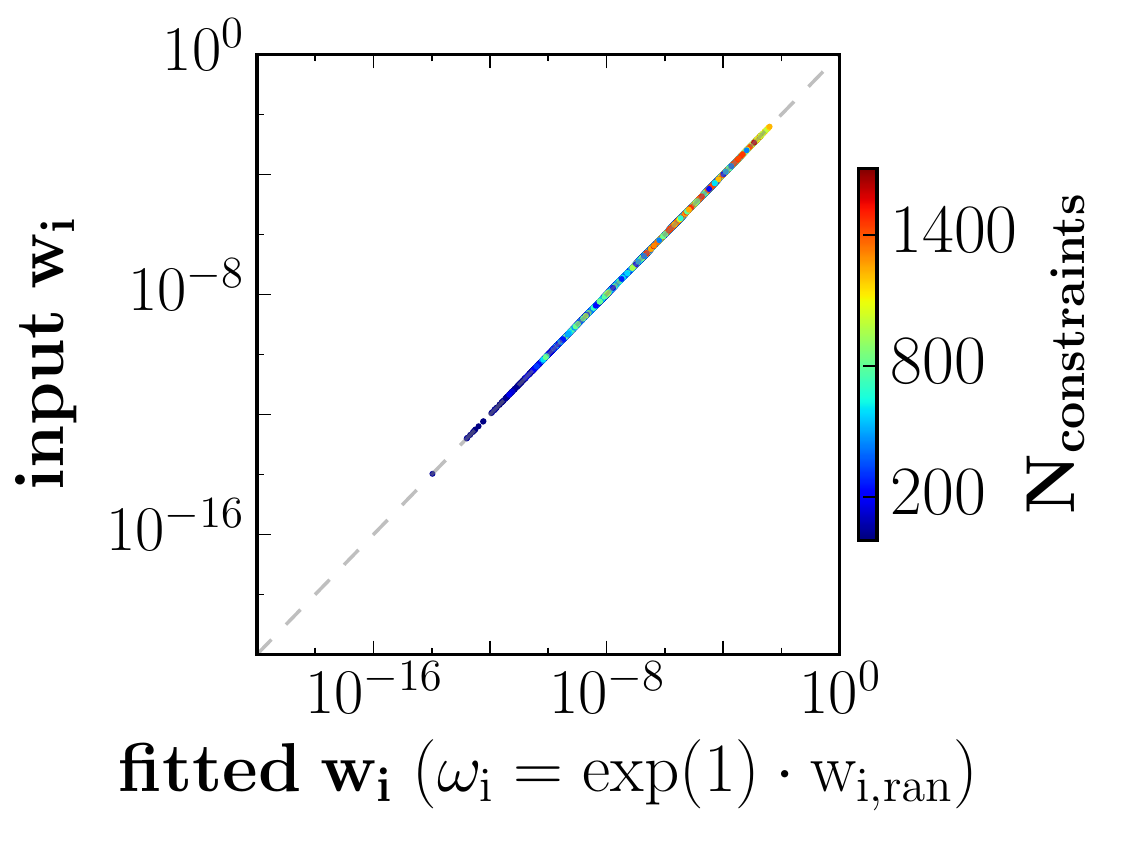}
    \caption{Same as Fig.~\ref{fig:Figure25}, but here we fitted the RANDOM model using $\omega_i=\mathrm{exp(1) \cdot w_{i,\mathrm{ran}}}$. With these bias factors the unique maximum of eq.~\ref{eq:costfunc} should occur at the original weights $w_{i,\mathrm{ran}}$ of the RANDOM model. And indeed, even though we start the fit with an initial guess where all the weights are equal, the true solution is reproduced with very high precision. This demonstrates the robustness and high accuracy of our optimization algorithm. The two different fits presented here and in Fig.~\ref{fig:Figure25} illustrate how the maximum entropy technique with variable $\omega_i$  can be used to sample different solutions that minimise the $\chi^2$. }
    \label{fig:Figure26}
\end{figure}

\texttt{SMART} is indeed able to fit the input data for both cases very well. We stopped the minimisation at $\bar{\chi^2}=2\times10^{-9}$ for $\omega_i=1$ and $\bar{\chi^2}=5\times10^{-9}$ for $\omega_i=exp(1) \cdot w_{i,\mathrm{ran}}$. 
Figure~\ref{fig:Figure25} shows the comparison between the true orbital weights of the RANDOM model and the recovered weights when using $\omega_i=1$ (case (i) above). As expected, these weights {\it differ} from the RANDOM weights, since the $\chi^2$ minimisation and, hence, the recovery of the entire phase-space distribution function, is non-unique. Orbits that contribute to a larger number of LOVSD data points and, in turn, are better constrained by the data have fitted weights that tend to be closer to the true ones. 
Figure~\ref{fig:Figure26} shows the orbital weights of the input model and the modelled fit using $\omega_i=\mathrm{exp(1) \cdot w_{i,\mathrm{ran}}}$.
As explained above, in this case we expect the recovered solution to exactly match with the RANDOM model.  Figure~\ref{fig:Figure26} demonstrates the high quality of our solver algorithm since the fitted weights indeed match exactly with the RANDOM weights ($\Delta \bar w=4\times10^{-9}$). 

The two different fits presented in Figs.~\ref{fig:Figure25} and \ref{fig:Figure26} illustrate how the maximum entropy technique with variable $\omega_i$  can be used to sample different solutions that are equivalent with respect to the $\chi^2$ minimisation. \\

\begin{figure}
    \centering
    \includegraphics[width=1.0\textwidth]{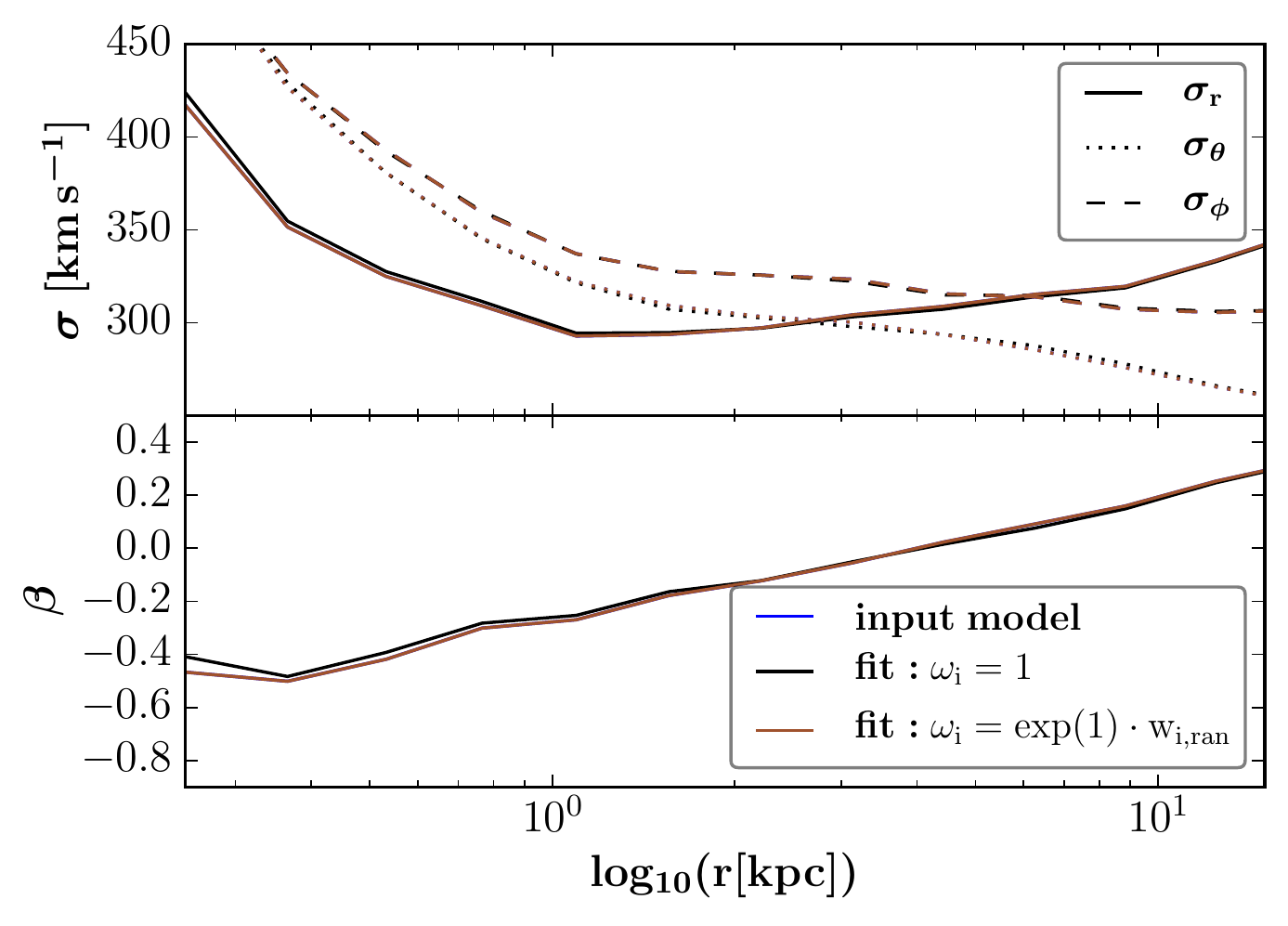}
    \caption{Internal velocity structure of the two fits presented in Fig.~\ref{fig:Figure25} (black lines) and \ref{fig:Figure26} (red lines). Despite the different orbital weight distributions, the anisotropy in the second order velocity moments is almost equally well recovered in both cases. This provides evidence that when using the full information contained in the entire LOSVDs, the remaining degeneracies in the recovery of the distribution function of triaxial galaxies (i.e. the differences in the detailed population of the various orbits) have only little impact on the 'macroscopic' internal galaxy properties like the anisotropy $\beta$. This, in turn, is crucial for a precise mass reconstruction.}
    \label{fig:Figure27}
\end{figure}

In Fig.~\ref{fig:Figure27} we compare the 'macroscopic' internal velocity structure of these two fits. Even though the detailed population of the orbits (i.e. the exact phase-space distribution function) of both models differ, they both reproduce the internal moments of the RANDOM model very well ($rms_\sigma=5\times10^{-3}$ for $\omega_i=1$ and $rms_\sigma=3\times10^{-5}$ for $\omega_i=\mathrm{exp(1) \cdot w_{i,\mathrm{ran}}}$). 
As further discussed in Section~\ref{sec:The quasi-uniqueness of the anisotropy reconstruction when fitting full LOSVDs}, this suggests, that \textit{when fitting the full LOSVDs} the remaining degeneracies in the recovery of the distribution function of a triaxial galaxy (with known normalized densities) show only little effect on the 'macroscopic' galaxy parameters of interest. 


\section{Orbit Fraction Recovery when testing the Uniqueness of the Anisotropy Recovery}
\label{sec:Remaining Orbit Fractions of Experiment to testing the Uniqueness of the Anisotropy Recovery}
\begin{figure*}
    \centering
    \includegraphics[width=0.9\textwidth]{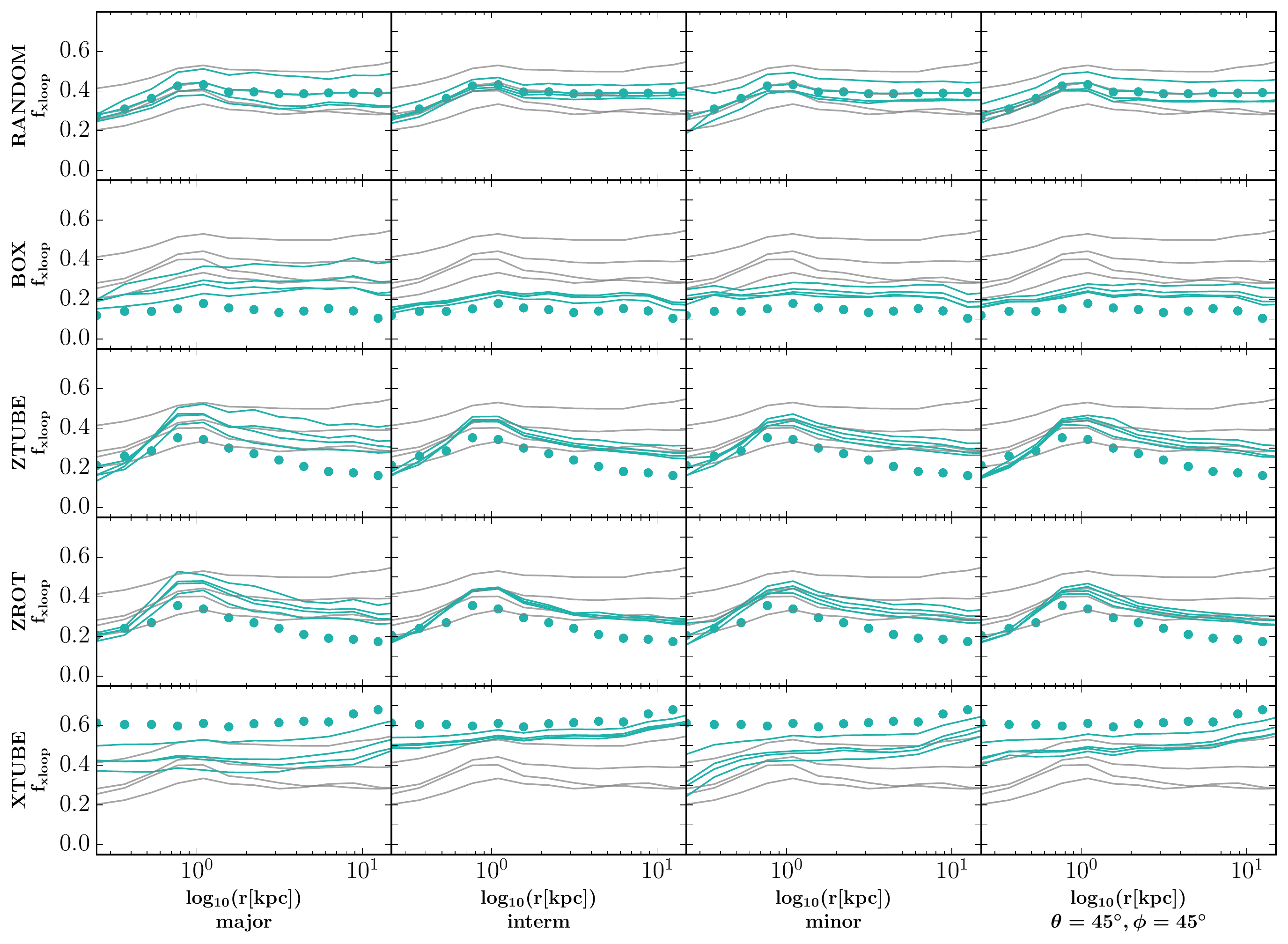}
    \caption{Recovery of the $x$-loop orbit fractions of different input toy models by models with different entropy methods. We here show the same analysis as in Figure~\ref{fig:Figure13} and Figure~\ref{fig:Figure14} but now for the reconstruction of the fraction of orbits classified as $x$-tubes. Independent of the tested projection, the $x$-loop fractions of the individual input toy models are well recovered by the models using different entropy methods.}
    \label{fig:Figure28}
\end{figure*}

\begin{figure*}
    \centering
    \includegraphics[width=0.9\textwidth]{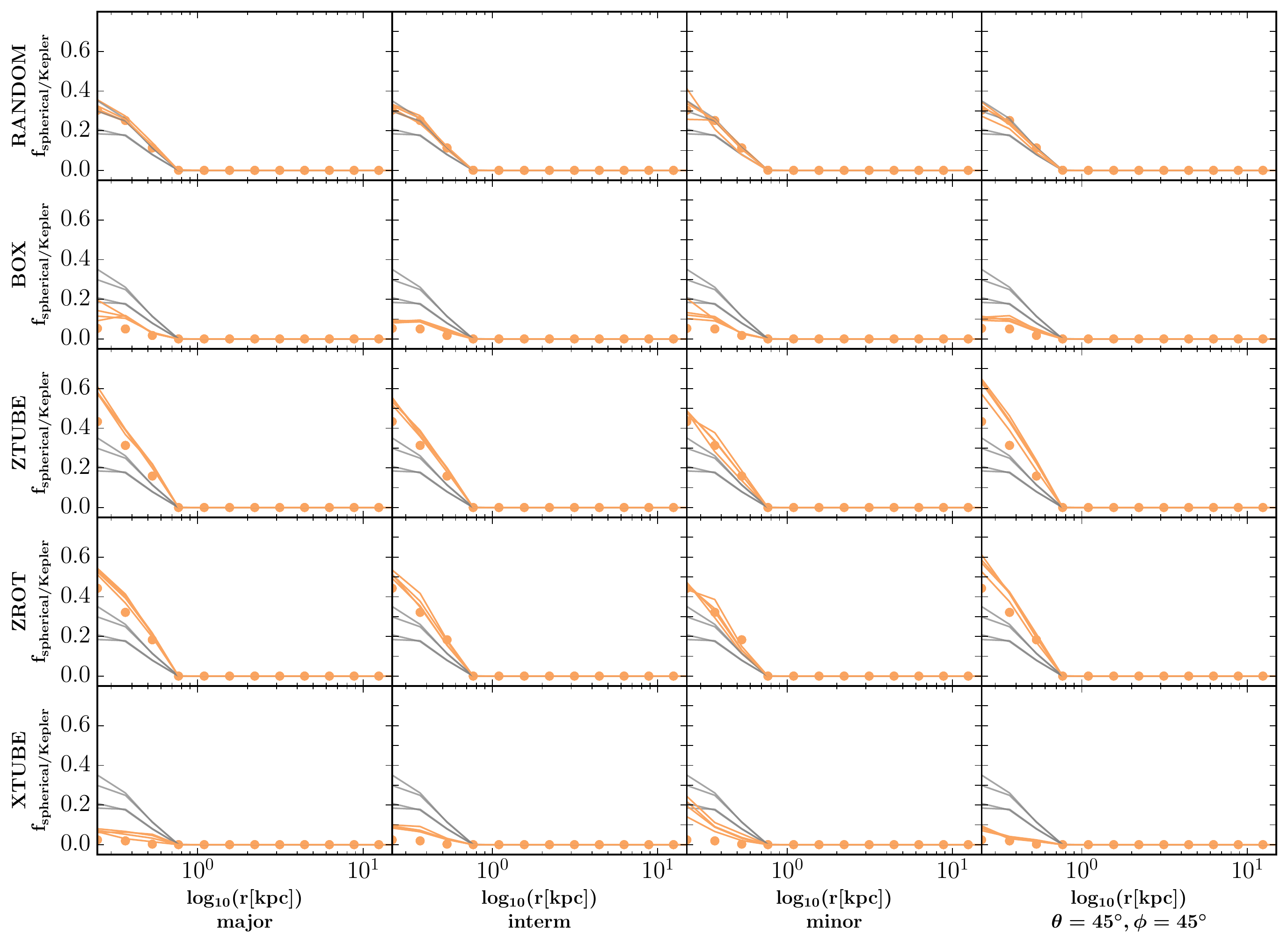}
    \caption{Recovery of the box/chaotic orbit fractions of different input toy models by models with different entropy methods. For a more detailed caption description see Fig.~\ref{fig:Figure13} and~\ref{fig:Figure28}.}
    \label{fig:Figure29}
\end{figure*}

\begin{figure*}
    \centering
    \includegraphics[width=0.9\textwidth]{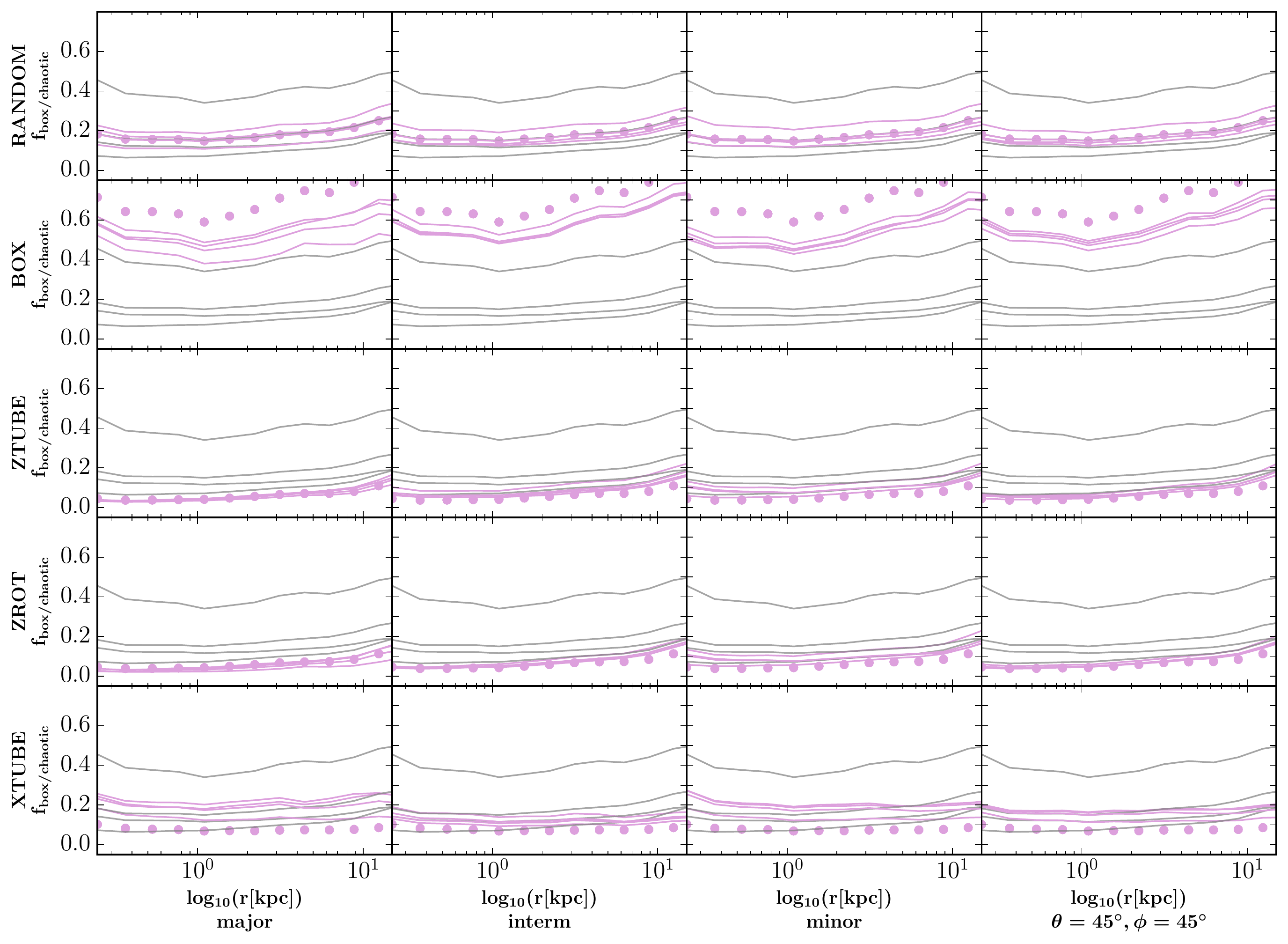}
    \caption{Recovery of the spherical/Kepler orbit fractions of different input toy models by models with different entropy methods. For a more detailed caption description see Fig.~\ref{fig:Figure13} and~\ref{fig:Figure28}.}
    \label{fig:Figure30}
\end{figure*}

We here show the remaining orbit fraction recovery plots for the analysis described in Section~\ref{sec:Experiment to testing the Uniqueness of the Anisotropy Recovery}. The recovery of the $z$-loops was already shown in Fig.~\ref{fig:Figure14} of that section.\\
Here, we plot the recoveries of the fraction of orbits classified as $x$-loops (see Fig.~\ref{fig:Figure28}), box/chaotic orbits (see Fig.~\ref{fig:Figure29}) as well as spherical/Kepler orbits (see Fig.~\ref{fig:Figure30}). Independent of the line of sight (different rows) or chosen entropy method (black lines), the orbit class fractions of the individual input toy-models (different rows) are well recovered. 


\bsp
\label{lastpage}
\end{document}